\documentclass[twocolumn, showpacs, amsmath, amssymb]{revtex4}

\usepackage[utf8]{inputenc}

\usepackage{latexsym}
\usepackage{pifont}

\usepackage{soulutf8}

\usepackage{array}
\usepackage{dcolumn}

\usepackage{stackrel}

\usepackage{ulem}
\usepackage{color}

\usepackage{amsmath}
\usepackage{amssymb}
\usepackage{amsfonts}
\usepackage{amsthm}
\usepackage{bbm}
\usepackage{bm}

\usepackage{braket}
 \usepackage{wasysym}

\usepackage{graphicx}
\graphicspath{{./Assets/GFX/}}

\usepackage[caption=false]{subfig}

\usepackage{fancyhdr}
 \pdfoutput=1

\begin{document}


\title{
	Progress and perspectives on composite laser-pulses spectroscopy \\
	for high-accuracy optical clocks.
}

\author{
	Thomas Zanon-Willette $^{1}$\footnote{E-mail address: thomas.zanon@upmc.fr\\},
	Rémi Lefevre $^{1}$,
	Rémi Metzdorff $^{1,2}$,
	Nicolas Sillitoe $^{1,2}$,
	Sylvain Almonacil $^{1,3}$, \\
	Marco Minissale $^{1,4}$,
	Emeric de Clercq $^{5}$,
	Alexey V. Taichenachev $^{6,7}$,
	Valeriy I. Yudin $^{6,7}$,
	Ennio Arimondo $^{8}$
}

\affiliation{
	$^{1}$ LERMA, Observatoire de Paris, PSL Research University, CNRS, Sorbonne Universités,
		UPMC Univ. Paris 06, F-75005, Paris, France \\
	$^{2}$ Present address: Laboratoire Kastler Brossel, UPMC-Sorbonne Universités, CNRS, ENS-PSL Research University,
		Collège de France, 4, place Jussieu, F-75252 Paris, France \\
	$^{3}$ Present address: Institut d'Optique Graduate School, 2 avenue Augustin Fresnel, 91127 Palaiseau Cedex, France \\
	$^{4}$ Present address: Aix Marseille Université, CNRS, PIIM UMR 7345, 13397 Marseille, France \\
	$^{5}$ LNE-SYRTE, Observatoire de Paris, PSL Research University, CNRS, Sorbonne Universités,
		UPMC Univ. Paris 06, 61 avenue de l'Observatoire, 75014 Paris, France \\
	$^{6}$ Novosibirsk State University, ul. Pirogova 2, Novosibirsk 630090, Russia \\
	$^{7}$ Institute of Laser Physics, SB RAS, pr. Akademika Lavrent'eva 13/3, Novosibirsk 630090, Russia \\
	$^{8}$ Dipartimento di Fisica "E. Fermi", Università di Pisa, Largo. B. Pontecorvo 3, 56122 Pisa, Italy
}

\thispagestyle{empty}

\begin{abstract}
	Probing an atomic resonance without disturbing it is an ubiquitous issue in physics. This problem is critical in
	high-accuracy spectroscopy or for the next generation of atomic optical clocks. Ultra-high
	resolution frequency metrology requires sophisticated interrogation schemes and robust protocols handling pulse length errors and residual	 frequency detuning offsets .
	This review reports recent progress and perspective in such schemes, using sequences of composite laser-pulses
	tailored in pulse duration, frequency and phase, inspired by NMR techniques and quantum information processing.
	After a short presentation of Rabi technique and NMR-like composite pulses allowing efficient compensation of
	electromagnetic field perturbations to achieve robust population transfers, composite laser-pulses are investigated
	within Ramsey's method of separated oscillating fields in order to generate non-linear compensation of probe-induced
	frequency shifts.
	Laser-pulses protocols such as Hyper-Ramsey (HR), Modified Hyper-Ramsey (MHR), Generalized Hyper-Ramsey (GHR)
	and hybrid schemes are reviewed. These techniques provide excellent protection against both probe induced
	light-shift perturbations and laser intensity variations. More sophisticated schemes generating synthetic
	frequency-shifts are presented. They allow to reduce or completely eliminate imperfect correction of probe-induced
	frequency-shifts even in presence of decoherence due to the laser line-width.
	Finally, two universal protocols are presented which provide complete elimination of probe-induced frequency shifts in
	the general case where both decoherence and relaxation dissipation effects are present by using exact analytic
	expressions for phase-shifts and the clock frequency detuning.
	These techniques might be applied to atomic, molecular and nuclear frequency metrology, mass spectrometry as well as
	precision spectroscopy.	
\end{abstract}

\maketitle

\tableofcontents

\section{Introduction}
\label{sec:I-Introduction}
The history of very high precision spectroscopy started in the 1930s with the atomic and molecular beam resonance method proposed
by I.I. Rabi to improve the resolution of nuclear moment measurements~\cite{Rabi:1938}. The original experiment was based on a
combination of inhomogeneous fields and a rotating field to flip nuclear spins and measure the magnetic moment of the species.
With enough interaction time and under a quasi-resonant irradiation, coherent oscillations can be achieved between targeted quantum
states which are manipulated by an external coherent field with tunable amplitude and frequency.
In such a case, the final two-level occupation probabilities can be controlled with an adequate selection of a radio-frequency
field and pulse duration. By scanning the frequency of the electromagnetic excitation around the exact resonance, a narrow
spectroscopic transition is observed which can be used to obtain a very narrow discriminator, stabilizing the frequency of a local
oscillator on the atomic frequency for instance, thus achieving an atomic clock. The Rabi technique provided plenty of information
not only on atomic and molecular structure, but also on nuclear properties~\cite{Rabi:1939}.

\indent To improve the frequency resolution, Ramsey designed a scheme, where he replaced the single oscillatory field
by a double microwave excitation pulse separated by a free evolution time~\cite{Ramsey:1950}. The probing electromagnetic field
perturbation on the atomic transition itself was reduced by averaging the probe induced frequency-shift over the entire sequence
of pulses separated in space or time~\cite{Ramsey:1956}. Such a technique has drastically impacted time and frequency metrology
with microwave atomic clocks since the 1950s~\cite{Essen:1955,Ramsey:1990,Vanier:1989}. This protocol also provides the highest resolution for evaluation and reduction of systematic frequency shifts perturbing
an atomic transition.

\indent Atomic optical clocks are today recognized to be ideal platforms for highly accurate frequency measurements, leading to very
stringent tests for physical theories and variations of the fundamental constants with time, and also for quantum simulation investigations, as reviewed in
~\cite{Ludlow:2015}. Depending on the selected atomic species used to achieve stable and accurate optical frequency standards,
single trapped ion clocks \cite{Rosenband:2008,Margolis:2009,Chou:2010-2} and neutral atoms lattice clocks
\cite{Ye:2008,Derevianko:2011,Katori:2011} have been characterized over many years, reducing systematic uncertainties to a value surpassing current microwave atomic frequency standards.
Very long storage time of Doppler and recoil-free quantum particles have been obtained using laser cooling techniques and a relative accuracy level below $10^{-18}$ will most likely be achieved in the near future. This uncertainty reduction
will be  obtained thanks to a combination of technological advances, but also to the development of ad-hoc protocols.

\indent  The present work reports the recent advances on the development of those protocols, where the interrogation process is composed by a sequence of laser pulse.  The basic idea is to drive the quantum system by a sequence of pulses whose composite action produces the planned target state. A tuning of the pulse parameters leads to the compensation of the quantum imperfections. Few composite pulse schemes were inspired by nuclear magnetic resonance (NMR)~\cite{Levitt:1982-1} and quantum information processing~\cite{Vandersypen:2005,Braun:2014}.
As tested in the experiments of refs.~\cite{Taichenachev:2009,Yudin:2010,Huntemann:2012b,Huntemann:2016,Hobson:2016}, this approach improves the performance of optical clocks
based on ultra-narrow atomic transitions, by relaxing the sensitivity to clock interrogation disturbances.

\indent Even if a clock optical transition has a very narrow linewidth, the clock interrogation process may limit the final accuracy. The limitations have different sources as the laser probe frequency/intensity instabilities, light-shifts associated to the very weak excitation of additional optical levels of the atom/ion, decoherence and relaxation of the probed atomic system. In weakly allowed transitions with fermionic species, the frequency light-shift  is usually small and does not represent an
important contribution at the relative level of accuracy presently reached~\cite{Hinkley:2013,Nicholson:2015}.
However, for clocks operating on strongly forbidden transitions and very long natural lifetimes,  or  for clocks using less stable local oscillators, shorter and more intense pulses are a necessity and may generate an important light-shift of the clock transition. These light-shifts represent a non-negligible issue for clocks either based on a single trapped ion, or on bosonic
neutral atoms with forbidden dipole transitions activated by mixing a static magnetic field with a single
laser~\cite{Taichenachev:2006,Barber:2006}, magic-wave induced transition in even isotopes~\cite{Ovsiannikov:2007}, or an E1-M1
two-photon laser excitation~\cite{Santra:2005,Zanon-Willette:2006,Zanon-Willette:2014}. In order to eliminate those systematic frequency-shift induced by the probe laser below the $10^{-18}$ accuracy level, it is necessary to develop new and very robust spectroscopic techniques schemes.

\indent This elimination of systematic frequency shifts based on  new  spectroscopic techniques is characterized by an important historical evolution. In NMR, the key issue is to tackle systematic effects responsible for imperfect rotations of the nuclear spins, because of the use of a non uniform electromagnetic field for instance~\cite{Levitt:1982-1}. Composite rotations manipulating the quantum system have been extensively developed to get rid of dual imperfections from pulse length error and resonant offset detuning. The composite pulse approach has been also applied in quantum computation to correct imperfect operations on qubits. It was theoretically investigated for a scalable quantum computer, based on trapped electrons
in vacuum, where qubits are encoded in the external (cyclotron motion) and internal (spin) degrees of freedom~\cite{Stortini:2005}.
This approach allowed manipulation of the cyclotron motion without modifying the spin evolution. In~\cite{Dunning:2014} several
sequences of NMR-type composite pulses were applied to manipulate a thermal cold atom cloud for interferometric applications.
In~\cite{LIn:2016} composite pulse sequences have been applied to laser and microwave excitation of trapped ions to produce
entanglement. In ref.~\cite{Vitanov:2015} the Ramsey interrogation of ytterbium trapped ions was modified by adding a central
off-resonant approximate $\pi$ pulse, in a way reminiscent of the spin-echo technique. Playing with the relative detuning between
applied pulses, this composite-like method eliminated small-to-moderate fluctuations in detuning, thereby greatly enhancing the
fringe contrast in the presence of laser detuning drifts.

\indent The optical-clock protocols are the results of a theoretical effort of deriving ad-hoc time-dependent Hamiltonians compensating the clock limitations listed above. Historically the first clock synthetic Hamiltonian is the well-known Ramsey protocol with separate oscillating fields creating an interference pattern in the clock spectral response. Within the last few year, some new protocols were derived in order to bypass the limitations previously listed and also to match the potential  high accuracy of the optical clock new generation.  In analogy with NMR techniques presented above, a composite pulse Ramsey (R) spectroscopy, denoted as hyper-Ramsey spectroscopy
(HR), modified hyper-Ramsey spectroscopy (MHR) or generalized hyper-Ramsey (GHR) including laser phase steps have been introduced  in
frequency metrology in order to provide highly efficient correction of the clock light-shift induced by the probing
laser~\cite{Yudin:2010,Zanon:2015,Zanon-Willette:2016a,Zanon-Willette:2016b, Hobson:2016}.
Additional perturbations associated with decoherence due to finite laser spectral width and atomic relaxation by spontaneous
emission are also corrected by composite pulses, and in particular by those based on a phase-step during the pulse sequence~\cite{Yudin:2016}.

\indent This effort may be considered as a rewriting of composite pulses for optical clocks.
However the goal is quite different: a resolution increase for NMR, a better accuracy for clocks. This difference implies
an accurate control of state populations in the first case. The second case instead requires a precise determination of the free evolution
of the quantum system. Therefore the basic tools are similar, but the final protocols different. Let's also point out the strong similarity between the design of time-dependent Hamiltonians for optical clocks and
the realization of artificial magnetism for ultra-cold neutral atoms, as reviewed in~\cite{Dalibard:2011}.
In this case the atomic center-of-mass is controlled by applying an Hamiltonian with proper space-dependence, and the successive application of different Hamiltonian may improve the target of reaching a specific final state. For the optical-clock case, different time-dependent Hamiltonians are applied within a  sequence designed for a very accurate measure of the quantum state under exploration. The sequence target is to improve the accuracy and robustness of the measurement itself.

\indent A complementary approach, denoted as synthetic protocol, is based on independent and parallel measurements of several clock-frequency shifts for different free evolution times and an appropriate combination of those measurements to generate the so-called synthetic frequency-shift, that produces the optical clock frequency with a high immunity to the laser probe perturbations. This approach, presenting large  advantages in presence of an atomic decoherence, may be considered as a different composite pulse protocol, where a different parameter of the Ramsey's interrogation scheme is properly varied: the free evolution time. The a posteriori treatment of the independent measurements applied within the synthetic protocol constitutes also an element of the phase-step protocols.

 \indent Let's mention here few original features of the composite pulse strategy for the optical clocks. As  key point in the determination of the clock proper frequency by eliminating all the  shifts produced by different source, the resonant frequency of an isolated atomic or molecular transition is a symmetry point (isolated meaning without a perturbation produced by the presence of a neighbouring transition). That symmetry is associated to the unperturbed atomic response and this symmetry feature characterizes the Ramsey's free evolution time within the pulse sequence. While the laser excitation and interrogation parts of the composite pulse sequence excite the atoms with different parameters (frequency, phase, and so on),  the probe signal produces an atomic response  "averaged"  over all the pulse sequence.  The central symmetry point should appear in that response: the composite pulse protocol reaches this target applying properly chosen parameters.

\indent In most atomic frequency standards, the laser probe is stabilized to the atomic transition by a frequency modulation
technique. Probe-induced shifts introduce a distortion of the absorption line-shape  and modify the clock operating frequency. In order to eliminate the asymmetry effect on the true clock position, a phase-step modulation was proposed and tested in~\cite{Ramsey:1951,Morinaga:1989,Klipstein:2001,Letchumanan:2004,Letchumanan:2006}. The phase-step composite pulse protocol introduces a similar but more sophisticated approach: the error signal is given by the difference between two different transition probabilities in presence of  phase steps within the composite pulse sequence. Here a technical feature introduced in order to improve the signal  quality is translated into the construction of ad-hoc Hamiltonians.

\indent Even if the probe laser is pre-stabilized on a high-finesse Fabry-Perot cavity,  the resulting finite line-width of the laser is often limited by thermal noise \cite{Numata:2004,Ludlow:2007}. For the clock atoms this limiting line-width represents a dephasing process which deteriorates the clock interrogation, reduces the contrast and compromises the robustness of any error signal. Fast improvements in the design of very high finesse Fabry-Perot cavities used to stabilize clock lasers should offer in the future very narrow line-widths below
a few 100~mHz \cite{Jiang:2011,Kessler:2012,Amairi:2013} for a new generation of frequency standards. Actually, this decoherence issue has been treated within the contest of the composite pulse protocols. Let's point out that the decoherence produced by the laser line-width dephasing does not influence the free evolution time, that represents the most important element in the precise recovery of the clock transition frequency. Instead it influences the atomic evolution within the excitation periods. The operation of most composite protocols is heavily compromised by the decoherence presence. However the  construction of ad-hoc Hamiltonian protocols produces a very large reduction of its role, greatly compensating the laser probe shifts and the dependence on the laser excitation parameters, i.e., intensity and interaction time. This is an important result because decoherence are usually considered as a strong  limit on the reachable accuracy, and because the decoherence role is circumvent by Hamiltonian interactions. This result is a part of the present large interest into the control over dissipative processes and to the realization that the coupling to the environment can be manipulated to drive the system into desired quantum states~\cite{Breuer:2007,Mueller:2012}.

\begin{figure}[t!!]
	\center
	\resizebox{\linewidth}{!}{
		\includegraphics[angle=0]{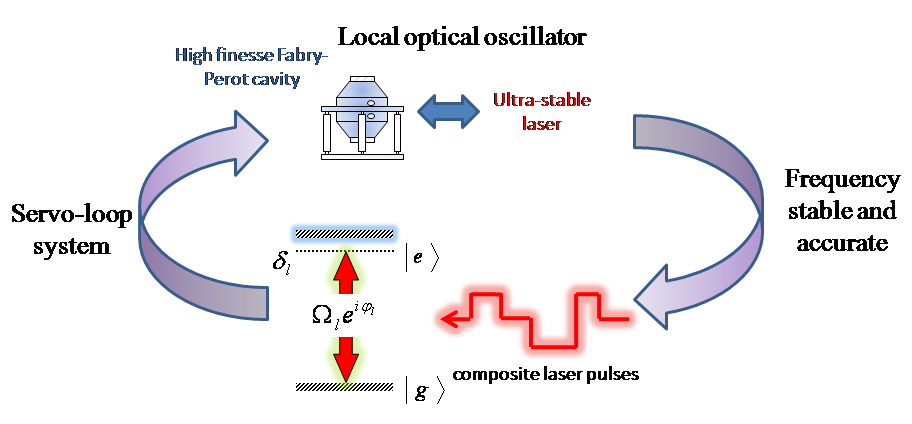}
	}
	\caption{
		Principle of an optical-clock local-oscillator interrogation protocol with the clock transition probed by a nearly
		monochromatic laser, pre-stabilized on a high finesse optical cavity. The optical interrogation is applied to the
		$|\textup{g}\rangle\Leftrightarrow|\textup{e}\rangle$ two-level quantum system, typically a weakly allowed or strongly
		forbidden transition, of a single ion or an ensemble of neutral atoms. An error signal is generated to lock the laser frequency on the
		atomic/ionic transition.}

	\label{fig:I-Optical-Clocks-Principle}
\end{figure}

\indent This work reviews the theoretical and experimental efforts performed so far on the optical clock composite pulse schemes for a robust compensation of features such as light-shifts, laser probe frequency and intensity induced instabilities, decoherence and relaxation. Section II, after introducing the wave-function formalism for a two-level atomic system in coherent interaction with a laser, investigates the NMR-like composite pulses and their simplest R and HR composite pulse counterparts. The benefits and limits of the R and HR schemes  are presented. Section III emphasizes the laser phase manipulation of each individual pulse through laser steps as a key parameter to improve clock operations. Probing schemes like HR, MHR, GHR containing phase-steps are described. Section IV presents optical Bloch-equations describing coherent interaction between laser and atoms, including several dissipative
processes disrupting clock operation. It is shown how  the combination of GHR error signals can provide immunity to both decoherence and relaxation. Section V discusses the synthetic frequency approach. A recent experimental modification of the Ramsey configuration, denoted as auto-balanced Ramsey spectroscopy, is also briefly discussed.
Section VI reports two recent experimental implementations based on HR and MHR composite laser-pulses protocols on a single
$^{171}$Yb$^{+}$ ion~\cite{Huntemann:2012b} and on $^{88}$Sr bosonic atoms optically trapped at a magic
wavelength~\cite{Hobson:2016}, respectively.  Conclusions and perspectives terminate our review.

\section{Ramsey and Hyper-Ramsey interrogation schemes}
\label{sec:II-Interrogation-Schemes}

\indent A frequency standard, shown in Fig.~\ref{fig:I-Optical-Clocks-Principle}, requires locking and stabilizing the phase or the
frequency of an external oscillator to some atomic or ionic transition by means of laser spectroscopy. Higher frequencies and
ultra-narrow lines improve measurement precision. This is the reason why the field of frequency metrology has moved over several
decades from microwave transitions to optical narrow lines \cite{Ludlow:2015}.
Weakly allowed or forbidden clock transitions are now widely investigated for the next generation of high-accuracy frequency standards
based on a single trapped ion \cite{Margolis:2009} and neutral atoms in optical lattices \cite{Katori:2003,Ye:2008,Derevianko:2011}.
Depending on the nature of the quantum absorbers, i.e fermionic or bosonic particles \cite{Akatsuka:2008}, using either a single
stabilized clock laser \cite{Ovsiannikov:2007}, a combination of static magnetic field and laser
\cite{Taichenachev:2006,Barber:2006,Baillard:2007,Kulosa:2015} or several oscillating laser fields may be exploited to probe the
atomic transition \cite{Santra:2005,Zanon-Willette:2006,Zanon-Willette:2014}.
In the following we consider a two-level system interacting with a single EM field without loss of generality.

\begin{figure}[t!!]
	\center
	\resizebox{0.4\linewidth}{!}{
		\includegraphics[angle=0]{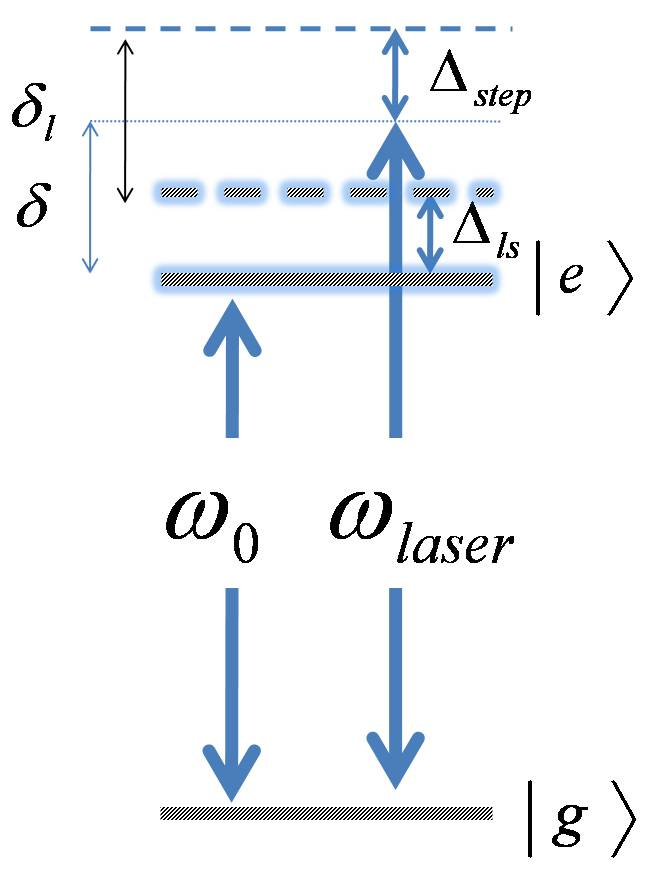}
	}
	\caption{
		Two-states description of energy levels. The energy splitting between the ground state and the excited state denoted by its
		natural frequency $\omega_0$ is probed by a laser at $\omega_{\textup{laser}}$. The detuning between the laser and the natural
		Bohr frequency of the system is $\delta=\omega_{\textup{laser}}-\omega_0$. The natural frequency $\omega_0$ is also altered by
		the probe-induced frequency shift $\Delta_{\textup{ls}}$ produced by off-resonant excitation to external states. Some composite pulse protocols include a
		laser frequency step $\Delta_{\textup{step}}$ used to generate an effective pre-compensation of the light-shift contribution from those off-resonant states.  The reduced detuning including frequency-shift compensation is written as $\delta_l=\delta-\Delta_l$, where $\Delta_l=\Delta_{\textup{ls}}-\Delta_{\textup{step}}$ is the residual uncompensated light-shift defined in the text.
	}
	\label{fig:IIA-Composite-Pulses-Parameters}
\end{figure}

\indent A Doppler recoil-free atomic clock as depicted in Fig.~\ref{fig:IIA-Composite-Pulses-Parameters} can be implemented as a
two-level quantum system with energy splitting $\hbar\omega_0$.
We start with a typical description of coherent atom-light interaction where external perturbations like decoherence, collisions
between particles and all atomic level relaxations are neglected. The two-level system is probed by a clock laser at frequency
$\omega_{\textup{laser}}$.
 Within rotating wave approximation (RWA), the interaction is governed by the complex Rabi frequency $\Omega_{l}e^{i\varphi_{l}}$ including the laser phase $\varphi_{l}$, where the $l-$th subscript will denote the number of the applied laser pulse in an interrogation pulse sequence.
The clock laser detuning $\delta$ from the unperturbed transition frequency is:
\begin{equation}
	\delta = \omega_{\textup{laser}} - \omega_{0}.
\end{equation}
\indent A two-level system is a good approximation for most spectroscopic investigations, but is not suitable for high-resolution
spectroscopy in the case of optical clocks. With a two-level transition, the virtual excitation of external non-resonant states
leads to light shifts of energy levels~\cite{Cohen-Tannoudji:2011}. We define $\Delta_{\textup{ls}}$ as the probe-induced frequency shift
altering atomic energies while the probe laser is switched on. It is proportional to the laser probe intensity.
It can sometimes be useful to compensate for that light shift by stepping the laser frequency $\omega_{\textup{laser}}$ during
pulses by a fixed amount $\Delta_{\textup{step}}$. Therefore we introduced $\delta_l$ as the effective total detuning:
\begin{equation}
	\delta_l = \omega_{\textup{laser}} + \Delta_{\textup{step}} - (\omega_{0} + \Delta_{\textup{ls}}) = \delta - \Delta_l.
	\label{eq:II-Delta-Definition}
\end{equation}
The above equation introduces the $\Delta_{l}$ residual uncompensated frequency shift as the difference between the external laser
probe-induced frequency shift and the laser frequency step used to cancel it (see Fig.~\ref{fig:IIA-Composite-Pulses-Parameters}).
Because the $\delta$ and $\Delta_{\textup{step}}$ are very small compared to the detuning of the non-resonant states, $\Delta_{\textup{ls}}$  is constant over
the whole clock interrogation process if the probe laser intensity is constant.

\indent It is important here to stress the operational difference between a laser spectroscopy experiment and an optical (or
microwave) clock. In a spectroscopic investigation, the $\omega_{\textup{laser}}$ frequency is controlled independently from the
atom (or sample) to be explored. Within a clock, a servo-loop system is applied as depicted in the upper left of
Fig.~\ref{fig:I-Optical-Clocks-Principle} producing the probe laser oscillation at clock frequency $\omega_{\textup{clock}}$.
The purpose of the servo-loop system, including the composite pulse interrogation, is to compensate light shifts and produce
$\omega_{\textup{clock}}=\omega_0$, that is a zero offset $\Delta_{\textup{off}}=\omega_{\textup{clock}}-\omega_0$.
An imperfect (or real!) clock operates with an arbitrary small offset, or equivalently the atoms are probed by a laser which
misses resonance, that is $\delta \ne 0$. In the following we investigate several protocols aiming to cancel
$\Delta_{\textup{off}}$ even in presence of a  $\Delta_l$  probe-induced shift correction different from zero.

\subsection{Single or adjacent pulses: Rabi interrogation and NMR-like composite pulses for robust population transfer}
\label{sec:IIA-NMR}

\begin{figure*}[t!!]
	\center
	\subfloat[$\boldsymbol{180}_{(0)}$ Rabi single pulse]{
		\resizebox{0.45\linewidth}{!}{
			\includegraphics[angle=0]{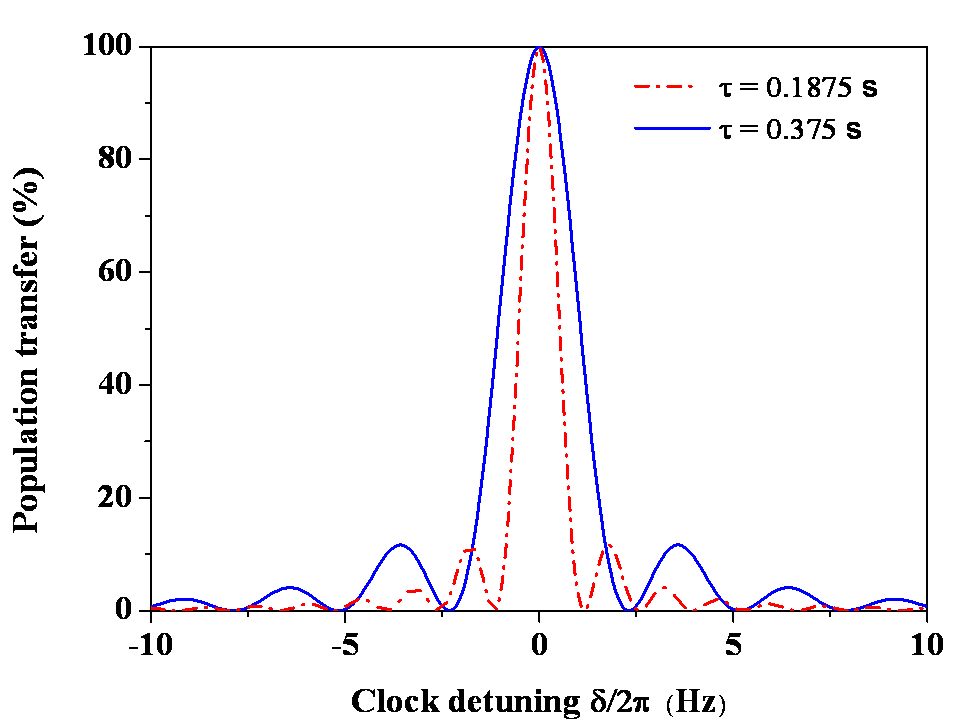}
		}
		\resizebox{0.45\linewidth}{!}{
			\includegraphics[angle=0]{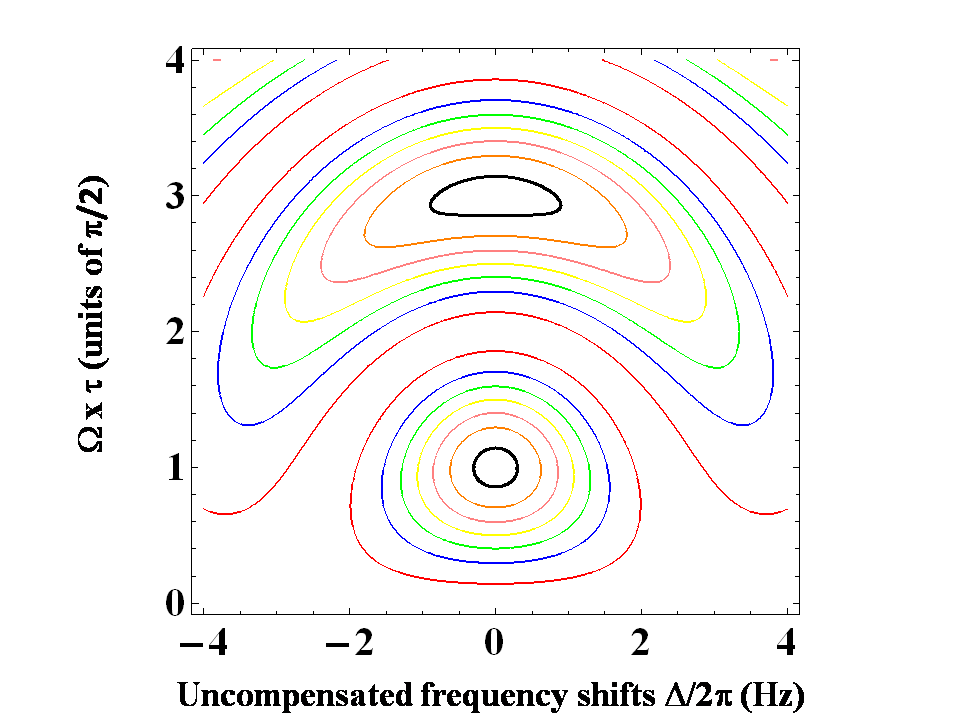}
		}
		\label{fig:IIA-Composite-Pulses-Levitt-a}
	} \\
	\subfloat[$\boldsymbol{90}_{(\frac{\pi}{2})} \boldsymbol{180}_{(0)} \boldsymbol{90}_{(\frac{\pi}{2})}$ Rabi composite pulse]{
		\resizebox{0.45\linewidth}{!}{
			\includegraphics[angle=0]{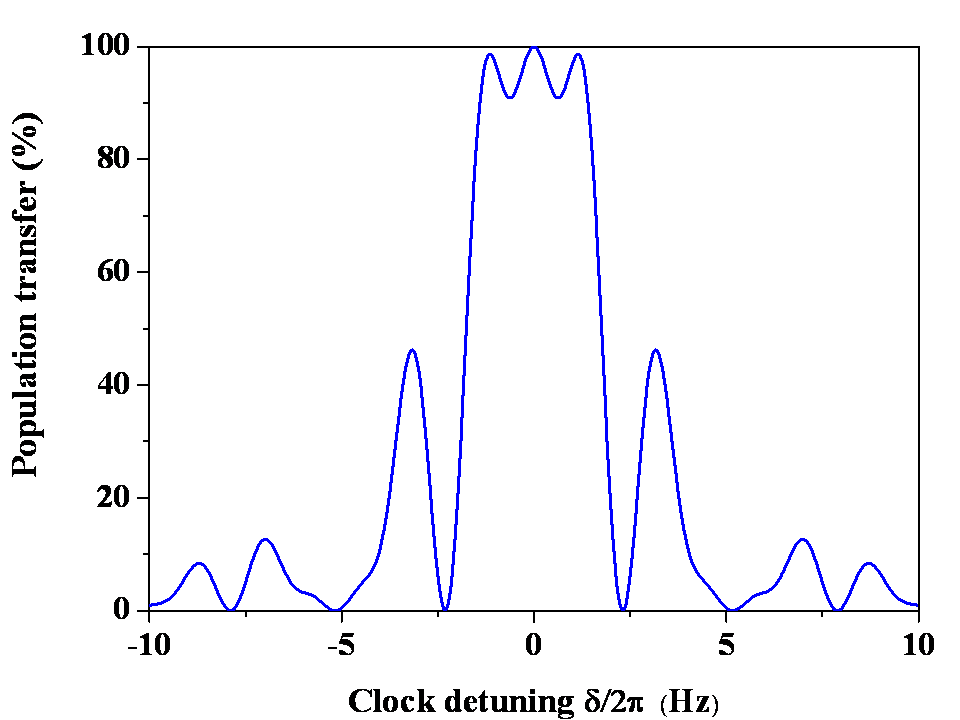}
		}
		\resizebox{0.45\linewidth}{!}{
			\includegraphics[angle=0]{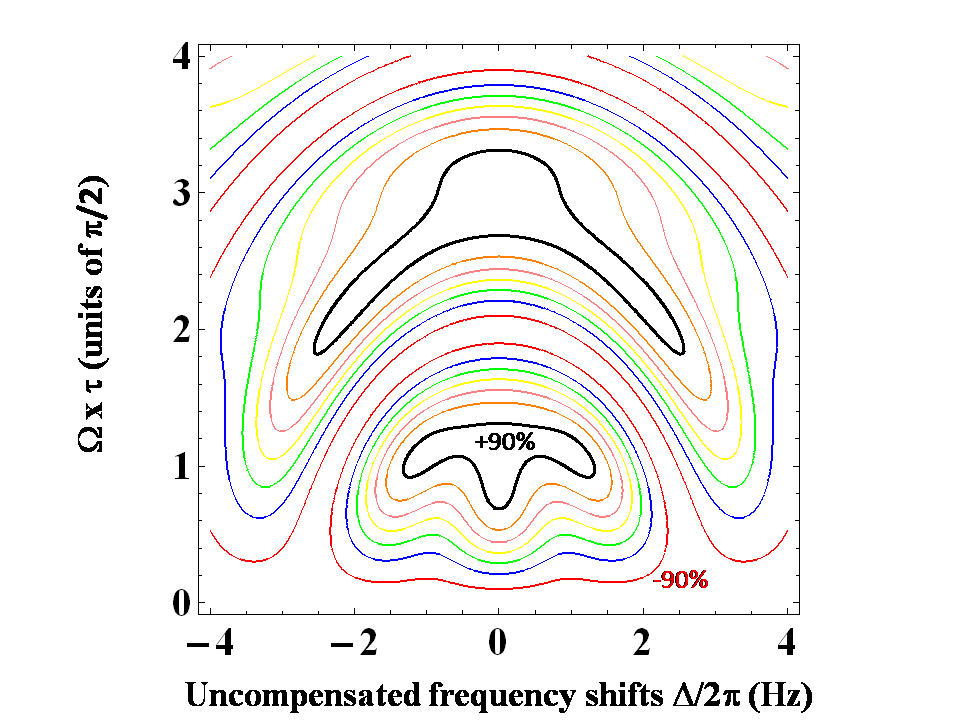}
		}
		\label{fig:IIA-Composite-Pulses-Levitt-b}
	} \\
	\subfloat[$\boldsymbol{90}_{(0)} \boldsymbol{360}_{(\frac{2\pi}{3})} \boldsymbol{90}_{(0)}$ Rabi composite pulse]{
		\resizebox{0.45\linewidth}{!}{
			\includegraphics[angle=0]{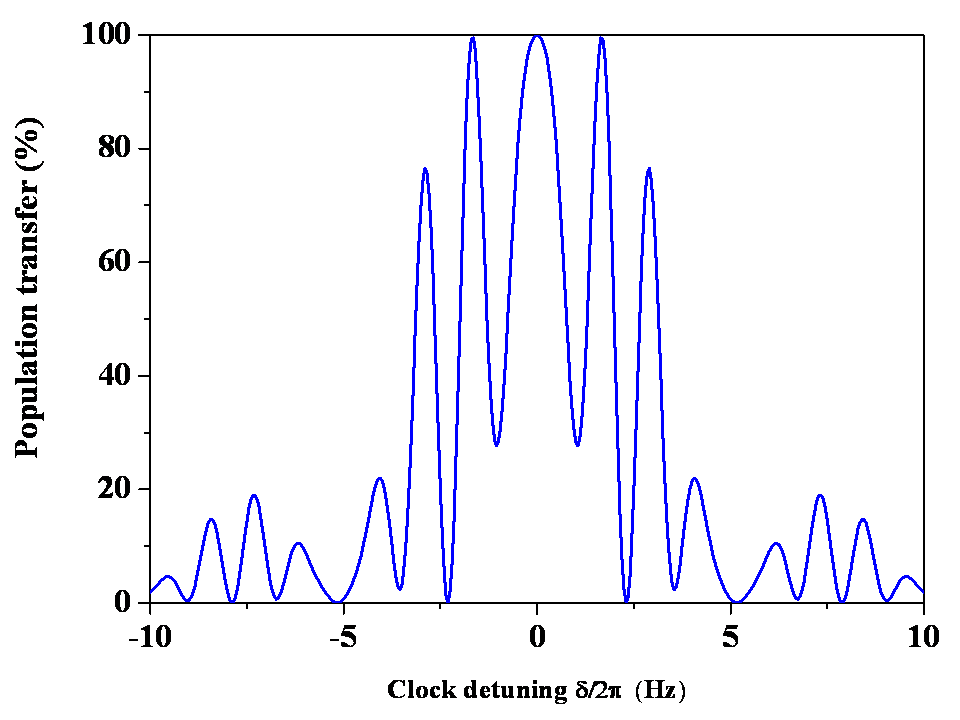}
		}
		\resizebox{0.45\linewidth}{!}{
			\includegraphics[angle=0]{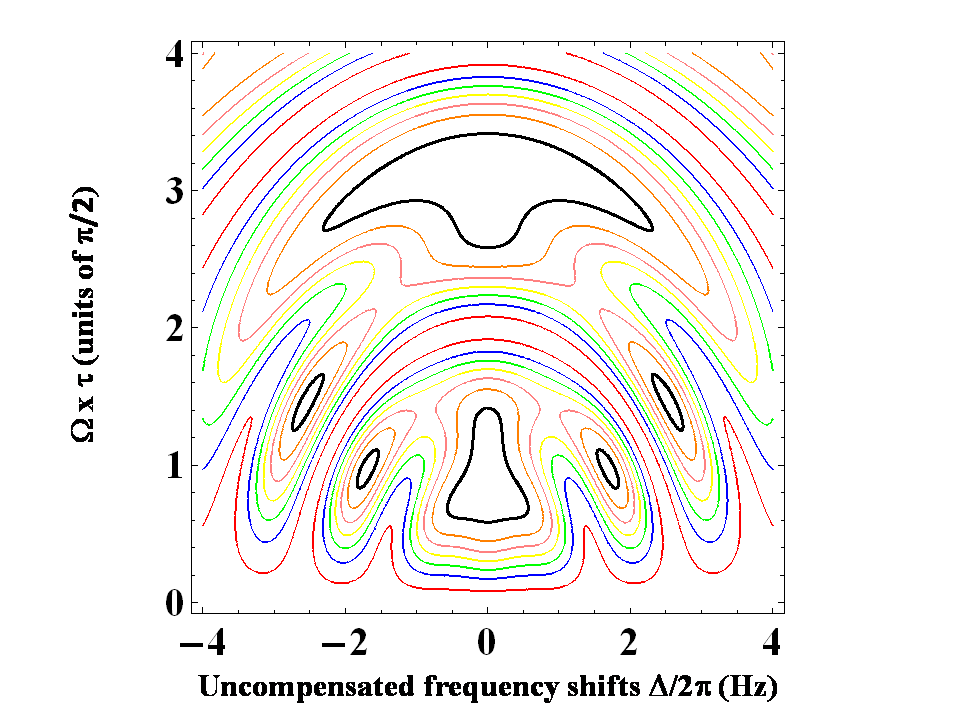}
		}
		\label{fig:IIA-Composite-Pulses-Levitt-c}
	}
	\caption{Rabi spectroscopy: on the first line based on single pulse and on the second and third lines based on NMR-like composite pulse. Left columns report population transfer efficiency; the right columns report  contour plots (solid lines in colour spanning from -$90\%$ to $90\%$ in steps of $30\%$) for the population inversion vs the reference pulse area (vertical axis) and the uncompensated frequency shift. Pulse duration and Rabi frequency fixed  to $\tau=3/16~$s and  $\Omega_l=\pi/2\tau$, respectively. The $\Delta_l$ detunings are all equal and denoted by $\Delta$. This applies also to similar analyses in the following.}
	\label{fig:IIA-Composite-Pulses-Levitt}
\end{figure*}

The superposition of $|g\rangle,|e\rangle$ clock states induced by a laser (or RF) pulse  is described by the following linear
combination:
\begin{equation}
	|\Psi(\theta_{l})\rangle = c_\textup{g}(\theta_{l}) |\textup{g}\rangle + c_\textup{e}(\theta_{l}) |\textup{e}\rangle,
\end{equation}
where $c_{\textup{g}}$ and $c_{\textup{e}}$ are probability amplitudes related to states $\ket{\textup{g}}$ and $\ket{\textup{e}}$
respectively. The $\theta_{l}$ parameter is here defined as the effective pulse area $ \omega_{l}\tau_{l}$:
\begin{equation}
	\theta_{l} = \omega_{l}\tau_{l} ,
\end{equation}
with $\tau_{l}$ being the laser pulse duration and where we introduce a generalized Rabi frequency
$\omega_{l}=\sqrt{\delta_{l}^{2}+\Omega_{l}^{2}}$ for convenience. If the total atom-laser interaction time is shorter than the damping times for the dissipation mechanisms discussed in Sec. IV, clock state dynamics are described by the following set of
Schr\"odinger's equations:
\begin{equation}
	\left\{	\begin{split}
		\dot{c}_{\textup{g}} &= i e^{i\varphi_{l}} \frac{\Omega_{l}}{2}c_{\textup{e}}, \\
		\dot{c}_{\textup{e}} &= i e^{-i\varphi_{l}} \frac{\Omega_{l}}{2}c_{\textup{g}} + i\delta_{l}c_{\textup{e}}.
	\end{split}	\right.
\end{equation}
Using the solution of Schr\"odinger's equation, the matrix solution for $c_{\textup{g},\textup{e}}(\theta_{l})$ transition amplitudes
can be written as:
\begin{equation}
	\begin{split}
		\left( \begin{array}{c}
			c_{\textup{g}}(\theta_{l}) \\
			c_{\textup{e}}(\theta_{l}) \\
		\end{array} \right)
		= \chi(\theta_{l}) \cdot \textup{M}(\theta_{l}) \cdot
		\left( \begin{array}{c}
			c_{\textup{g}}(0) \\
			c_{\textup{e}}(0) \\
		\end{array} \right),
	\end{split}
	\label{eq:IIA-Evolution-Matrix}
\end{equation}
including a phase factor of the form $\chi(\theta_{l})=\exp\left[-i\delta_{l}\frac{\tau_{l}}{2}\right]$. The wave-function evolution
driven by a pulse area $\theta_{l}$ is determined by a complex $2\times2$  interaction matrix as
\cite{Rabi:1945,Rabi:1954,Jaynes:1955}:
\begin{equation}
	\begin{split}
		\textup{M}(\theta_{l}) &=
		\left( \begin{array}{cc}
			\textup{M}_{+}(\theta_{l}) & e^{i\varphi_{l}} \textup{M}_{\dagger}(\theta_{l}) \\
			e^{-i\varphi_{l}} \textup{M}_{\dagger}(\theta_{l}) & \textup{M}_{-}(\theta_{l}) \\
		\end{array}	\right) \\
		&=
		\left( \begin{array}{cc}
			\cos\frac{\theta_{l}}{2} + i \frac{\delta_{l}}{\omega_{l}} \sin\frac{\theta_{l}}{2} &
				-i e^{i\varphi_{l}} \frac{\Omega_{l}}{\omega_{l}} \sin\frac{\theta_{l}}{2} \\
			-i e^{-i\varphi_{l}} \frac{\Omega_{l}}{\omega_{l}} \sin\frac{\theta_{l}}{2} &
				\cos\frac{\theta_{l}}{2} - i \frac{\delta_{l}}{\omega_{l}} \sin\frac{\theta_{l}}{2} \\
		\end{array} \right).
	\end{split}
	\label{eq:IIA-Spinor-Matrix}
\end{equation}
Applying the above matrix with initial conditions $c_{\textup{g}}(0)=1,c_{\textup{e}}(0)=0$, a final complex amplitude is obtained,
leading to the well-known Rabi transition probability $\textup{P}_{|\textup{g}\rangle\mapsto|\textup{e}\rangle}$:
\begin{equation}
	\textup{P}_{|\textup{g}\rangle \mapsto |\textup{e}\rangle} = \frac{\Omega_{l}^{2}}{\omega_{l}^{2}} \sin^{2}\frac{\theta_{l}}{2}.
\end{equation}

In Rabi's original experiment \cite{Rabi:1938,Rabi:1939}, a thermal molecular beam passes through a coil excited by a radio frequency
(RF) field with interaction time $\tau$. The time of interaction $\tau$ and the field amplitude are chosen such that the product
$\Omega\tau=\pi$, a so-called $\pi$ pulse, corresponds to a $\pi$ pulse area. If the radio frequency is tuned to the transition, at
time $t=\tau$ all particles with $c_g(0)=1$ are detected in their excited state. In this configuration, the single field has to be
perfectly controlled and homogeneous in the interrogation zone to achieve good sensitivity. Fig.~\ref{fig:IIA-Composite-Pulses-Levitt-a} shows the Rabi spectrum associated with the single pulse excitation scheme obtained by scanning the frequency detuning. As shown by the Fourier transforms of Fig.~\ref{fig:IIA-Composite-Pulses-Levitt-a}, the
Rabi line-shape exhibits a decreasing width with increasing excitation time.
The resolution of this spectroscopy   is limited by the flight time of particles through the coil. The FWHM width of a $\pi$ pulse is
$0.8/\tau$ (in Hz). An increase of the $\tau$ duration is possible to a certain extent, but for experimental reasons, it becomes difficult
to maintain the appropriate microwave field for long periods or large pulse areas.  It is important to note that in this case, the clock
frequency-shift of the transition probability is always linearly dependent on the residual uncompensated part $\Delta_{\textup{ls}}$
of the light-shift.  In modern optical ion and lattice clocks, the Rabi spectroscopy has been
mostly implemented to obtain very high resolution measurements of various metrological clock transitions with low systematics
\cite{Ludlow:2015}.

\begin{table}[b!!]
	\centering
	\caption{
		Examples of composite pulses proposed in NMR to increase net magnetization of nuclear spins while compensating for RF field
		variation $\delta\Omega$ and frequency offset error $\delta\Delta$ in detuning. Pulse area $\boldsymbol{\theta_l}$ is given
		in degrees and phase-steps $\varphi_{l}$ are indicated in subscript-brackets with radian unit. The standard Rabi frequency for
		all pulses is $\Omega=\pi/2\tau$ where $\tau$ is the pulse duration reference. Note that in order to compare sensitivity of various protocols to pulse defects in Fig.~\ref{fig:IIA-Composite-Pulses-Levitt}, the single Rabi $\pi$ (180\ensuremath{^\circ}) pulse is computed as two adjacent $\pi/2$ (90\ensuremath{^\circ}) pulses with duration $\tau$.}
	\renewcommand{\arraystretch}{2.5}
	\begin{tabular}{|c||c|c|}
		\hline
		Pulses area $\boldsymbol{\theta_l}$$_{(\varphi_{l})}$ & $\delta\Omega/\Omega$ & $\delta\Delta/\Delta$ \\
		\hline
		$\boldsymbol{180}_{(0)}$ & low & low \\
		\hline
		$\boldsymbol{90}_{(\frac{\pi}{2})} \boldsymbol{180}_{(0)} \boldsymbol{90}_{(\frac{\pi}{2})}$ & low & medium \\
		\hline
		$\boldsymbol{90}_{(0)} \boldsymbol{360}_{(\frac{2\pi}{3})} \boldsymbol{90}_{(0)}$ & medium & low \\
		\hline
	\end{tabular}
	\label{tab:IIA-Composite-Pulses-Rabi}
\end{table}

\indent If several adjacent pulses are used, the single matrix of Eq.~\eqref{eq:IIA-Evolution-Matrix} is replaced by a product of
several matrices in order to explore various sequences based on NMR-like composite pulse excitations.
Tab.~\ref{tab:IIA-Composite-Pulses-Rabi} reports examples of Rabi composite $\pi$ pulses inducing a robust population transfer between
two targeted quantum states required either for offset detuning or RF field compensation~\cite{Levitt:1986}. The Table first column introduces a compact indication of the composite pulse composition. The lineshapes for different NMR-like composite pulses are shown in Fig.~\ref{fig:IIA-Composite-Pulses-Levitt-b} and Fig.~\ref{fig:IIA-Composite-Pulses-Levitt-c}.
The corresponding contour plots in the right column of Fig.~\ref{fig:IIA-Composite-Pulses-Levitt} evidence the optimized robustness of the population inversion by using
several adjacent pulses optimized against  uncompensated residual frequency-shifts in Fig.~\ref{fig:IIA-Composite-Pulses-Levitt-b}, and against important pulse area
variations in Fig.~\ref{fig:IIA-Composite-Pulses-Levitt-c}. 
\subsection{Pulses with an interleaved free evolution time: Ramsey and Hyper-Ramsey schemes}
\label{sec:IIB-Composite-Pulses}

\begin{figure}[t!!]
	\center
	\resizebox{\linewidth}{!}{
		\includegraphics[angle=0]{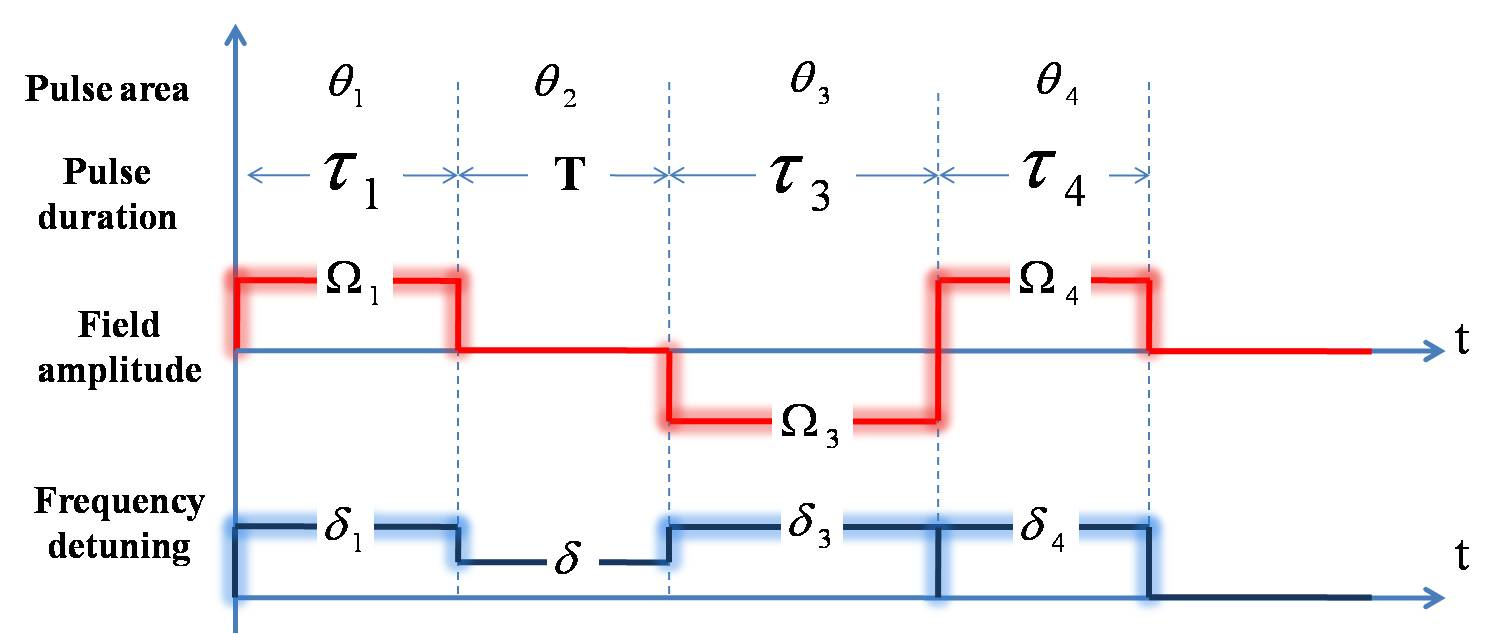}
	}
	\caption{
		Composite pulses in a general three-pulse interaction scheme acting on a two-level transition for the HR protocol. Pulses are
		labeled by $l = (1,2,3,4)$ where index $2$ denotes free evolution time. Each pulse is characterized by laser parameters as
		frequency detuning $\delta_l$, pulse duration $\tau_l$ and field Rabi frequency $\Omega_l$.
	}
	\label{fig:IIB-HR}
\end{figure}

Within the original R spectroscopy configuration, atoms are probed by two successive pulses separated by a free evolution
time $\textup{T}$ \cite{Ramsey:1950}. During free evolution time $\textup{T}$ where the probe laser is switched-off, the matrix given by Eq.~\eqref{eq:IIA-Spinor-Matrix} does not contain the light-shift from
external off-resonant states. The free-evolution transfer matrix reduces to:
\begin{equation}
	\begin{split}
		\textup{M}(\delta\textup{T}) =
		\left( \begin{array}{cc}
			e^{i\delta\textup{T}/2} & 0 \\
			0 & e^{-i\delta\textup{T}/2} \\
		\end{array} \right),
	\end{split}
\end{equation}
because no additional laser frequency step $\Delta_{\textup{step}}$ is applied during free evolution time.

\indent In the HR interrogation three pulses are applied, with a free evolution time after the first or second pulse \cite{Yudin:2010}, see an example in Fig.~\ref{fig:IIB-HR} with the free evolution time applied after the first pulse.  Note that another composite sequence with pulse order reversal can be also used as proposed in \cite{Zanon:2015}. The coherent population transfer
$\textup{P}_{|\textup{g}\rangle\mapsto|\textup{e}\rangle}$ induced by such pulse sequences is given by a simple product of matrices,
each of them being individually tailored in frequency, duration and phase. If pulses are labeled by $l = (1,2,3,4)$, the corresponding transition probability is given by:
\begin{equation}
	\textup{P}_{|\textup{g}\rangle \mapsto |\textup{e}\rangle} =
		| \langle\textup{e}| \textup{M}(\theta_{\textup{4}}) \textup{M}(\theta_{\textup{3}})
		\textup{M}(\theta_{\textup{2}}) \textup{M}(\theta_{\textup{1}}) |\textup{g}\rangle |^{2},
	\label{eq:IIB-Peg-GHR-Matrices}
\end{equation}
where we introduce $\theta_{\textup{2}}=\delta\textup{T}$. The composite  pulse sequence may
includes a laser phase-step during each pulse which can be manipulated to control the resonance shape. This applies to the HR-$\pi$ protocol of Fig.~\ref{fig:IIB-HR} where the pulse sequence includes a laser phase sign inversion during the second interaction. The transition probability describing the coherent population transfer between atomic states  depends on pulse areas and phase jumps over the entire laser probing sequence. By scanning the $\delta$ detuning between the laser and the two-level resonant frequency, a HR resonance
is constructed containing information about perturbations induced by the laser probe on the line-shape.\\
\indent Ignoring for simplicity the additional phase step, the $P_{|g\rangle\mapsto|e\rangle}$  expression can be written in a compact form as:
\begin{equation}
	P_{|g\rangle\mapsto|e\rangle} = A+B\cos(\delta\textup{T}+\Phi),
	\label{eq:IIB-Prob-Transition-Canonical}
\end{equation}
where $\delta$ is the clock frequency detuning during free evolution time. The envelopes are given by
\begin{equation}
	\begin{split}
		A &= \alpha^{2} \left[ 1 + \beta(\Phi)^2 \right], \\
		B &= 2 \alpha^{2} \beta(\Phi),
	\end{split}
	\label{eq:IIB-Enveloppes}
\end{equation}
and the phase
\begin{equation}
	\beta(\Phi) = \beta \sqrt{1 + \tan^2\Phi}.
\end{equation}
The envelopes $\alpha$, $\beta$ and the $\Phi$ phase driving the resonance amplitude are given in Appendix \ref{sec:VIIA-Envelopes-HR}. This formula is valid only for $\pm \pi$ phase jumps within the pulse sequence. However note that the general form  valid for arbitrary phase steps, described in the following by Eq.~\eqref{eq:IIIA-Error-Signal-Canonical_2},
has the same structure of the above one. Using Eq.~\eqref{eq:IIB-Prob-Transition-Canonical}, and its generalization in presence of  phase-steps derived in ref.~\cite{Zanon:2015}, the population transfer efficiency and the frequency-shift affecting the resonance can both be evaluated accurately under
various experimental laser pulse conditions including R and HR schemes \cite{Ramsey:1950,Yudin:2010}.

\indent When the second pulse area vanishes, i.e $\theta_3=0$, {\it i.e.} $\textup{M}(\theta_{\textup{3}})=\mathbbm{1}$, and $\Omega_1=\Omega_4=\Omega$, $\omega_1=\omega_4=\omega$, $\theta_1=\theta_4=\theta$, the generalized transition probability takes the following form:
\begin{equation}
	\begin{split}
		P_{|g\rangle\mapsto|e\rangle} &= 2\frac{\Omega^{2}}{\omega^{2}} \sin^{2}\frac{\theta}{2}
			\left( \cos^{2}\frac{\theta}{2} + \frac{\delta^{2}}{\omega^{2}} \sin^{2}\frac{\theta}{2} \right) \\
		&\times \left[ 1 + \cos(\delta\textup{T} + \Phi) \right],
	\end{split}
\label{eq:IIB-Prob-Transition-Ramsey}
\end{equation}
with $A=B$ in Eq.\eqref{eq:IIB-Prob-Transition-Canonical}, and for the phase
\begin{equation}
		\Phi = \arctan \left[
				\frac{2\frac{\delta}{\omega} \tan\frac{\theta}{2}}{1 - \left( \frac{\delta}{\omega} \right)^2 \tan^2\frac{\theta}{2}}
			\right]
		= 2 \arctan \left[ \frac{\delta}{\omega} \tan\frac{\theta}{2} \right].
\end{equation}
By applying a trigonometrical transformation, we recover the standard expression for the transition probability derived by Ramsey
in 1950 for a spin $1/2$ interacting with a radio-frequency field \cite{Ramsey:1950,Ramsey:1956} as:
\begin{equation}
	\begin{split}
		P_{|g\rangle \mapsto |e\rangle} = 4 \frac{\Omega^{2}}{\omega^{2}} \sin^{2}\frac{\theta}{2} &
			\left[ \cos\left( \frac{\delta\textup{T}}{2} \right) \cos\frac{\theta}{2} \right. \\
		& \left. -\frac{\delta}{\omega} \sin\left( \frac{\delta\textup{T}}{2} \right) \sin\frac{\theta}{2} \right]^{2}.
	\end{split}
	\label{eq:IIB-Prob-Transition-Ramsey-1950}
\end{equation}
This expression established by Ramsey \cite{Ramsey:1950} was the initial version of the method of separated oscillating fields in
molecular beams.

A remarkable information is the frequency-shift generated by Eq.~\eqref{eq:IIB-Prob-Transition-Ramsey}
that determines the central fringe position sensitivity to a detuning fluctuation. From a geometrical point of view, this Ramsey phase-shift is exactly two times the Euler angle accumulated by a Bloch's vector
projection of rotating components in the complex plane using a two dimensional Cauley-Klein representation of the spin 1/2 rotational group \cite{Rabi:1945}.

\begin{figure*}[t!!]
	\center
	\subfloat[Ramsey (\textup{R}) spectroscopy]{
		\resizebox{0.45\linewidth}{!}{
			\includegraphics[angle=0]{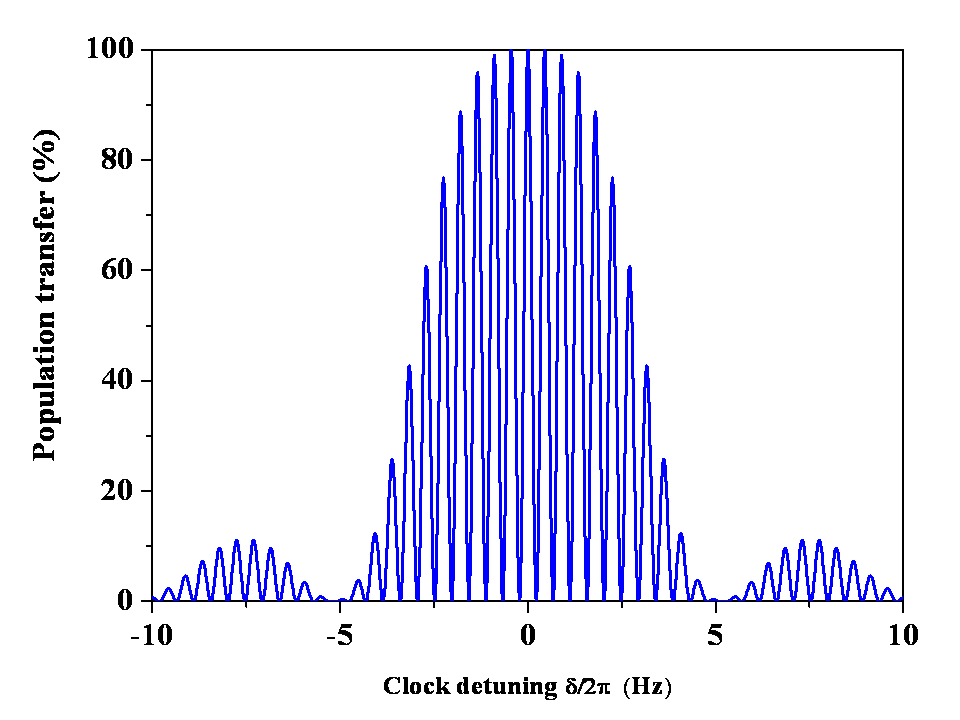}
		}
		\resizebox{0.45\linewidth}{!}{
			\includegraphics[angle=0]{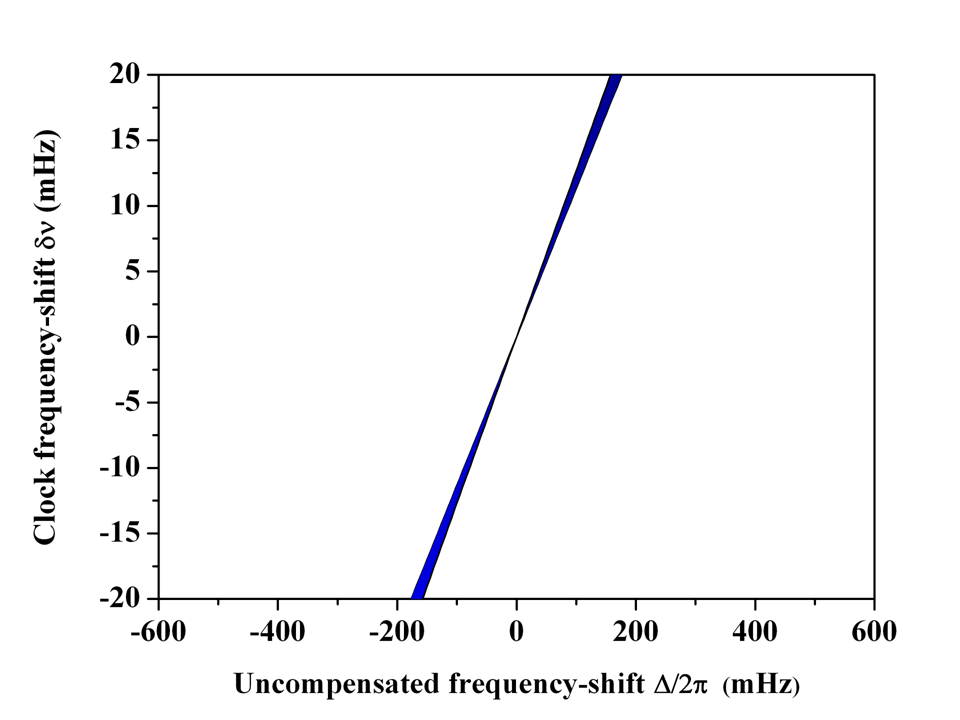}
		}
		\label{fig:IIB-Composite-Pulses-R-HR-Pi-a}
	} \\
	\subfloat[\textup{HR} spectroscopy]{
		\resizebox{0.45\linewidth}{!}{
			\includegraphics[angle=0]{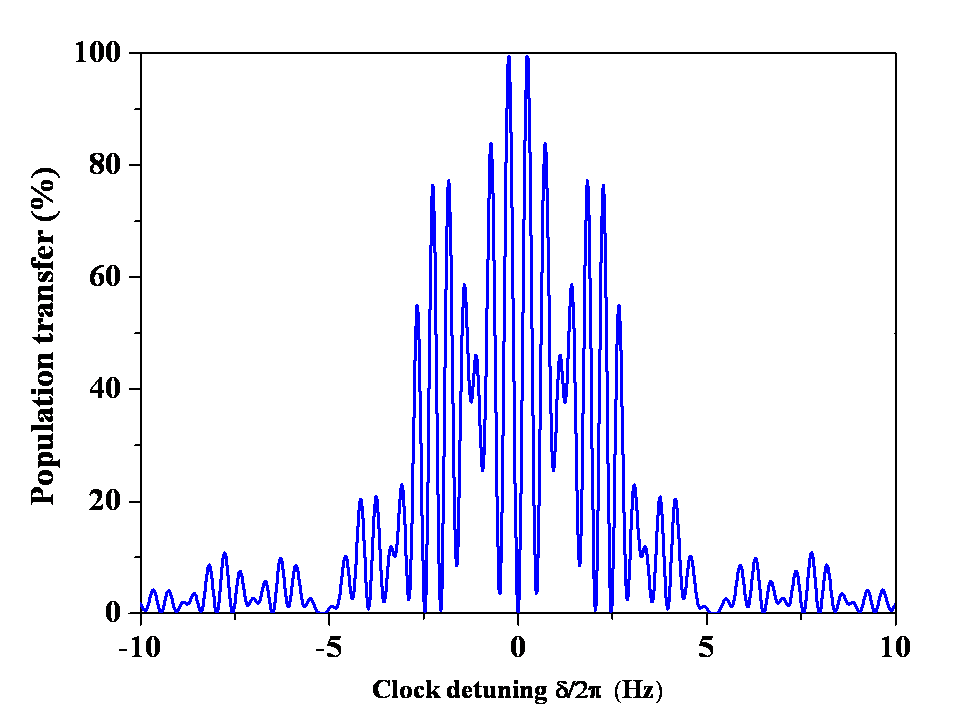}
		}
		\resizebox{0.45\linewidth}{!}{
			\includegraphics[angle=0]{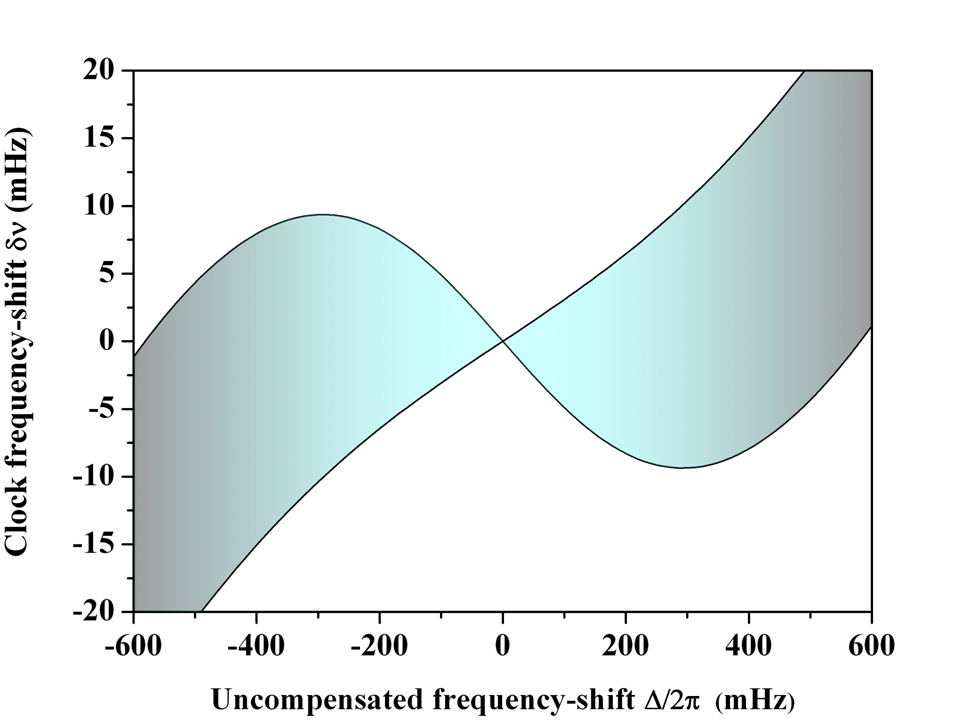}
		}
		\label{fig:IIB-Composite-Pulses-R-HR-Pi-b}
	} \\
	\subfloat[\textup{HR}-$\pi$ spectroscopy]{
		\resizebox{0.45\linewidth}{!}{
			\includegraphics[angle=0]{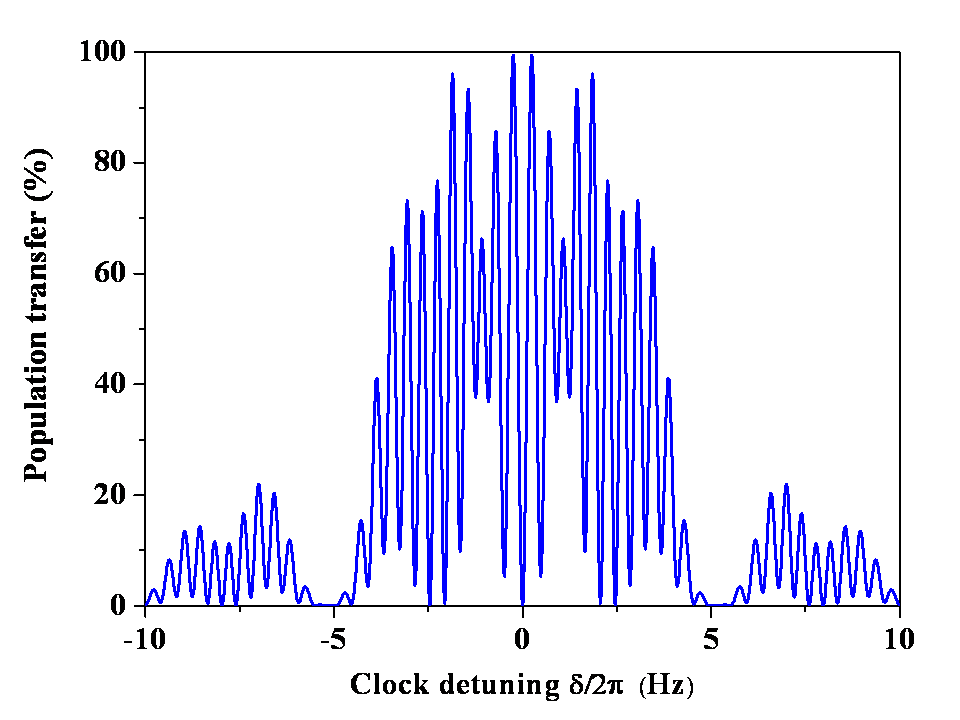}
		}
		\resizebox{0.45\linewidth}{!}{
			\includegraphics[angle=0]{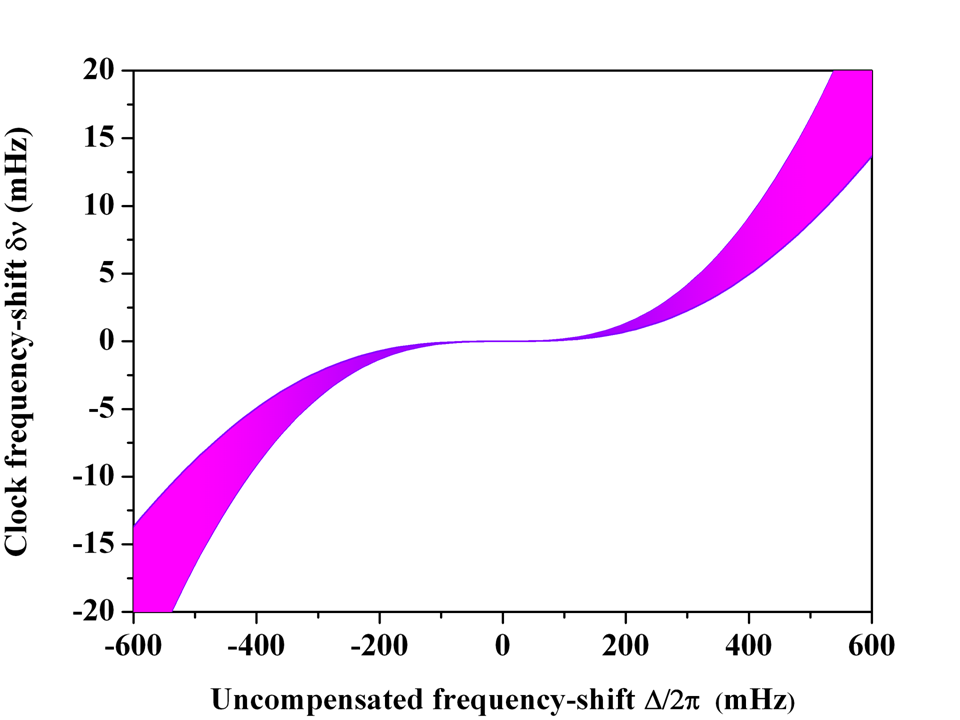}
		}
		\label{fig:IIB-Composite-Pulses-R-HR-Pi-c}
	}
	\caption{
		Left column shows the lineshapes of the transition probability for different composite pulses. Right column shows the frequency lock-point sensitivity against uncompensated residual light-shifts, whith pulse area variation set to $\Delta\theta/\theta=\pm10\%$ (shadow regions). For all graphs $\tau=3/16~$s, $\Omega=\pi/2\tau$, $\textup{T}=2~$s and
		light shift $\Delta_{ls}=0$ for convenience. The $\Delta_{l}$ uncompensated frequency shift on the horizontal axis of the right plots is denoted by $\Delta$.
	}
	\label{fig:IIB-Composite-Pulses-R-HR-Pi}
\end{figure*}

\indent The separated oscillating fields method invented by Ramsey and presented in Fig.~\ref{fig:IIB-Composite-Pulses-R-HR-Pi-a}
effectively reduces the clock sensitivity to light-shift effects. Interference fringes in the population transfer, as shown in
Fig.~\ref{fig:IIB-Composite-Pulses-R-HR-Pi-a}, are observed versus the clock laser detuning and the central feature is used to lock
the local oscillator to the atomic or molecular transition. It has been widely applied in high precision measurements for atomic clocks
based on atomic beams crossing a microwave cavity twice \cite{Vanier:1989,Ramsey:1990,Essen:1955} and was extended to Zacharias-type
fountain geometries where laser cooled atoms are thrown up vertically \cite{Kasevich:1989,Clairon:1991}. In the last device, a cold
atomic cloud experiences a first $\pi/2$ pulse ($\theta=\Omega\tau=\pi/2$) during its rise when passing through a microwave
resonator, then freely evolves without light interaction during its free launch and free fall. Finally it undergoes a second $\pi/2$
pulse in the same cavity before detection.
The resolution of such a clock configuration is only limited by the atomic cloud time of flight $\textup{T}$ between microwave interactions
\cite{Campbell:2011}. The resonance width (in Hz) is $1/(2\textup{T})$ when $\textup{T}\gg\tau$. The reduction of the Ramsey clock frequency-shift
reported in Fig.~\ref{fig:IIB-Composite-Pulses-R-HR-Pi-a} by the factor $\propto\tau/\textup{T}$ was observed in molecular beam experiments
with RF fields \cite{Shirley:1963,Fabjan:1972,Code:1971,Greene:1978}.\\
\indent Despite its great resolution, the original Ramsey method remains
too sensitive to perturbations from the optical probe laser field itself. Some spatial laser beam  configurations where proposed
in the 1980's by Bordé canceling first-order Doppler-shifts to observe optical Ramsey fringes \cite{Borde:1983}.
However, if external AC Stark-shifts are not reduced or potentially eliminated, the Ramsey central fringe is pulled away from resonance and the fringes themselves become asymmetric around the
maximum \cite{Marrocco:1998}. This asymmetry was observed for single ion clock using an ultra-narrow electric electric octupole (E3)
optical transition \cite{Margolis:2009} and in some alkaline-earth neutral bosonic clocks with completely forbidden transitions
\cite{Ludlow:2015}.

\indent Few techniques were proposed to solve such a problem. A spectroscopic laser probe configuration based on a pulsed EIT
(Electromagnetically Induced Transparency)/Raman two-photon excitation suggested in 2006 introduces internal ac Stark-shifts of a
three-level system in order to counteract the light-shift contribution from external off-resonant states. This approach restores the
Ramsey fringes at the unperturbed clock frequency \cite{Zanon-Willette:2006}.
A modified Ramsey method was also proposed to cancel the overall light-shift with deliberate application of a laser frequency step
during light pulses \cite{Taichenachev:2009}. However, all these methods require exact knowledge of the light-shift correction or
an excellent control of laser power variations to efficiently compensate frequency shifts.

\indent The HR spectroscopy was proposed in 2010 to relax the constraint on laser power control and to eliminate the  probe induced frequency-shifts \cite{Yudin:2010}. The scheme is based on pulses that can have different lengths, frequencies, and possibly phase inversion. The initial version is based on a sequence of two different pulses, a first $\pi/2$ pulse of length $\tau$ as in the Ramsey's technique, and a second $3\pi/2$ pulse of length $3\tau$, tailored in two parts $2\tau$ and $\tau$. A laser frequency step for a basic pre-compensation of the light-shift is also introduced during pulses to correct the expected external light-shift from off-resonant states of the probed two level system. The transition probability describing the HR resonance, reported in Fig.~\ref{fig:IIB-Composite-Pulses-R-HR-Pi-b},  has a fringe inversion at the resonance because of the $2\pi$ pulse area. A discriminator slope to lock the laser frequency is obtained by a $\pm \pi/2$ phase modulation on one of the pulses. Even if the phase is discussed in detail within the following SubSection, let's point out here the presence of a small sensitivity to pulse area variation, limiting the method's efficiency. This result is shown by the shaded area on the right panel of Fig.~\ref{fig:IIB-Composite-Pulses-R-HR-Pi-b}. That sensitivity can be compensated by applying a $\pi$ laser phase step during the $2\tau$-length pulse, as in the HR-$\pi$ composite pulse  of  Fig.~\ref{fig:IIB-HR}. Therefore the HR-$\pi$ sequence can be seen as an echo pulse \cite{Yudin:2010}. The important result is a strong non-linear cubic dependence of the central fringe frequency shift with the uncompensated light-shift $\Delta_l$, see on the right in Fig.~\ref{fig:IIB-Composite-Pulses-R-HR-Pi-c}. This drastically reduces residual uncompensated light-shift contribution to a very low order.

\subsection{Canonical form of the clock frequency-shift}
\label{sec:IIC-Clock-Frequency-Shift}

This subsection examines the clock-frequency shift which impacts the Ramsey interference pattern as described by
Eq.~\eqref{eq:IIB-Prob-Transition-Canonical}. A good approximation of the shift for the central fringe extremum is given by the following simple relation:
\begin{equation}
	\delta\nu \sim -\frac{\Phi|_{\delta \rightarrow 0}}{2\pi \textup{T}},
	\label{eq:IIC-Clock-Frequency-Shift}
\end{equation}
It is thus possible to
eliminate the frequency shift of the central fringe by engineering $\Phi$ with special choices of laser step frequency, pulse duration,
and phase inversion.

\begin{figure}[t!!]
	\centering
	\resizebox{1.1\linewidth}{!}{
		\includegraphics[angle=0]{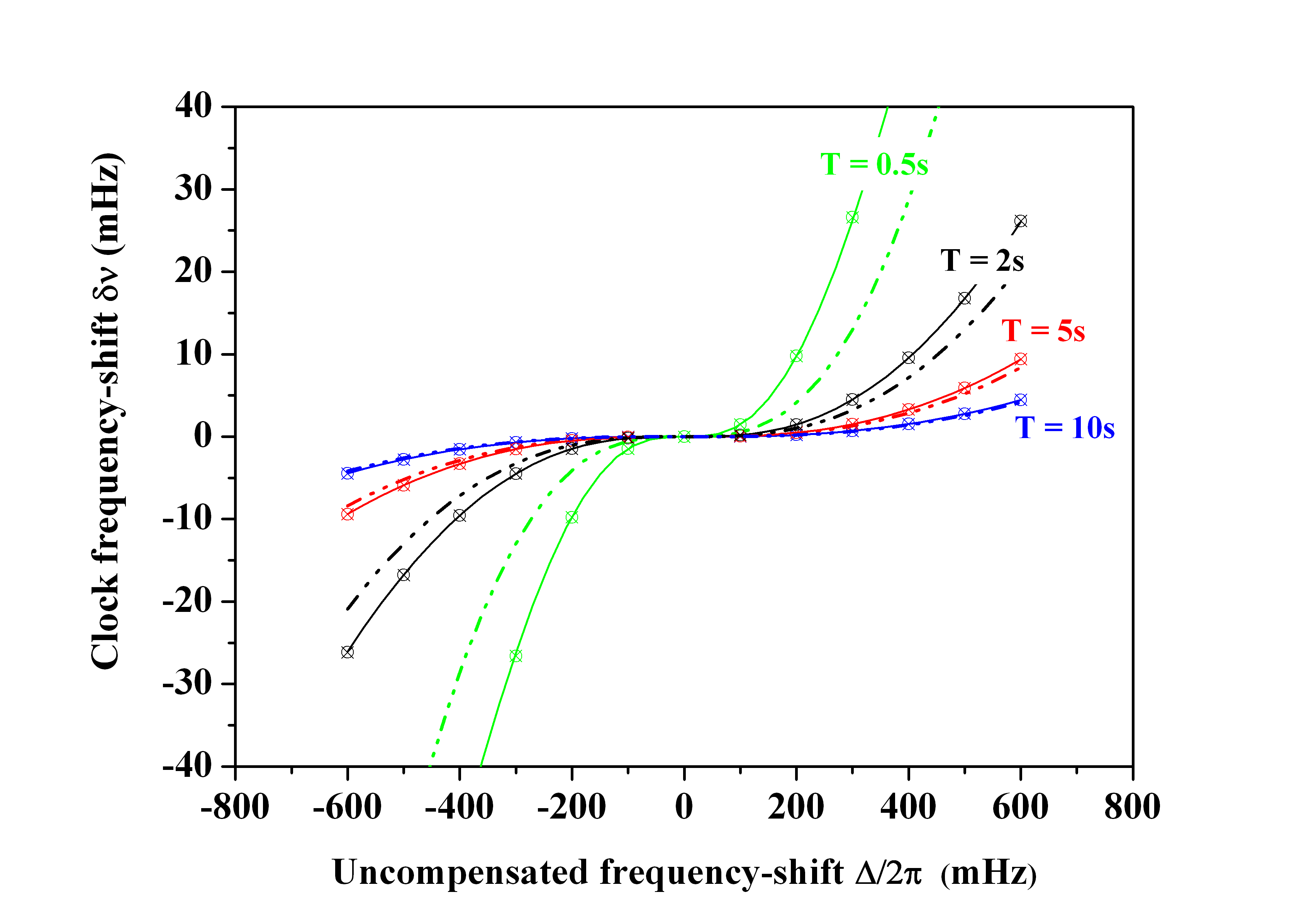}
	}
	\caption{
		(Color online) Comparison between clock frequency shifts for the HR-$\pi$ protocol computed with Eq.~\eqref{eq:IIC-Clock-Frequency-Shift}
		(dash-dotted line) and Eq.~\eqref{eq:IIC-Clock-Frequency-Shift-High-Order} (solid line) versus residual uncompensated frequency
		shifts $\Delta/2\pi$. Dots represent the numerical tracking of the central fringe extremum. Parameter as in
		Fig.~\ref{fig:IIB-Composite-Pulses-R-HR-Pi} except for $\textup{T}$.
	}
	\label{fig:IIC-Clock-Frequency-Shifts-HR}
\end{figure}

\indent A more sophisticated expression for the composite clock-frequency shift than Eq.~\eqref{eq:IIC-Clock-Frequency-Shift} is needed
if the line-shape is perturbed by weak distortions due to decoherence, as derived in Section~\ref{sec:IVC-Dissipation-Effects}.
The central fringe frequency-shift $\delta\nu$ is thus calculated by applying a first-order expansion to
Eq.~\eqref{eq:IIB-Prob-Transition-Canonical} around the unperturbed frequency clock detuning $\delta$ of the resonance. The result takes
the form:
\begin{equation}
	\delta\nu \approx -\frac{\Phi|_{\delta \rightarrow 0}}
		{ 2\pi \left( \textup{T} + \partial_{\delta} \Phi|_{\delta \rightarrow 0} \right) },
	\label{eq:IIC-Clock-Frequency-Shift-High-Order}
\end{equation}
where $\partial_{\delta}$ is the partial derivative with respect to the unperturbed clock detuning $\delta$.
The main term $\Phi$ is modified by two high-order phase-shifts as follows:
\begin{equation}
	\Phi \rightarrow \Phi + \Psi + \Theta,
\end{equation}
where
\begin{subequations}
	\begin{align}
		\Psi&= -\arctan \left[ \frac{\partial_{\delta}B}{B (\textup{T} + \partial_{\delta}\Phi)} \right],
		\label{eq:IIC-Phase-Shift-High-Order-Term-A} \\
		\Theta&= \arcsin \left[
			\frac{ \partial_{\delta}A }{
				\sqrt{ \left( \partial_{\delta}B \right)^{2} + \left( B (\textup{T} + \partial_{\delta}\Phi) \right)^{2} }
			}
		\right].
		\label{eq:IIC-Phase-Shift-High-Order-Term-B}
	\end{align}
\end{subequations}
The high-order expressions given by Eq.~\eqref{eq:IIC-Phase-Shift-High-Order-Term-A} and Eq.~\eqref{eq:IIC-Phase-Shift-High-Order-Term-B} account for a possible distortion of the line-shape when the free evolution time $\textup{T}$ is not very large compared to each pulse
duration $\tau_{l}$ ($l=\textup{1,3,4}$). This is shown in Fig.~\ref{fig:IIC-Clock-Frequency-Shifts-HR} for the HR-$\pi$ protocol and various Ramsey free
evolution times and a fixed pulse duration.

\indent The analytical expression of the clock frequency-shift  for the two-pulse R protocol is written in a simplified expression as \cite{Zanon:2015,Zanon-Willette:2016a}:
\begin{equation}
	\Phi = \arctan \left[ \frac{\delta_{1}}{\omega_{1}} \tan\frac{\theta_{1}}{2} \right]
		+ \arctan \left[ \frac{\delta_{4}}{\omega_{4}} \tan\frac{\theta_{4}}{2} \right].
	\label{eq:IIC-Phase-Shift-HR-Reduced}
\end{equation}
and in an alternative expression,
\begin{equation}
	\Phi = \arctan \left[ \frac{
		\frac{\delta_{1}}{\omega_{1}} \tan\frac{\theta_{1}}{2} + \frac{\delta_{4}}{\omega_{4}} \tan\frac{\theta_{4}}{2}
	}{
		1 - \frac{\delta_{1}\delta_{4}}{\omega_{1}\omega_{4}} \tan\frac{\theta_{1}}{2}
 \tan\frac{\theta_{4}}{2}
	}\right],
\end{equation}
to be compared to the following one for the three-pulse scheme.

In the case of three different pulse areas, the HR and HR-$\pi$ phase shifts are derived in \cite{Yudin:2010,Zanon:2015}. Following \cite{Abramowitz:1968}, they can be rewritten into a closed form solution as:
\begin{equation}
	\begin{split}
		\Phi &= \arctan \left[
			\frac{ 	
				\frac{\delta_{3}}{\omega_{3}} \tan\frac{\theta_{3}}{2}
 + \frac{\delta_{4}}{\omega_{4}} \tan\frac{\theta_{4}}{2}
			}{
				1 - \left( \frac{\delta_3\delta_4 + \Omega_3\Omega_4}{\omega_3\omega_4} \right) \tan\frac{\theta_{3}}{2}
\tan\frac{\theta_{4}}{2}
			}
		\right] \\
		&+ \arctan \left[ \frac{\delta_{1}}{\omega_{1}} \tan\frac{\theta_{1}}{2}
\right]
		+ \arctan \left[ \frac{\delta_{34}}{\omega_{34}} \tan\frac{\theta_{34}}{2}
 \right]
			,
	\end{split}
	\label{eq:IIC-Phase-Shift-GHR-Reduced}
\end{equation}
where the reduced notation of Eq.~\eqref{eq:reducednotation} was inserted within the last term.\\

\indent For the R, HR and HR-$\pi$ protocols the clock frequency phase-shift based on Eq.~\eqref{eq:IIC-Clock-Frequency-Shift-High-Order} is plotted on the right column of  Fig.~\ref{fig:IIB-Composite-Pulses-R-HR-Pi} versus  the uncompensated residual frequency shift $\Delta$. Here and in following figures they are plotted in the case of $\Delta_{l}=\Delta$ during each laser pulse.
In the R case, the clock frequency shift is linearly dependent  as shown in
Fig.~\ref{fig:IIB-Composite-Pulses-R-HR-Pi-a}. As in Fig.~\ref{fig:IIB-Composite-Pulses-R-HR-Pi-b} for the HR technique based
on the combination $\theta_1=\pi/2$ and $\theta_4=3\pi/2$, to be inserted in Eq.~\eqref{eq:IIC-Phase-Shift-HR-Reduced}, the clock becomes non linear. A relative variation of $\pm10\%$ for all pulse
areas affects the protocol and requires a careful control to avoid significant shifts. Fig.~\ref{fig:IIB-Composite-Pulses-R-HR-Pi-c} presents the clock frequency-shifts of the HR-$\pi$  technique  including a laser phase inversion coupled with a $\pi$ pulse. By inserting in Eq.~\eqref{eq:IIC-Phase-Shift-GHR-Reduced} the pulse areas
$\theta_1=\theta_4=\pi/2$ and $\theta_3=\pi$ while fixing the intermediate laser field phase
to $\pi$~($\Omega\rightarrow-\Omega$), a very good compensation of the $\pm10\%$ relative pulse area variation is obtained.
These plots clearly show that composite pulses are really efficient to extend the region where both pulse area variations and residual
light-shifts are simultaneously rejected to a very low level of perturbations. HR spectroscopy is now implemented in single ion clocks
based on ultra-narrow transitions \cite{Huntemann:2012b,Huntemann:2016}, as presented in Section~\ref{sec:VI-Clock-Implementations}.\\
\indent Fig.~\ref{fig:IIC-Clock-Frequency-Shifts-HR} shows the comparison between clock frequency shifts computed from Eq.~\eqref{eq:IIC-Clock-Frequency-Shift},
Eq.~\eqref{eq:IIC-Clock-Frequency-Shift-High-Order} and the numerical tracking of the extremum of the central fringes. The high-order corrections given by Eq.~\eqref{eq:IIC-Clock-Frequency-Shift-High-Order} are in very good  agreement with numerical trackings for all free evolution times. On the contrary, differences exist between
the results based on Eq.~\eqref{eq:IIC-Clock-Frequency-Shift} analytical expression and  those of the  numerical tracking, becoming more pronounced when the free evolution time is comparable to the pulse duration.

\section{Composite Ramsey spectroscopy with phase-step protocols}
\label{sec:III-Phase-Step-Protocols}

\begin{table}[b!!]
	\centering
	\caption{
		Composite phase-step interrogation protocols neglecting the dissipative processes.
			Reverse composite pulse protocols are denoted by a $\dagger$ prefix. Free evolution appears at index $l=2$, denoted by $\delta\textup{T}$.
			Pulse area $\theta_{l}$ is given in degrees and the two $(\varphi_{l+},\varphi_{l-})$ values of the phase-step  are indicated in subscript-brackets with radian units. The protocol name is that reported in the previous literature without an effort for an uniform notation.
	}
	\renewcommand{\arraystretch}{2.0}
	\begin{tabular}{|c|c|c|}
		\hline
		protocols & composite pulses $\boldsymbol{\theta_{l}}$ $_{(\varphi_{l+},\varphi_{l-})}$
			 \\
		\hline
		R & \begin{tabular}{c}
				$\boldsymbol{90}_{(\frac{\pi}{2},-\frac{\pi}{2})} \dashv \delta\textup{T} \vdash \boldsymbol{90}_{(0,0)}$ \\
				$(\dagger)~\boldsymbol{90}_{(0,0)} \dashv \delta\textup{T} \vdash \boldsymbol{90}_{(-\frac{\pi}{2},\frac{\pi}{2})}$
			\end{tabular}
			 \\
		\hline
		HR-$\pi$ & \begin{tabular}{c}
				$\boldsymbol{90}_{(\frac{\pi}{2},-\frac{\pi}{2})} \dashv \delta\textup{T} \vdash
					\boldsymbol{180}_{(\pi,\pi)} \boldsymbol{90}_{(0,0)}$ \\
				$(\dagger)~\boldsymbol{90}_{(0,0)} \boldsymbol{180}_{(\pi,\pi)} \dashv \delta\textup{T} \vdash
					\boldsymbol{90}_{(-\frac{\pi}{2},\frac{\pi}{2})}$
			\end{tabular}
			  \\
		\hline
		MHR & \begin{tabular}{c}
				$\boldsymbol{90}_{(\frac{\pi}{2},0)} \dashv \delta\textup{T} \vdash
					\boldsymbol{180}_{(\pi,\pi)} \boldsymbol{90}_{(0,-\frac{\pi}{2})}$ \\
				$(\dagger)~\boldsymbol{90}_{(-\frac{\pi}{2},0)} \boldsymbol{180}_{(\pi,\pi)} \dashv \delta\textup{T} \vdash
					\boldsymbol{90}_{(0,\frac{\pi}{2})}$
			\end{tabular}
			 \\
		\hline
		GHR$(\frac{\pi}{4})$ & \begin{tabular}{c}
				$\boldsymbol{90}_{(0,0)} \dashv \delta\textup{T} \vdash \boldsymbol{180}_{(\frac{\pi}{4},-\frac{\pi}{4})}
					\boldsymbol{90}_{(0,0)}$ \\
				$(\dagger)~\boldsymbol{90}_{(0,0)} \boldsymbol{180}_{(-\frac{\pi}{4},\frac{\pi}{4})} \dashv \delta\textup{T} \vdash
					\boldsymbol{90}_{(0,0)}$
			\end{tabular}
			 \\
		\hline
		GHR$(\frac{3\pi}{4})$ & \begin{tabular}{c}
				$\boldsymbol{90}_{(0,0)} \dashv \delta\textup{T} \vdash \boldsymbol{180}_{(3\frac{\pi}{4},-3\frac{\pi}{4})}
					\boldsymbol{90}_{(0,0)}$ \\
				$(\dagger)~\boldsymbol{90}_{(0,0)} \boldsymbol{180}_{(-\frac{3\pi}{4},\frac{3\pi}{4})} \dashv \delta\textup{T} \vdash
					\boldsymbol{90}_{(0,0)}$
			\end{tabular}
			 \\
		\hline
	\end{tabular}
	\label{tab:III-Composite-Pulses-Protocols}
\end{table}

In most atomic frequency standards, the laser probe is stabilized to the atomic transition by a standard frequency modulation
technique applied at the half-height of the central Ramsey fringe. But if some AC Stark-shifts are present due to non-resonant
atomic states, the line-shape is distorted and shifted from the correct clock frequency, leading also to errors
and instabilities in the frequency lock point. The proper strategy to eliminate the asymmetry effect on the true position of the central fringe and to generate a robust and stable lock-point for the local laser probe oscillator is the phase-step modulation, as proposed and tested in \cite{Ramsey:1951,Morinaga:1989,Letchumanan:2006,Letchumanan:2004}. In addition this phase modulation
technique produces an error signal with enhanced immunity to potential offset variations \cite{Morinaga:1989,Klipstein:2001}.
A corresponding  composite pulse approach is characterized by the presence of  appropriate phase-step modulations within specific areas of the pulse sequence. The measured signal is based on a difference of properly chosen generalized transition probabilities following the application of  the $\pm\varphi_{l}$ phase-step modulation. The signal becomes anti-symmetric with respect to the clock laser detuning $\delta$. This Section presents the composite pulse phase-step protocols of Table~\ref{tab:III-Composite-Pulses-Protocols}. Frequency lock points generated from these configurations are well protected against large laser pulse area variations and potential errors in the frequency shift compensations, because they decouple the
unperturbed frequency measurement from laser intensity variations \cite{Hobson:2016,Zanon-Willette:2016b}.
\begin{figure}[t!!]
	\center
	\subfloat[\textup{GHR}$(\varphi_{l})$]{
		\resizebox{\linewidth}{!}{
			\includegraphics[angle=0]{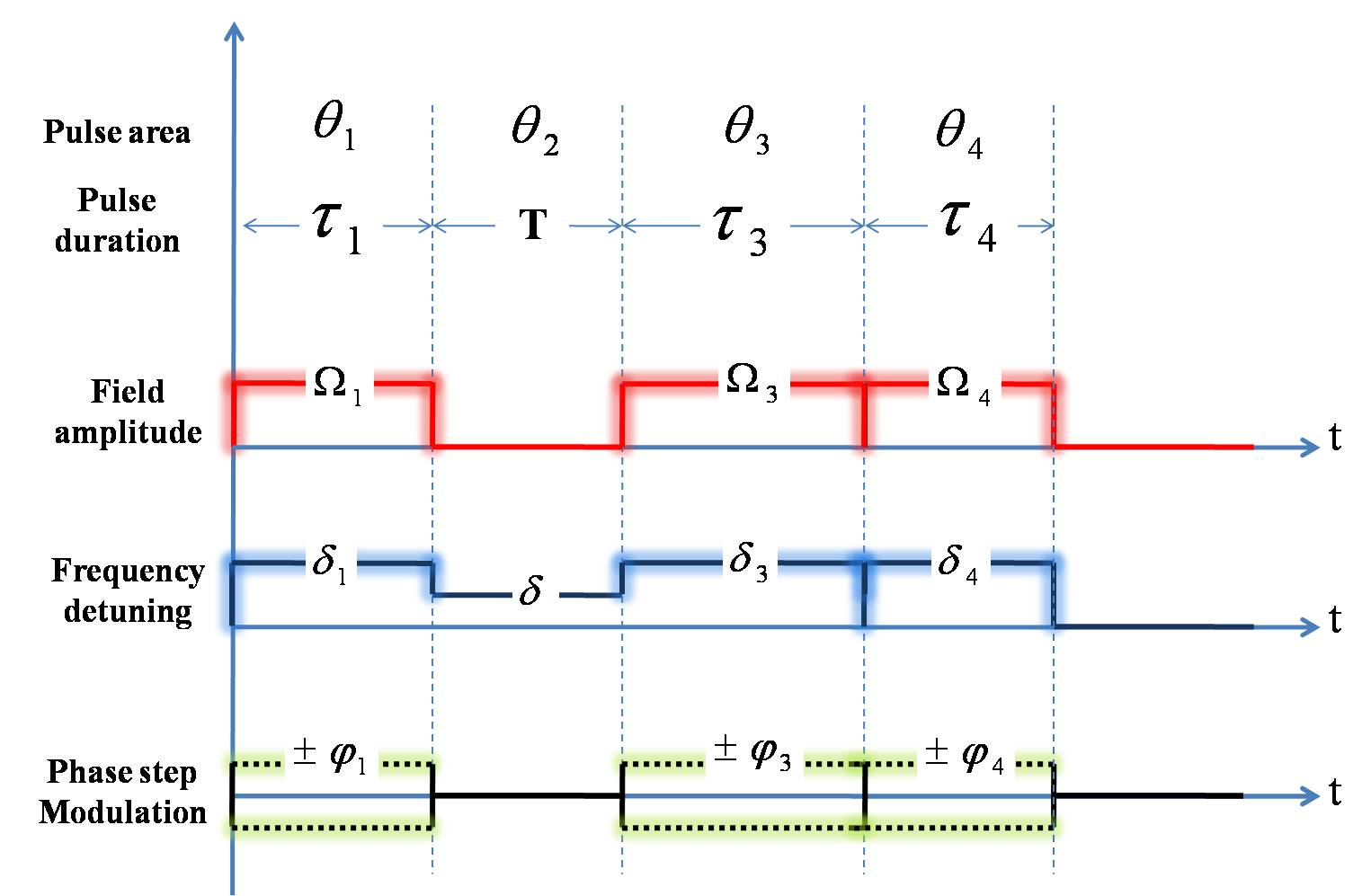}
		}
	} \\
	\subfloat[\textup{GHR}$^{\dagger}(\varphi_{l})$]{
		\resizebox{\linewidth}{!}{
			\includegraphics[angle=0]{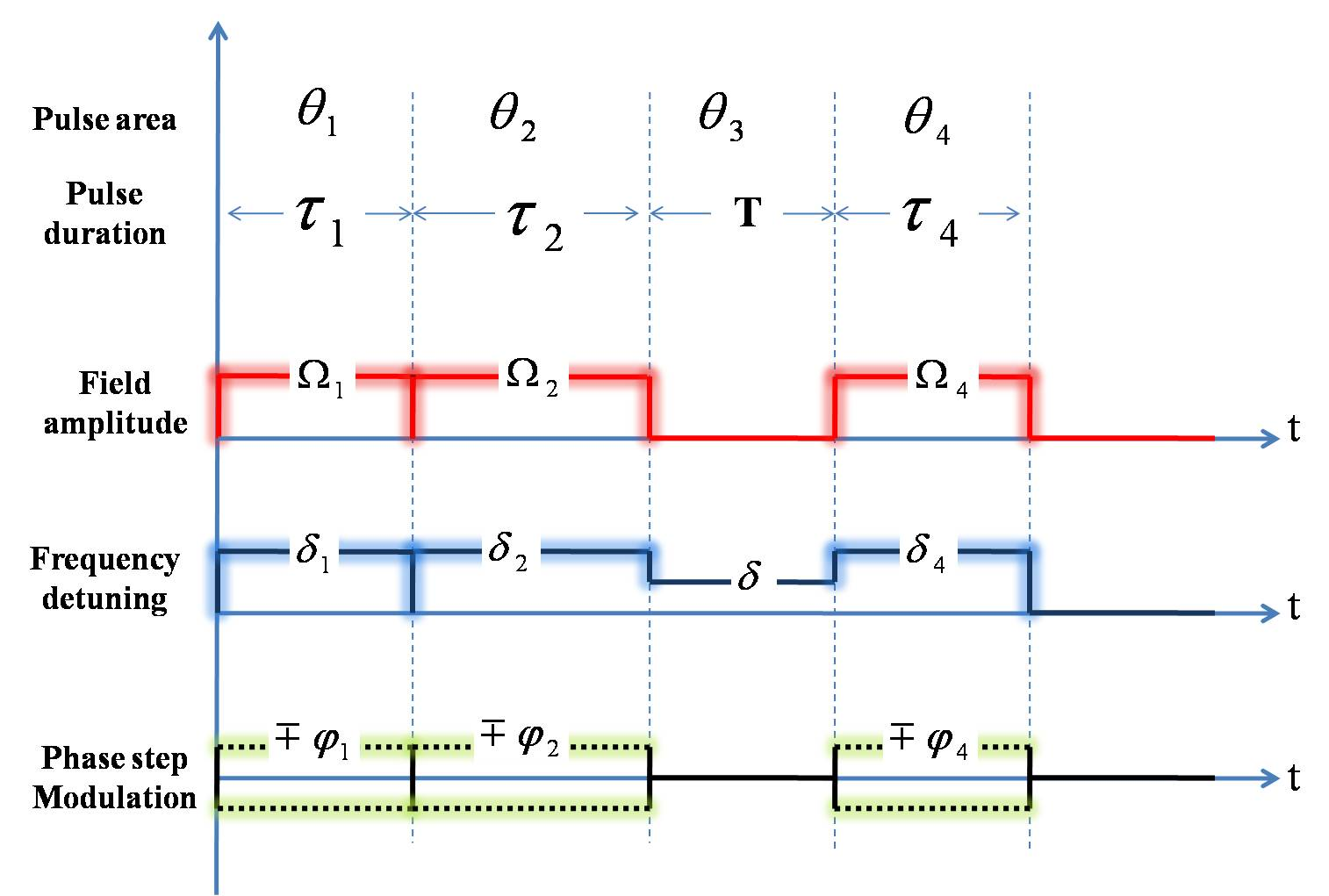}
		}
	}
	\caption{
		Composite pulses in a general three-pulse interaction scheme acting on a two-level transition for GHR$(\varphi_{l})$ and GHR$^{\dagger}(\varphi_{l})$ protocols reported in Table~\ref{tab:III-Composite-Pulses-Protocols}. Pulses are labeled by $l = (1,2,3,4)$. Each pulse is characterized by laser parameters: frequency detuning $\delta_l$, pulse
		duration $\tau_l$, field Rabi frequency $\Omega_l$ and laser phase $\varphi_l$.
	}
	\label{fig:III-Composite-Pulses-GHR-1-2}
\end{figure}

\subsection{Transition probabilities including laser phase-steps and error signal definition}
\label{sec:IIIA-Probability-Transitions}

\indent Within a GHR sequence, the second or first Ramsey pulse is divided into subsections with
individual manipulation of frequency, duration and laser phase, as shown in Fig.~\ref{fig:III-Composite-Pulses-GHR-1-2} for two types
of laser pulsed sequences, GHR$(\varphi_{l})$ and GHR$^{\dagger}(\varphi_{l})$. The calculations of different error signals require to explicitly include the $\varphi_{l}$ $(l=\textup{1,3,4,\, or} \; l=\textup{ 1,2,4})$ laser phase dependence, produced by the phase steps, within the matrix elements $\textup{M}(\theta_{l})$ in Eq.~\eqref{eq:IIA-Spinor-Matrix}. The GHR$(\varphi_{l})$ transition probability is expressed by:
\begin{equation}
	\textup{P}(\{\varphi_{l}\})_{|\textup{g}\rangle \mapsto |\textup{e}\rangle} =
		| \alpha_{\{\textup{l}\}} |^{2} \left| 1 + \beta_{\{\textup{l}\}} e^{-i(\delta T - \Phi_{\{\textup{l}\}})} \right|^{2},
	\label{eq:IIIA-Prob-Transition-GHR}
\end{equation}
where the envelopes $\alpha_{\{\textup{l}\}}$ and  $\beta_{\{\textup{l}\}}$ of the two sequences are given in Appendix~\ref{sec:VIIB-Envelopes-GHR}, taking into account both laser phases and initial atomic preparation. The composite phase-shift $\Phi_{\{\textup{l}\}}$ represents the atomic phase accumulated by the wave-function during the $l$-th laser interrogation sequence.
It is expressed for two different pulse sequences as follows.

\indent $\bullet$ For the GHR$(\varphi_{l})$ sequence:
\begin{equation}
	\Phi_{\{\textup{l}\}} = \textup{Arg} \left[
		\frac{
			\textup{M}_{\dagger}(\theta_1) e^{-i\varphi_1} c_{\textup{g}}(0) + \textup{M}_{-}(\theta_1) c_{\textup{e}}(0)
		}{
			\textup{M}_{+}(\theta_1) c_{\textup{g}}(0) + \textup{M}_{\dagger}(\theta_1) e^{i\varphi_1} c_{\textup{e}}(0)
		}
		\cdot \frac{
			\textup{M}_{-}(\theta_3,\theta_4)
		}{
			\textup{M}_{\dagger}(\theta_4,\theta_3)
		}
	\right],
	\label{eq:IIIA-Phase-Shift-GHR1}
\end{equation}

\indent $\bullet$ And for the GHR$^{\dagger}(\varphi_{l})$ sequence:
\begin{equation}
	\Phi_{\{\textup{l}\}} =  \textup{Arg} \left[
		\frac{
			\textup{M}_{\dagger}(\theta_1,\theta_3) c_{g}(0) + \textup{M}_{-}(\theta_1,\theta_3) c_{e}(0)
		}{
			\textup{M}_{+}(\theta_1,\theta_3) c_{g}(0) + \textup{M}_{\dagger}(\theta_3,\theta_1) c_{e}(0)
		}
		\cdot \frac{
			\textup{M}_{-}(\theta_4)
		}{
			\textup{M}_{\dagger}(\theta_4)
		}
	\right],
\end{equation}
in both cases, reduced matrix components are given in Appendix~\ref{sec:VIIB-Envelopes-GHR}.

\indent In the case of phase-step protocols, the dispersive-shape of the error signal $\Delta\textup{E}$ is computed by
taking the difference between two spectroscopic signals $\textup{P}(\varphi_{\{\textup{l}\}})_{|\textup{g}\rangle\mapsto|\textup{e}\rangle}$ with opposite phase $\varphi_{\{\textup{l}\}}^{+}$ and $\varphi_{\{\textup{l}\}}^{-}$ as:
\begin{equation}
		\Delta E =	\textup{P}_{|\textup{g}\rangle\mapsto|\textup{e}\rangle}(\varphi_{\{\textup{l}\}}^{+})
			- \textup{P}_{|\textup{g}\rangle\mapsto|\textup{e}\rangle}(\varphi_{\{\textup{l}\}}^{-}).
				\label{eq:IIIA-Error-Signal-Canonical}
\end{equation}
It may be written as
\begin{equation}
				\Delta E = \tilde{\textup{A}} + \tilde{\textup{B}}(\tilde{\Phi}) \cos( \delta\textup{T} + \tilde{\Phi} ),
	\label{eq:IIIA-Error-Signal-Canonical_2}
\end{equation}
where
\begin{subequations}
	\begin{align}
		\widetilde{\textup{A}} &= \textup{A}_{(\varphi_{\{\textup{l}\}}^{+})} - \textup{A}_{(\varphi_{\{\textup{l}\}}^{-})}, \\
		\widetilde{\textup{B}}(\widetilde{\Phi}) &= \left[
			\textup{B}(\Phi)_{(\varphi_{\{\textup{l}\}}^{+})} \cos\Phi_{(\varphi_{\{\textup{l}\}}^{+})}
			- \textup{B}(\Phi)_{(\varphi_{\{\textup{l}\}}^{-})} \cos\Phi_{(\varphi_{\{\textup{l}\}}^{-})}
		\right] \nonumber \\
		&\times \sqrt{ 1 + \tan^2\widetilde{\Phi}}.
	\end{align}
\end{subequations}
and
\begin{equation}
	\widetilde{\Phi} = \arctan\left[
		\frac{
			\textup{B}(\Phi)_{(\varphi_{\{\textup{l}\}}^{+})} \sin\Phi_{(\varphi_{\{\textup{l}\}}^{+})}
			- \textup{B}(\Phi)_{(\varphi_{\{\textup{l}\}}^{-})} \sin\Phi_{(\varphi_{\{\textup{l}\}}^{-})}
		}{
			\textup{B}(\Phi)_{(\varphi_{\{\textup{l}\}}^{+})} \cos\Phi_{(\varphi_{\{\textup{l}\}}^{+})}
			- \textup{B}(\Phi)_{(\varphi_{\{\textup{l}\}}^{-})} \cos\Phi_{(\varphi_{\{\textup{l}\}}^{-})}
		}
	\right],
\end{equation}
with A and B$(\Phi)$ related to $\alpha_{\{l\}}$ and $\beta_{\{l\}}$ by Eq.~\eqref{eq:IIB-Enveloppes}.

\indent This laser frequency stabilization scheme synthesizes an anti-symmetric error signal, i.e., a dispersion line-shape, to lock the laser frequency to
the center of the unperturbed clock transition. The frequency lock point shift $\delta\widetilde{\nu}$ from the error signal due to an imperfect
light-shift compensation is directly given by the relation:
\begin{equation}
	\Delta\textup{E}|_{\delta=\delta\widetilde{\nu}} = 0.
	\label{eq:IIIA-Error-Signal-Lock-Condition}
\end{equation}
Using Eqs.~\eqref{eq:IIIA-Error-Signal-Canonical_2} and~\eqref{eq:IIIA-Error-Signal-Lock-Condition}, the analytical form of the frequency-shifted lock-point is
\begin{equation}
	\delta\widetilde{\nu} = \frac{1}{2\pi\textup{T}} \left(
		-\widetilde{\Phi}|_{\delta\rightarrow0} \pm \arccos\left[
			-\frac{
				\widetilde{\textup{A}}|_{\delta\rightarrow0}
			}{
				\widetilde{\textup{B}}(\widetilde{\Phi})|_{\delta\rightarrow0}
			}
		\right]
	\right).
	\label{eq:IIIA-Shift-DeltaNu}
\end{equation}
This expression is similar to the rotation parametrization applied in quantum computing to achieve robust cancelations of systematic errors \cite{Bando:2013,Shaka:1983}.

\subsection{Error signals of R and HR schemes}
\label{sec:IIIB-Error-Signals-Standard}

\begin{figure*}[t!!]
	\centering
	\subfloat[Ramsey (\textup{R})]{
		\resizebox{0.45\linewidth}{!}{
			\includegraphics[angle=0]{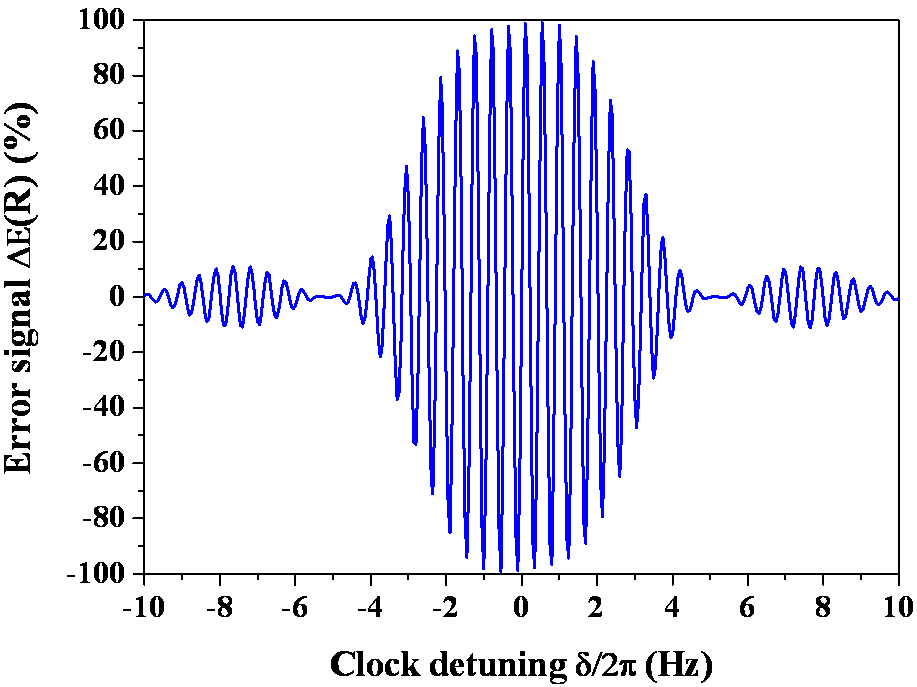}
		}
		\resizebox{0.45\linewidth}{!}{
			\includegraphics[angle=0]{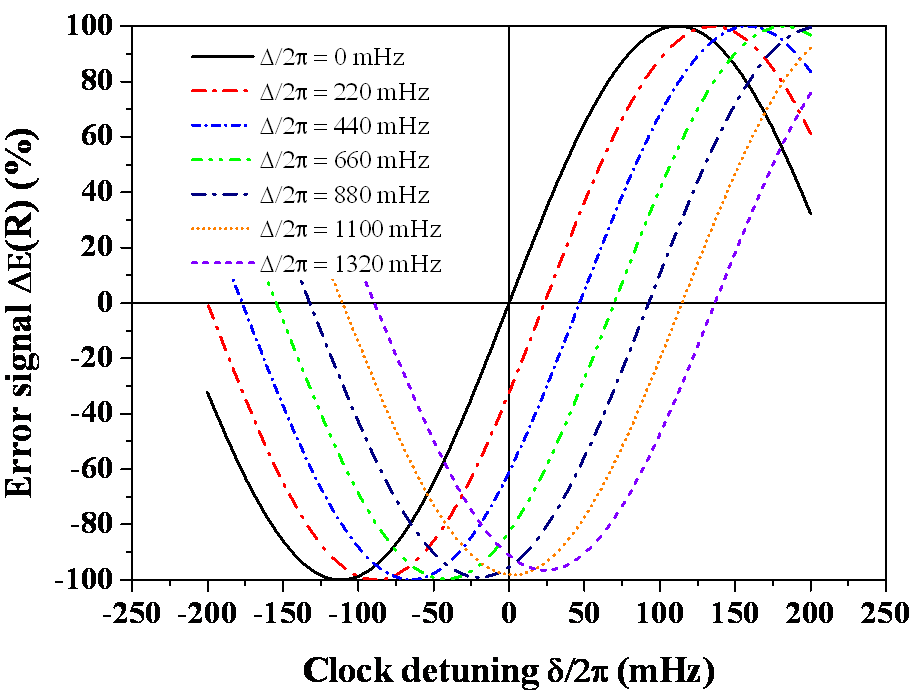}
		}
		\label{fig:IIIB-Phase-Modulation-R-HR-a}
	} \\
	\subfloat[Hyper-Ramsey (\textup{HR}-$\pi$)]{
		\resizebox{0.45\linewidth}{!}{
			\includegraphics[angle=0]{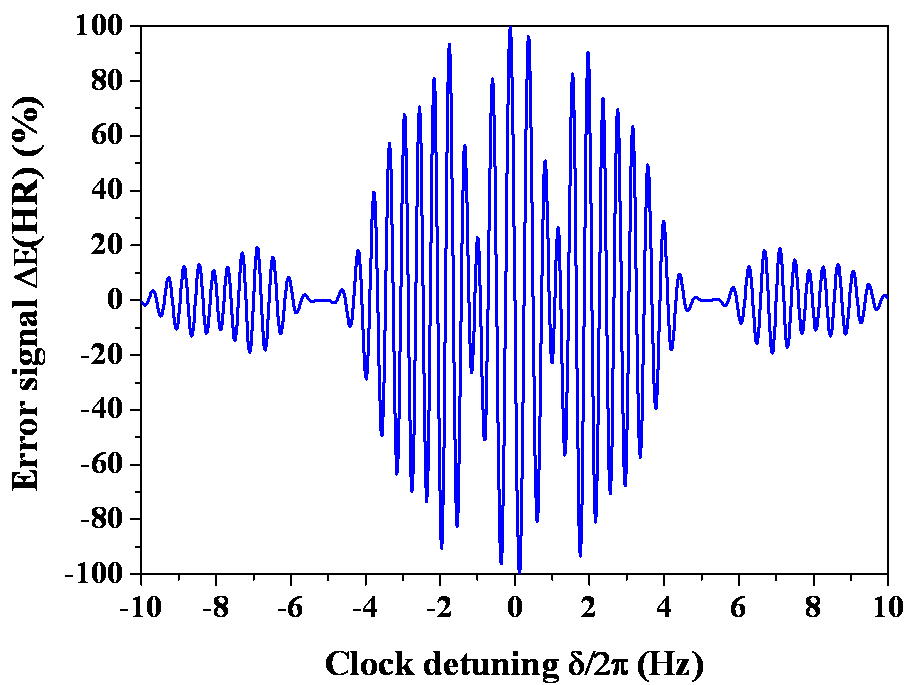}
		}
		\resizebox{0.45\linewidth}{!}{
			\includegraphics[angle=0]{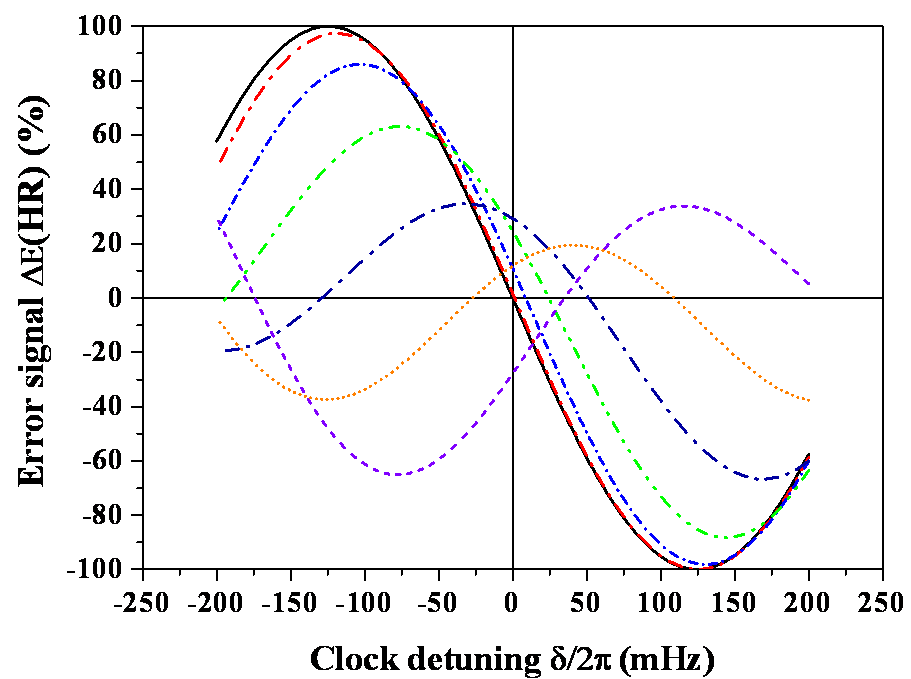}
		}
		\label{fig:IIIB-Phase-Modulation-R-HR-b}
	}\\
   \subfloat[Modified Hyper-Ramsey (\textup{MHR)}]{
		\resizebox{0.45\linewidth}{!}{
			\includegraphics[angle=0]{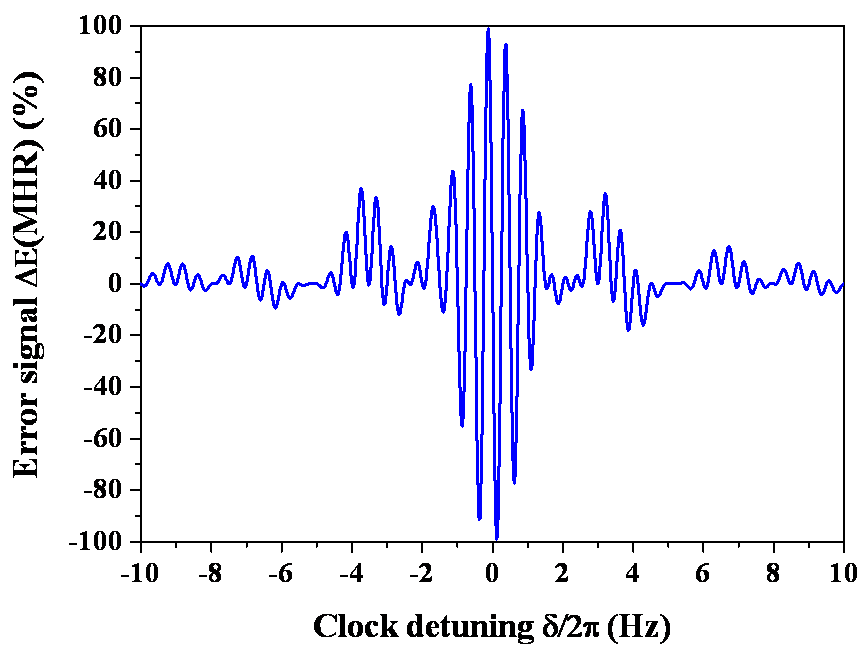}
		}
		\resizebox{0.45\linewidth}{!}{
			\includegraphics[angle=0]{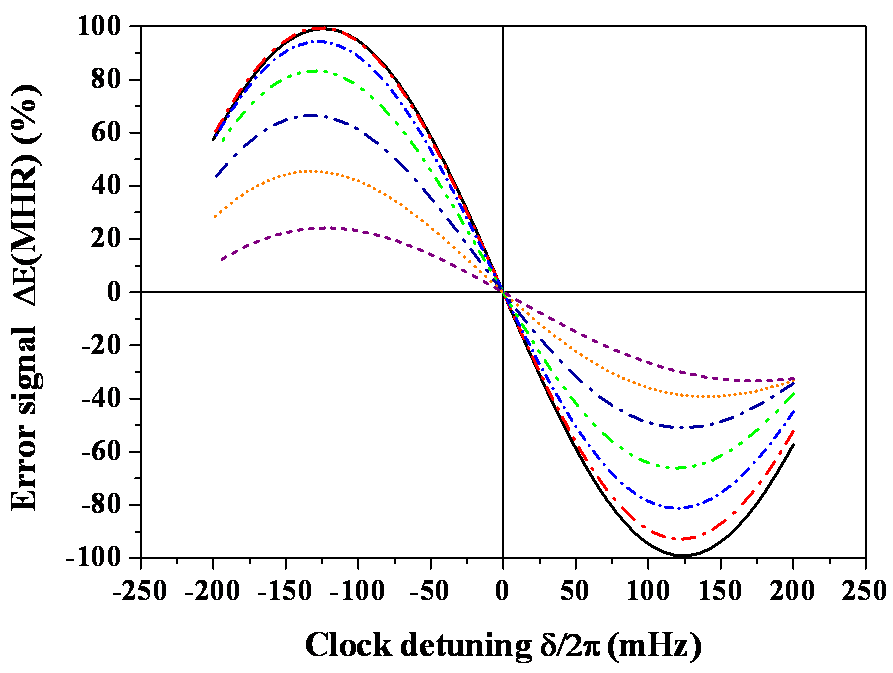}
		}
		\label{fig:IIIB-Phase-Modulation-R-HR-c}
	}
	\caption{
		Left column shows error signals against clock frequency detuning in presence of uncompensated residual light-shifts. Right
		column shows the generated frequency lock point sensitivity against uncompensated residual light-shifts. Other parameters are
		the same as in Fig.~\ref{fig:IIB-Composite-Pulses-R-HR-Pi}.
	}
	\label{fig:IIIB-Phase-Modulation-R-HR}
\end{figure*}

This Subsection, as well the next one, reviews the robustness of different dispersive errors signals to some
residual light-shifts and pulse area variations. Some relevant $\pm\pi/2$ phase-step protocols are reported in
Table~\ref{tab:III-Composite-Pulses-Protocols} for application in R spectroscopy~\cite{Ramsey:1951,Letchumanan:2004,Letchumanan:2006} and HR spectroscopy~\cite{Yudin:2010, Huntemann:2012b,Huntemann:2016}.

\indent The first laser phase-step configuration based on the R protocol in Tab.~\ref{tab:III-Composite-Pulses-Protocols}  was initially proposed in \cite{Ramsey:1951}. Following Eq.~\eqref{eq:IIIA-Error-Signal-Canonical}, the error signal is produced by combining two Ramsey transition probabilities in presence of $\pm\pi/2$ phase shifts
\begin{equation}
		\Delta\textup{E}_{\textup{R}} =
			\textup{P}(\pi/2,0)_{|\textup{g}\rangle \mapsto |e\rangle} -
			\textup{P}(-\pi/2,0)_{|\textup{g}\rangle \mapsto |\textup{e}\rangle}
		\label{eq:IIIB-Error-Signal-R}.
\end{equation}
This approach leads to dispersive-like resonance lineshapes with increased sensitivity to detect the clock resonance
frequency. This phase-step modulation was applied to single ion clock devices in order to produce better control of the frequency
lock point stabilizing the local laser oscillator \cite{Letchumanan:2004,Letchumanan:2006}. The error signal of the R phase-step protocol is plotted on the top left part
Fig.~\ref{fig:IIIB-Phase-Modulation-R-HR-a}. Its associated dispersive signal is plotted for several values of the uncompensated
residual light-shift on the top right part.

\indent The error signal of the second phase-step configuration based on the HR-$\pi$ protocol, presented in Fig.~\ref{fig:IIIB-Phase-Modulation-R-HR-b},
is based on the same $\varphi_1=\pm\pi/2$ phase-steps in presence of an additional phase reversal $\varphi_3=\pi$ of the laser field during
the intermediate pulse. Following Eq.~\eqref{eq:IIIA-Error-Signal-Canonical}, the error signal is
\begin{equation}
			\Delta\textup{E}_{\textup{HR}} =
			\textup{P}(\pi/2,\pi,0)_{|\textup{g}\rangle \mapsto |e\rangle} -
			\textup{P}(-\pi/2,\pi,0)_{|\textup{g}\rangle \mapsto |\textup{e}\rangle}.
		\label{eq:IIIB-Error-Signal-HR}
\end{equation}
The main advantage of such a protocol is to generate a frequency lock point which is driven by a cubic non-linear
sensitivity to the uncompensated residual light-shifts leading to a much better control of the frequency discriminant as shown on the right
of Fig.~\ref{fig:IIIB-Phase-Modulation-R-HR-b}.\\

\indent The HR phase-step protocol was proposed in \cite{Yudin:2010} and experimentally implemented on a single ion $^{171}\textup{Yb}^{+}$ clock in
\cite{Huntemann:2012b} to strongly reduce the residual probe-induced frequency shift by four orders of magnitude.
Recently, using this spectroscopic technique \cite{Huntemann:2016}, the single ion clock achieved a $1.1\times10^{-18}$ systematic relative uncertainty of probe induced shifts. This protocol still suffers from small residual light-shifts when exploring a wider range of uncompensated frequency offsets. The HR protocol has a residual uncompensated
frequency-shift of around 1-2~mHz over a 220~mHz residual offset which may compromise the access to a  clock fractional accuracy below $10^{-18}$.

\subsection{Error signals of MHR and GHR schemes}
\label{sec:IIIC-Error-Signals-Non-Standard}

\begin{figure*}[t!!]
	\centering
	\subfloat[\textup{GHR}($\pi/4$)]{
		\resizebox{0.425\linewidth}{!}{
			\includegraphics[angle=0]{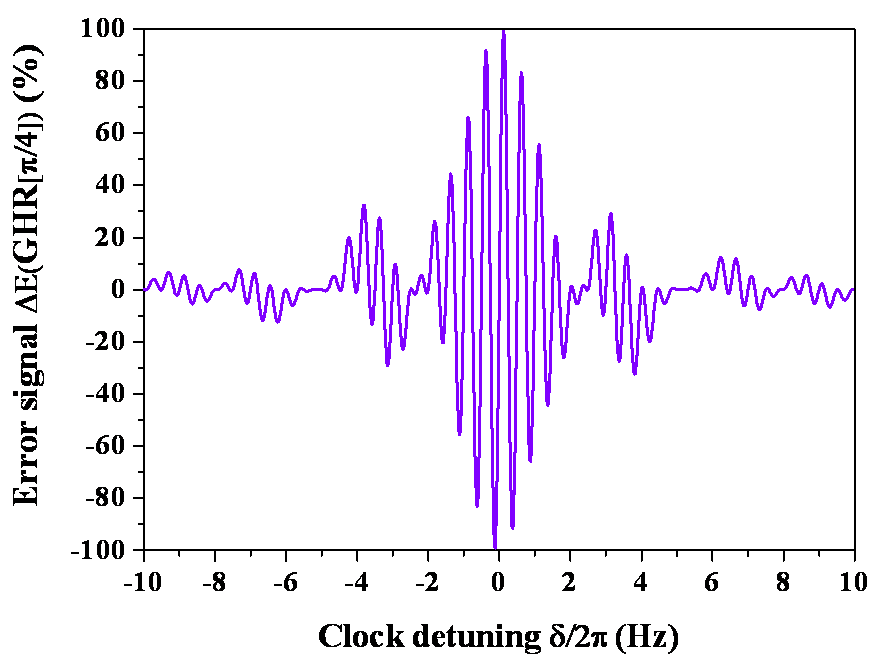}
		}
		\resizebox{0.425\linewidth}{!}{
			\includegraphics[angle=0]{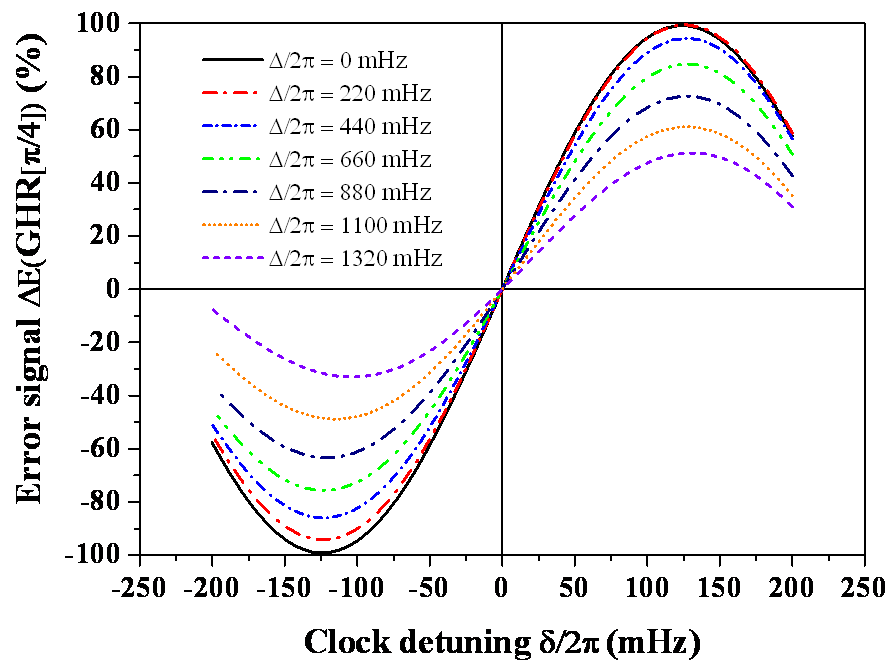}
		}
		\label{fig:IIIC-Phase-Modulation-MHR-GHR-a}
	} \\
	\subfloat[\textup{GHR}($3\pi/4$)]{
		\resizebox{0.425\linewidth}{!}{
			\includegraphics[angle=0]{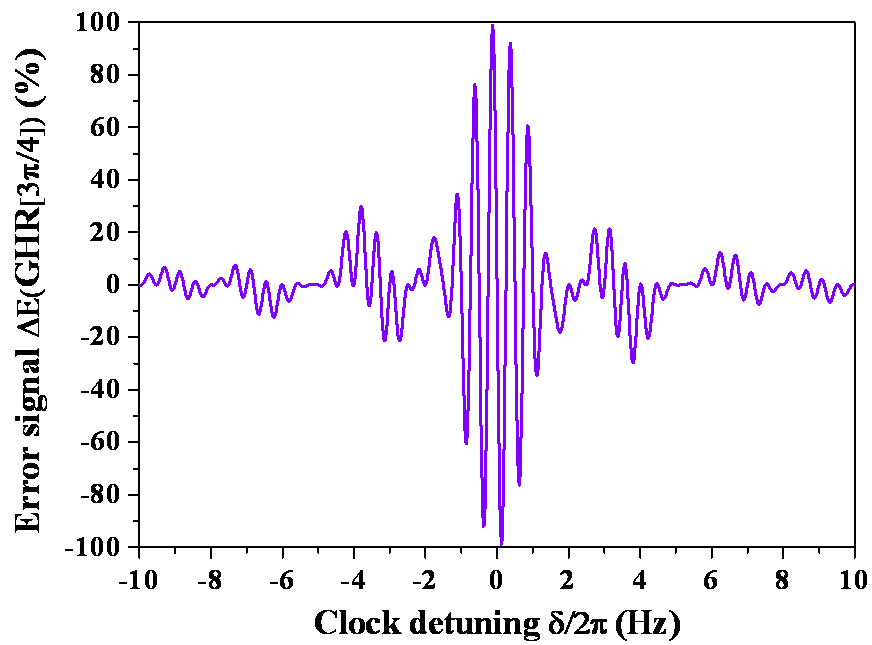}
		}
		\resizebox{0.425\linewidth}{!}{
			\includegraphics[angle=0]{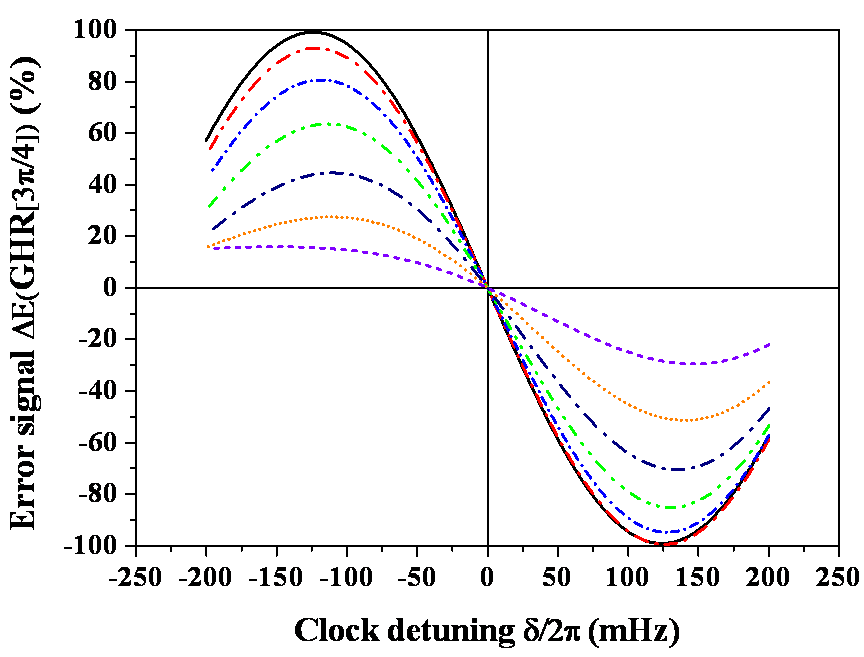}
		}
		\label{fig:IIIC-Phase-Modulation-MHR-GHR-b}
	}\\
	\subfloat[\textup{GHR}($\pi/4,3\pi/4$)]{
		\resizebox{0.425\linewidth}{!}{
			\includegraphics[angle=0]{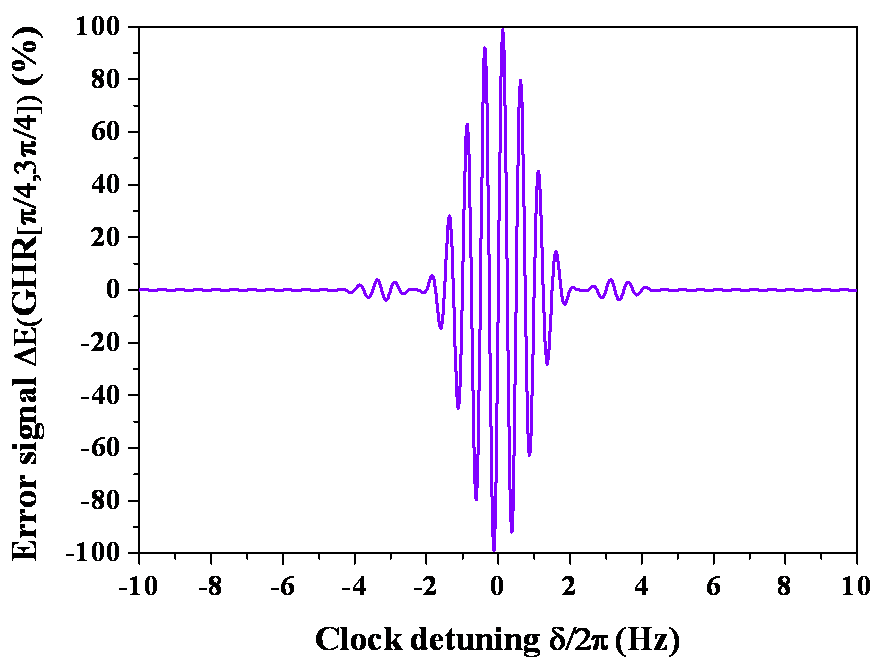}
		}
		\resizebox{0.425\linewidth}{!}{
			\includegraphics[angle=0]{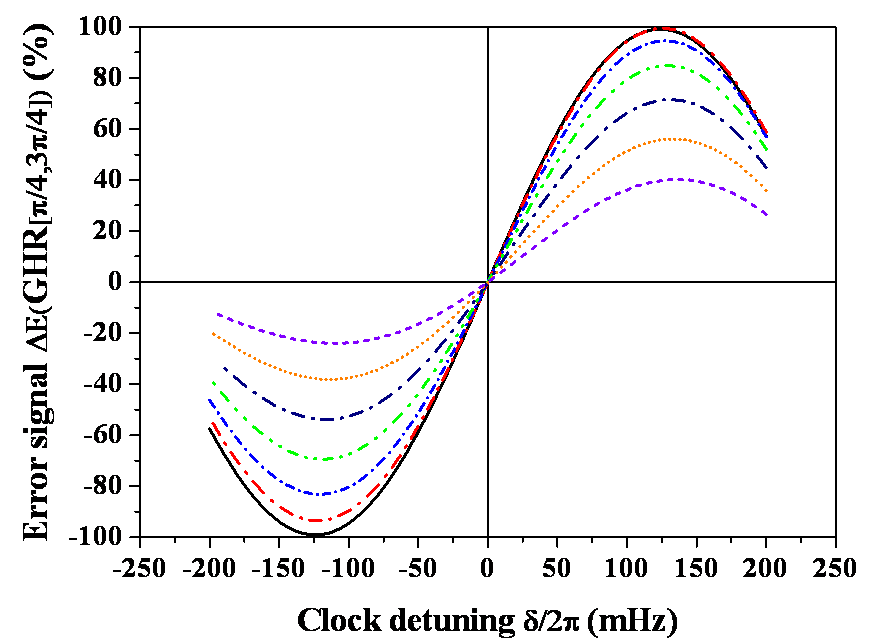}
		}
		\label{fig:IIIC-Phase-Modulation-MHR-GHR-c}
	}
	\caption{
		Left column shows error signals against clock frequency detuning $\delta$ for GHR$(\pi/4)$, GHR$(3\pi/4)$ and GHR$(\pi/4,3\pi/4)$ phase-step protocols. Right column shows associated frequency lock point sensitivity for various uncompensated residual light-shifts $\Delta$. Parameters are as in Fig.~\ref{fig:IIB-Composite-Pulses-R-HR-Pi}.
	}
	\label{fig:IIIC-Phase-Modulation-MHR-GHR}
\end{figure*}

\indent Spectroscopic schemes using different phase-step protocols have been recently introduced in order to completely eliminate residual
light-shift corrections on the central fringe over large uncompensated residual light-shifts.
These are the MHR \cite{Hobson:2016} and GHR \cite{Zanon-Willette:2016b} schemes.
These new dispersive error signals are centered at the unperturbed atomic resonance with steep discriminants which are impervious
to variations in laser probe induced clock frequency shifts.\\
\indent The error signals are built from a combination of different transition probabilities as in~\cite{Hobson:2016,Zanon-Willette:2016b}:
\begin{subequations}
	\begin{align}
		\Delta\textup{E}_{\textup{MHR}} &=
			\textup{P}(\pi/2,\pi,0)_{|\textup{g}\rangle \mapsto |\textup{e}\rangle}
			- \textup{P}(0,\pi,-\pi/2)_{|\textup{g}\rangle \mapsto |\textup{e}\rangle},
		\label{eq:IIIC-Error-Signal-MHR} \\
		\Delta\textup{E}_{\textup{GHR}(\varphi_3)} &=
			\textup{P}(0,+\varphi_3,0)_{|\textup{g}\rangle \mapsto |\textup{e}\rangle}
			- \textup{P}(0,-\varphi_3,0)_{|\textup{g}\rangle \mapsto |\textup{e}\rangle}.
		\label{eq:IIIC-Error-Signal-GHR}
	\end{align}
\end{subequations}
\indent The MHR protocol described by Eq.~\eqref{eq:IIIC-Error-Signal-MHR} is based on a superposition of two HR-$\pi$ transition
probabilities. For error signal generation, the phase-step modulation is obtained by interleaving a HR-$\pi$ protocol from
Tab.~\ref{tab:III-Composite-Pulses-Protocols}, where $\varphi_1=\pi/2$ during the first pulse, with a HR-$\pi$ protocol
where an opposite phase $\varphi_4=-\pi/2$ is used during the last pulse.
The calculated dispersive error signal shape is presented in Fig.~\ref{fig:IIIB-Phase-Modulation-R-HR-c} along its frequency
lock point response to residual probe-induced shifts.
This non standard protocol was the first to synthesize an error signal yielding full immunity to residual probe light-shifts and great
robustness to pulse area errors originated from laser power variations. It has been successfully tested in a neutral atom optical lattice clock
based on magnetically induced spectroscopy \cite{Hobson:2016} demonstrating suppression of a sizable $2\times10^{-13}$ probe Stark
shift to below $10^{-16}$ even with very large errors in shift compensation.\\
\indent The GHR($\pi/4$) and GHR($3\pi/4$) protocols from Tab.~\ref{tab:III-Composite-Pulses-Protocols}, proposed in
\cite{Zanon-Willette:2016b} and shown in Fig.~\ref{fig:IIIC-Phase-Modulation-MHR-GHR-a} and Fig.~\ref{fig:IIIC-Phase-Modulation-MHR-GHR-b}, respectively, use a single either
$\varphi_3=\pm\pi/4$ or a $\varphi_3=\pm3\pi/4$ phase-step modulation during the intermediate pulse.
The associated error signals computed using Eq.~\eqref{eq:IIIC-Error-Signal-GHR} are  presented in the same figure. The plots of their
lock point sensitivity shown on the right in Fig.~\ref{fig:IIIC-Phase-Modulation-MHR-GHR-a} and
Fig.~\ref{fig:IIIC-Phase-Modulation-MHR-GHR-b}, evidence that they are fully protected from errors in residual probe-induced
light-shifts.
Another stabilization scheme can be generated by combining the error signals of the GHR($\pi/4$) and GHR($3\pi/4$)
protocols, presenting opposite slopes of the error signal. This hybrid scheme denoted GHR($\pi/4,3\pi/4$) is defined by the
following normalized difference between two error signals:
\begin{equation}
	\begin{split}
		\Delta\textup{E}(\pi/4,3\pi/4)&=
			\frac{1}{2} \left( \Delta\textup{E}_{\textup{GHR}(\pi/4)} - \Delta\textup{E}_{\textup{GHR}(3\pi/4)} \right).
	\end{split}
	\label{eq:GHR-Pi-3Pi}
\end{equation}
We have reported the corresponding error signal shape $\Delta\textup{E}_{\textup{GHR}(\pi/4,3\pi/4)}$ in Fig.~\ref{fig:IIIC-Phase-Modulation-MHR-GHR-c}.
A combination of such protocols will be demonstrated to be efficient in presence of decoherence and relaxation (see \ref{sec:IV-Dissipation}). 
\subsection{Robustness of error signal slopes}
\label{sec:IIID-Error-Signal-Slopes}

\begin{figure*}[t!!]
	\centering
	\begin{minipage}[c]{0.425\linewidth}
		\subfloat[Ramsey (\textup{R})]{
			\resizebox{\linewidth}{!}{
				\includegraphics[angle=0]{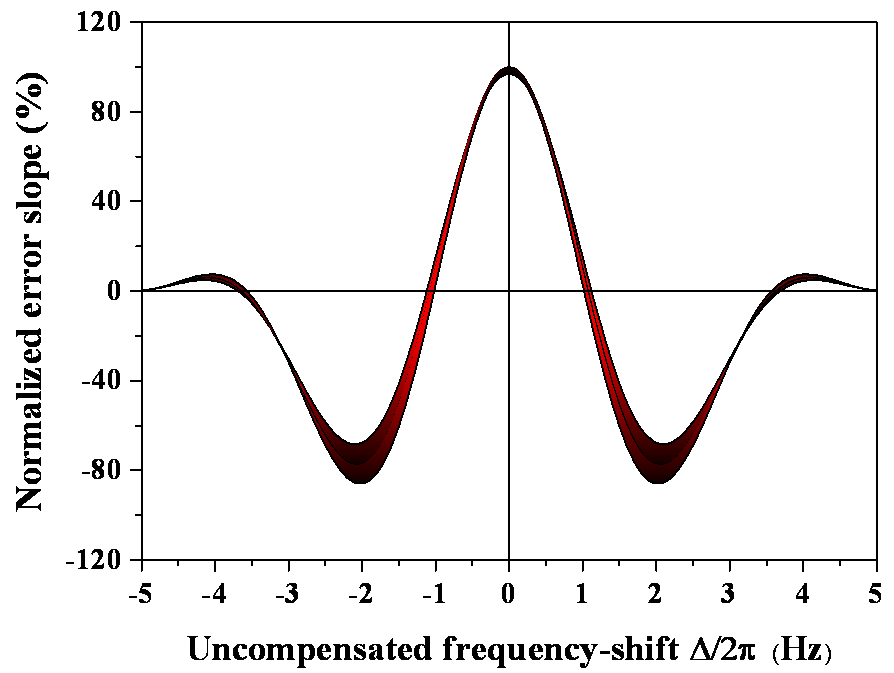}
			}
			\label{fig:IIID-Error-Slope-Comparison-a}
		} \\
		\subfloat[\textup{HR}-$\pi$]{
			\resizebox{\linewidth}{!}{
				\includegraphics[angle=0]{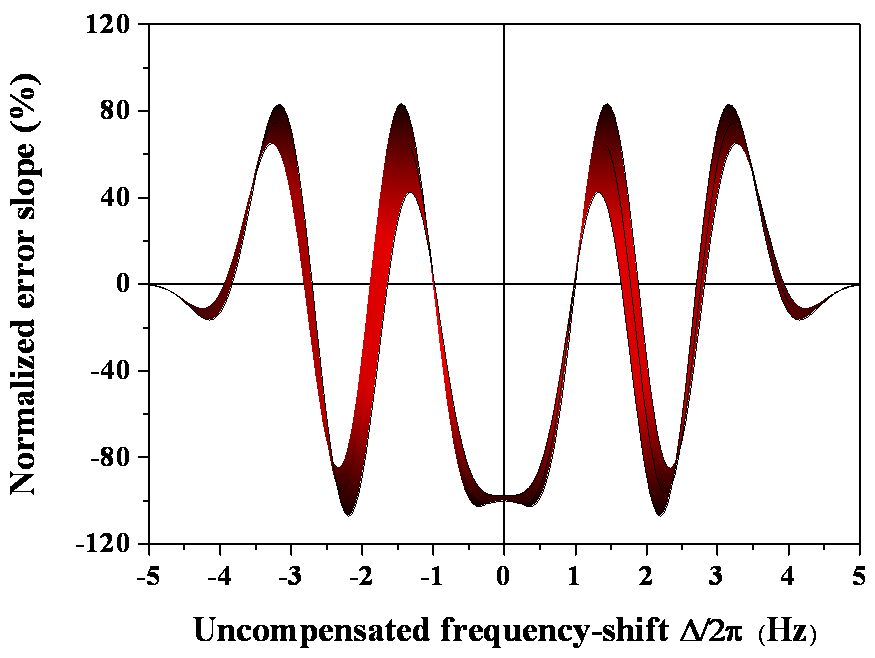}
			}
			\label{fig:IIID-Error-Slope-Comparison-b}
		} \\
		\subfloat[\textup{MHR}]{
			\resizebox{\linewidth}{!}{
				\includegraphics[angle=0]{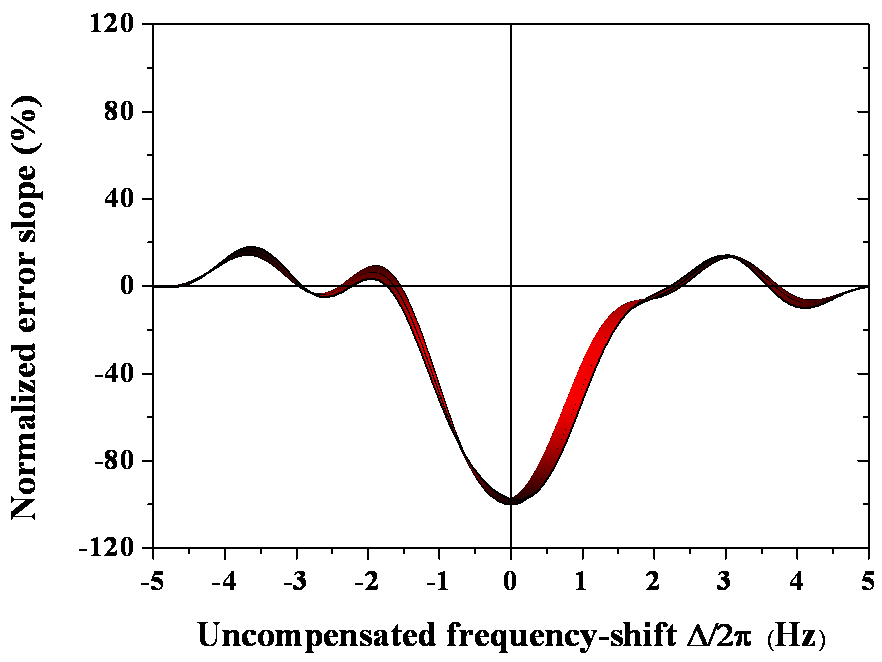}
			}
			\label{fig:IIID-Error-Slope-Comparison-c}
		}
	\end{minipage}
	\begin{minipage}[c]{0.425\linewidth}
		\subfloat[\textup{GHR}($\pi/4$)]{
		    \resizebox{\linewidth}{!}{
				\includegraphics[angle=0]{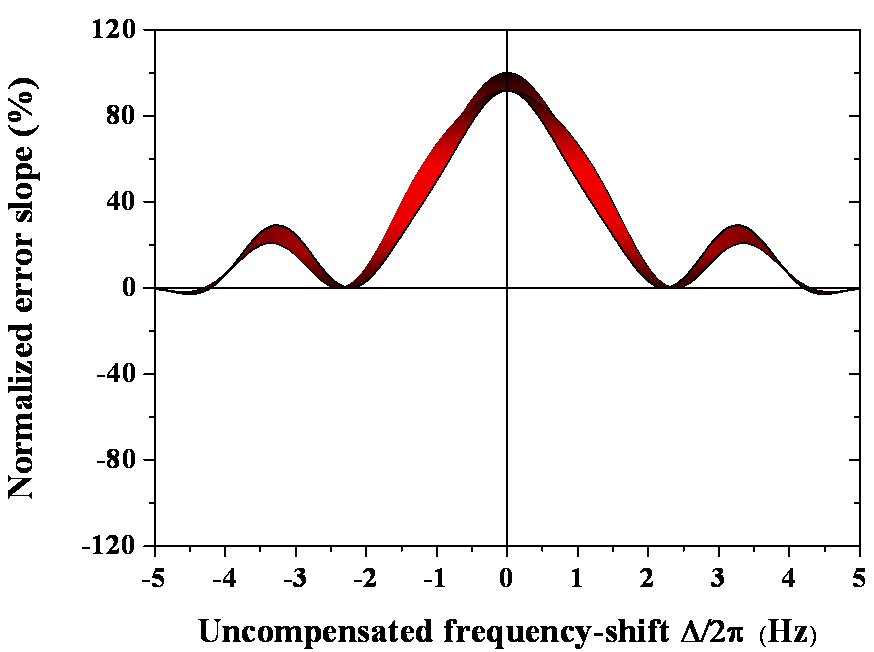}
			}
			\label{fig:IIID-Error-Slope-Comparison-d}
		} \\
		\subfloat[\textup{GHR}($3\pi/4$)]{
		    \resizebox{\linewidth}{!}{
				\includegraphics[angle=0]{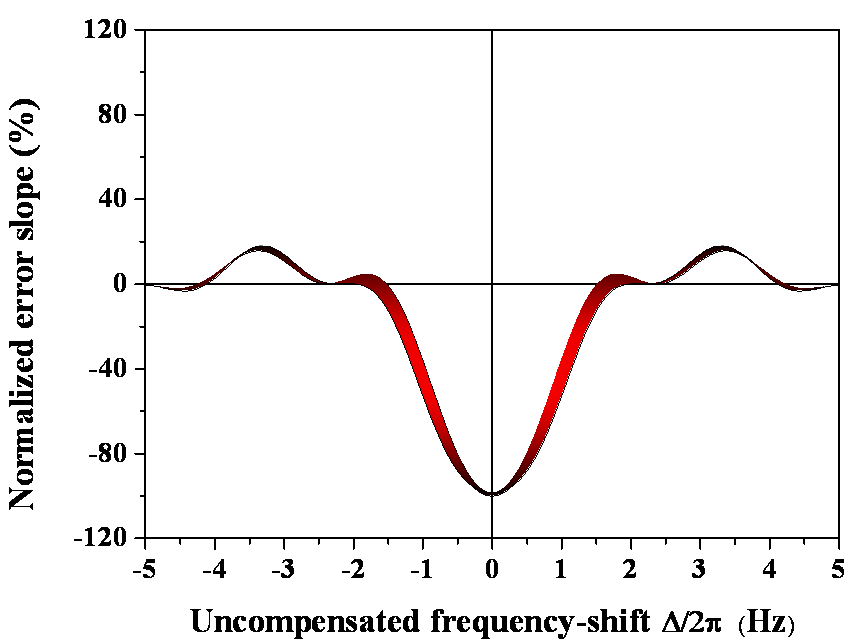}
			}
			\label{fig:IIID-Error-Slope-Comparison-e}
		} \\
		\subfloat[\textup{GHR}($\pi/4,3\pi/4$)]{
			\resizebox{\linewidth}{!}{
				\includegraphics[angle=0]{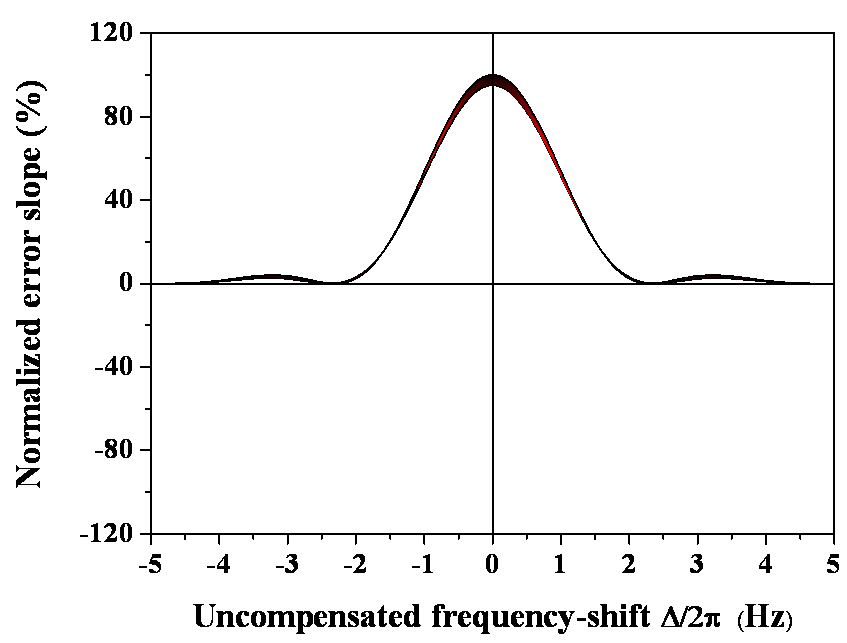}
			}
			\label{fig:IIID-Error-Slope-Comparison-f}
		}
	\end{minipage}
	\caption{
		Comparison of normalized error signal slopes at $\delta=0$ for the protocol investigated in Fig.~\ref{fig:IIIB-Phase-Modulation-R-HR} and
		Fig.~\ref{fig:IIIC-Phase-Modulation-MHR-GHR} against residual uncompensated light-shift $\Delta$ (solid lines) and in presence of a $\Delta\theta_l/\theta_l=\pm10\%$ pulse area variation (shadow region).}
	\label{fig:IIID-Error-Slope-Comparison}
\end{figure*}

\indent This subsection presents parameters useful to setup optimal working conditions in interrogation
protocols in order to achieve a very efficient lock of the local oscillator to the unperturbed optical clock transition \cite{Riis:2004}.
For simplicity, and because out of scope of this work, we neglect statistical fluctuation in the measured signals produced by external time-dependent perturbation or phase noise contribution \cite{Kabytayev:2014,Chen:2012}.\\
\indent We study the stability and robustness of error signal slopes at $\delta=0$ against residual uncompensated light-shift $\Delta_\textup{l}$
associated to small variations of pulse areas. The frequency stability
scales as the inverse of the slope and a reduced slope results in a clock instability.
Error signal slopes against residual shifts for all protocols are shown in Fig.~\ref{fig:IIID-Error-Slope-Comparison} for a
$\Delta\theta_l/\theta_l=\pm10\%$ pulse area variation.
Compared to standard error signals based on R protocol in Fig.~\ref{fig:IIID-Error-Slope-Comparison-a} and HR protocol in Fig.~\ref{fig:IIID-Error-Slope-Comparison-b}, the new protocols expand the possible range for the uncompensated shift, i.e., the range between two zero crossings of the error slope.
Both residual uncompensated offsets and pulse area variations are directly transferred to a slope
reduction of error signals with no change in their frequency lock point. However the MHR protocol shown in
Fig.~\ref{fig:IIID-Error-Slope-Comparison-c} presents a small shape asymmetry and a tiny deviation of the maximum slope value from $\Delta=0$.

Note that the GHR$(\pi/4,3\pi/4)$ protocol, shown in Fig.~\ref{fig:IIID-Error-Slope-Comparison-f},
eliminates unstable operation by an undesired sign inversion of the slope when the uncompensated light shift is too large.

\indent If the light shift is not constant over the pulse sequences, and therefore a large residual offset is present, a strong distortion or rotation of the error signal slope is originated (see for example Fig.~\ref{fig:IIIB-Phase-Modulation-R-HR-b}, Fig.~\ref{fig:IIIB-Phase-Modulation-R-HR-c} and Fig.~\ref{fig:IIIC-Phase-Modulation-MHR-GHR})
and it may compromise the lock point stability and degrade the clock operating condition.  Then, the laser compensation step $\Delta_\textup{step}$  should be checked and steered to the point where
$\Delta\simeq 0$.

\indent Various implementations have been experimentally tested. For the single ion clock frequency standard \cite{Huntemann:2012b},
a stabilization using the HR phase-step protocol scheme was combined with a second interleaved servo system where Rabi spectroscopy
with the same probe light intensity is used.
Then, the frequency difference between the two interrogation techniques  was used to control laser frequency step $\Delta_\textup{step}$.
This method ensures that slow drifts of the light-shift will not degrade its suppression.
It is also possible to use some clock frequency-shift symmetries offered by the MHR protocol with the uncompensated part of the
probe shift to design an efficient steering process to $\Delta=0$ \cite{Hobson:2016}.

\section{Protocols based on free evolution time combinations}
\label{sec:V-Synthetic-HR}

\indent This section reviews some combinations of R and HR interrogation protocols with different free evolution times.
The target is to generate a clock frequency-shift strongly protected against residual light-shifts over larger offset clock detunings.
Some non-linear clock frequency-shifts can be synthesized specifically to be extremely robust against decoherence and relaxation.

\subsection{Synthetic frequency protocol for HR spectroscopy}

\indent The synthetic shift technique reduces the sensitivity to pulse area variations and extends the non-linear efficiency of HR protocols to
larger uncompensated residual light-shifts. It is based on both independent and parallel measurements of several clock-frequency shifts for different free evolution times and
careful combination of those measurements to generate the so-called synthetic frequency-shift. This approach is more robust than
previous phase-step locking protocols, reducing both residual uncompensated light-shifts and laser power variations even in presence of
decoherence.

\indent The synthetic frequency method, discussed in detail in~\cite{Yudin:2016}, is based on a polynomial serie expansion of the clock's residual frequency shift on its
dependence to free evolution time $\textup{T}$ under frequency stabilization:
\begin{equation}
	\delta\nu_{\textup{T}} = \frac{\textup{A}_1}{\textup{T}} + \frac{\textup{A}_2}{\textup{T}^2} + ...
		+ \frac{\textup{A}_n}{\textup{T}^n} + ...,
	\label{eq:IVD-Shift-Serie-Expansion}
\end{equation}
The coefficients $\textup{A}_n$ depend on pulse parameters (durations, amplitudes, phases) and uncompensated  frequency shift
$\Delta_{l}$. This method was originally designed to allow suppression of the black-body radiation shift experienced by atomic clocks \cite{Yudin:2011}, but it can easily be
extended to handle arbitrary systematic shifts (Stark shift, Zeeman shift, and so on).

\indent The basic idea is to build a synthetic frequency using multiple HR sequences with specific choices of free evolution
times (but same pulse parameters, assuming we can enforce exact ratios between Ramsey free evolution times) to cancel contributions up to a given order in Eq.~\eqref{eq:IVD-Shift-Serie-Expansion}. For example, using two sequences with free evolution times $\textup{T}_1$ and $\textup{T}_2$, and stabilized frequencies $\nu_1$ and $\nu_2$ shifted by $\delta_1$ and $\delta_2$, respectively,   the synthetic frequency at the lowest order is defined as
\begin{equation}
	\nu_{\text{syn}} = \frac{\nu_1 - \varepsilon_{12}\nu_2}{1 - \varepsilon_{12}},
	\label{eq:IVD-Synthetic-Frequency}
\end{equation}
where $\varepsilon_{12} = \delta_1/\delta_2$ is the offset detuning ratio. Considering the case of
$\textup{T}_1 = \textup{T}$ and $\textup{T}_2 = \textup{T}/2$, it can be shows that the frequency shift can be written as:
\begin{equation}
	\delta\nu_{\text{syn}}^{(1)} = 2\delta\nu_{\textup{T}} - \delta\nu_{\textup{T}/2}.
	\label{eq:IVD-Syn-Order-2}
\end{equation}

\indent Similarly, using three different HR sequences with free evolution times
$\textup{T}_1 = \textup{T}$, $\textup{T}_2 = \textup{T}/2$, and $\textup{T}_3 = \textup{T}/3)$, the shift of the the synthetic frequency at the following order is given by:
\begin{equation}
	\delta\nu_{\text{syn}}^{(2)} = 3\delta\nu_{\textup{T}} - 3\delta\nu_{\textup{T}/2} + \delta\nu_{\textup{T}/3}.
	\label{eq:IVD-Syn-Order-3}
\end{equation}
\indent Following this pattern to higher orders, it results that the higher orders expansions follow binomial coefficient laws.
Using Eq.~\eqref{eq:A-tan-phi} for the HR phase shift, the clock frequency shift  is calculated  on the basis of Eq.~\eqref{eq:IIC-Clock-Frequency-Shift-High-Order}. Thus, under $|\delta/\Omega|^2 \ll 1$, the calculations show the following general character of dominating dependencies on $\delta/\Omega$:
\begin{equation}
		\delta\nu_{\textup{T}} \approx \frac{4}{\pi}
			\left( \frac{\delta}{\Omega} \right)^3
	\label{eq:IVD-Syn-HRA-Gen-1}
\end{equation}
without synthetic frequency approach, and
\begin{equation}
	\begin{split}
		\delta\nu^{(1)}_{\text{syn}} &\approx \frac{48}{\pi^2} \frac{4\tau}{\textup{T}}
			\left( \frac{\delta}{\Omega} \right)^5 ; \\
		\delta\nu^{(2)}_{\text{syn}} &\approx \frac{865}{\pi^3} \left( \frac{4\tau}{\textup{T}} \right)^2
			\left( \frac{\delta}{\Omega} \right)^7.
	\end{split}
	\label{eq:IVD-Syn-HRA-Gen}
\end{equation}
for the synthetic frequency approaches at different orders.
\begin{figure}[t!!]
	\center
	\resizebox{8cm}{!}{
		\includegraphics[angle=0]{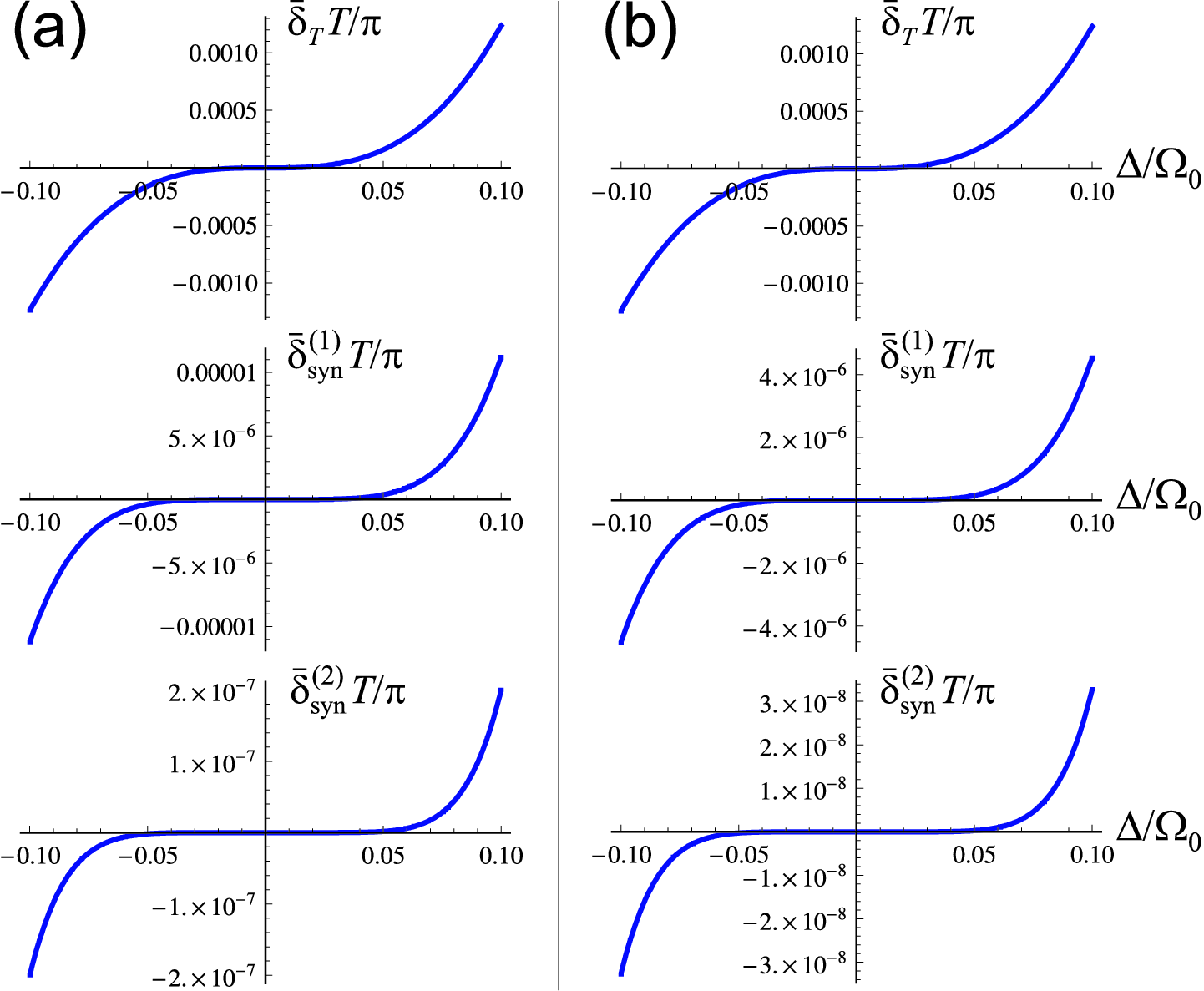}
	}
		\caption{
		(Color online) For the HR$-\pi$ protocol, dependencies vs $\delta/\Omega$ for the shift without synthetic frequency approach of Eq.~\eqref{eq:IVD-Syn-HRA-Gen-1}, denoted as
$\bar{\delta}_{\textup{T}}$, and using the first and second order synthetic frequency approaches of Eqs.~\eqref{eq:IVD-Syn-HRA-Gen}, denoted as
$\bar{\delta}^{(1)}_\text{syn}$, and $\bar{\delta}^{(2)}_\text{syn}$. $\Omega\tau$=$\pi/2$ in all cases, and $4\tau/\textup{T}$ is 0.25 in (a), and 0.1 in (b).}
	\label{fig:IVD-HRA-1}
\end{figure}

\indent Fig.~\ref{fig:IVD-HRA-1} reports the calculations for the above quantities. For the synthetic frequencies, higher-order (more than cubic) non-linearities appear. This character is not changed under variations of $\Omega$, $\tau$, and $\textup{T}$, i.e., we do not need the rigorous condition $\Omega\tau$=$\pi/2$. Because in real experiments the value of $\Omega$ can be controlled only
at the level of 1-10$\%$, this method can be very successful in atomic clock implementations.

\begin{figure}[t!!]
	\center
	\resizebox{8cm}{!}{
		\includegraphics[angle=0]{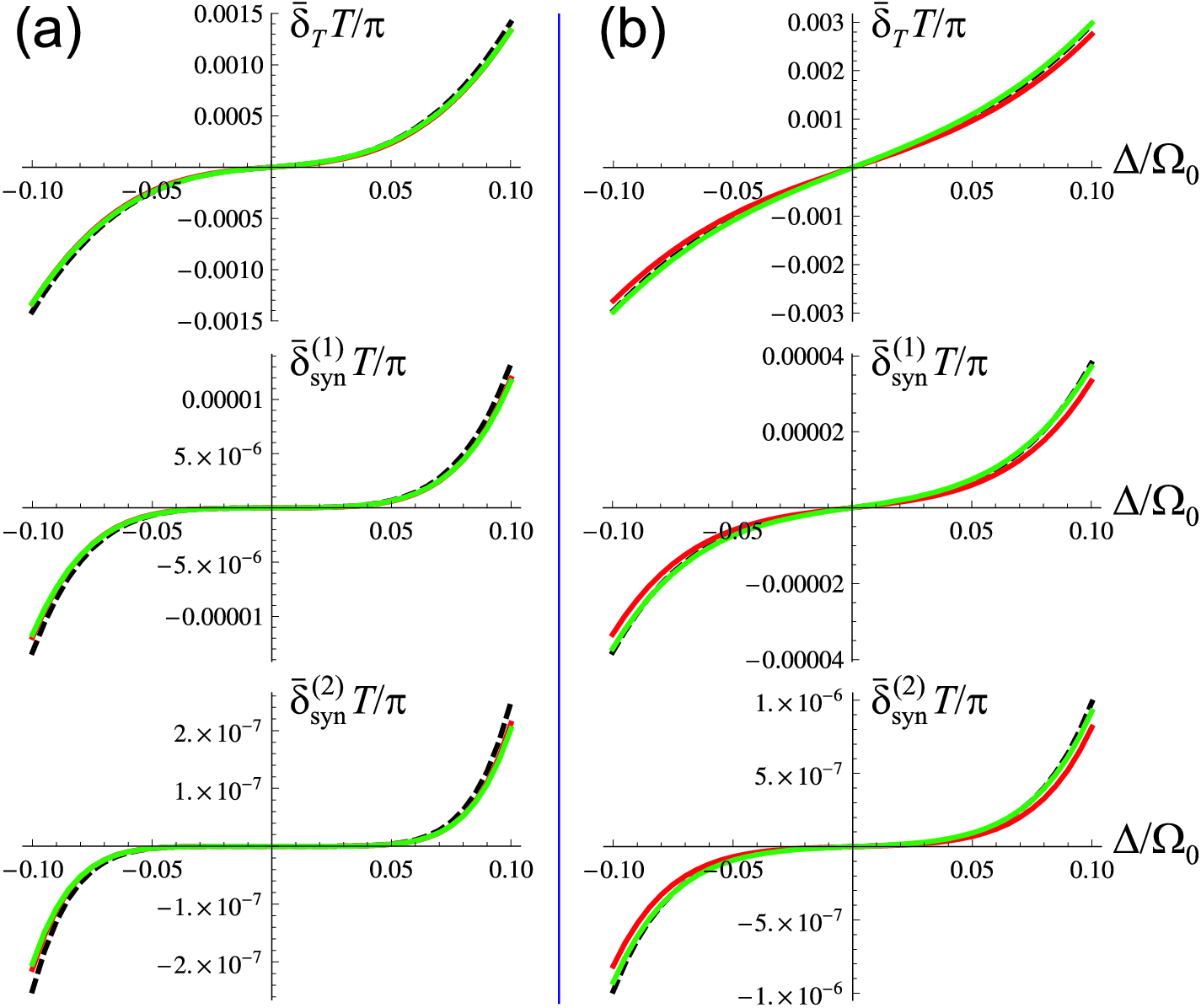}
	}
	\caption{
		(Color online) For the HR$-\pi$ protocol in presence of $\gamma_c$ decoherence, dependencies vs $\delta/\Omega$ for  the shift without synthetic frequency approach of Eq.~\eqref{eq:IVD-Syn-HRA-Gen-1}, denoted as $\bar{\delta}_{\textup{T}}$, and using the first and second order synthetic frequency approaches of Eqs.~\eqref{eq:IVD-Syn-HRA-Gen}, denoted as   $\bar{\delta}^{(1)}_\text{syn}$, and $\bar{\delta}^{(2)}_\text{syn}$. Parameters $4\tau/\textup{T}$=0.25; $\gamma_c$=0.01$\pi/\textup{T}$ in (a), $\gamma_c$=0.1$\pi/\textup{T}$, and for different Rabi frequencies: $\Omega\tau$=$\pi/2$ (black dashed lines); $\Omega \tau$=0.9$\pi/2$ (red lines); $\Omega \tau$=1.1$\pi/2$ (green lines).}
	\label{fig:IVD-HRA-Comb}
\end{figure}

\indent It is also very important to notice that the combination of the synthetic frequency protocol applied to the HR-$\pi$ scheme is quite stable to decoherence. Indeed, Fig.~\ref{fig:IVD-HRA-Comb}, showing graphs for the above clock shifts against the ratio $\delta/\Omega$ in the presence of decoherence described by the $\gamma_c$ parameter, demonstrates strong and robust suppression of the shift.

\indent The main advantage of the synthetic method is to reduce the decoherence perturbation without destroying an efficient light-shift compensation using the HR-$\pi$ protocol even when all pulse areas are modified by $\pm10\%$. The synthetic frequency protocol is also better by one to three orders of magnitude than MHR and GHR($\varphi_3$) protocols when laser-induced decoherence is considered.

\indent Apart from the combination with Ramsey and hyper-Ramsey spectroscopy for two-level systems, the synthetic frequency protocol
can be applied to the Ramsey spectroscopy of coherent population trapping (CPT) resonances (e.g., see \cite{Zanon:2005,Chen:2010,Blanshan:2015}). Note that CPT clocks are one of the prospective variants of compact RF clocks with relatively valuable metrological characteristics. Because the probe-induced shift for CPT-Ramsey resonance satisfies the
general dependence of Eq.~\eqref{eq:IVD-Shift-Serie-Expansion} on the free evolution interval $\textup{T}$ (see \cite{Hemmer:1989}, where the dependence $1/\textup{T}$ was found), one can expect good efficiency of the synthetic frequency protocol in this case too. The same approach can also be applied to so-called pulsed optical pumping (POP) clocks \cite{Micalizio:2012}. All these examples demonstrate the universality of the synthetic frequency protocol, which can be used in any type of clocks based on Ramsey spectroscopy.

\subsection{Auto-balanced Ramsey spectroscopy}

\indent A variant of the synthetic protocol approach, denoted as auto-balanced Ramsey spectroscopy, was very recently presented in ref.~\cite{Sanner:2017}. It is based on the combination of two Ramsey sequences, with short and long free evolution times, whence $\textup{T}$ is the control parameter.  The originality of the method is the use of two interconnected control loops. The first feedback loop uses the error signal provided by the short Ramsey sequence to lock an additional phase step correction between the Ramsey pulses, while the second loop locks the mean frequency from the error signal of the long Ramsey sequence. Notice that this sequence contains the phase step correction as an additional control parameter. To demonstrate the efficiency of the auto-balancing approach, the $^{171}$Yb$^{+}$ clock transition was experimentally probed with 3 different technical pulse defects. The auto-balanced Ramsey probing technique was thus able to recover the undisturbed clock transition against pulse areas delivered with 97$\%$ of the nominal intensity for the last 3~ms of their 15~ms on-time, against weak phase step excursion of the local laser oscillator and finally against phase lag. A final reduction by about $10^4$ of the light shift was experimentally observed in an $^{171}$Yb$^+$ ion clock operating on the E3 transition. 
\section{Composite laser-pulses protocols robust against dissipation}
\label{sec:IV-Dissipation}

\indent  This Section introduces new protocols dealing with the decoherence associated to the finite line-width of the probe laser,  which disturbs the
clock interrogation by reducing the contrast while also compromising the robustness of any error signal.
For the aimed $10^{-18}$ relative accuracy, the decoherence induced by clock laser line-width degrades the robustness of clock lock points for
GHR protocols of the last Section. This issue will be mitigated by fast improvements in the
design of very high finesse Fabry-Perot cavities used to stabilize clock lasers, thus offering very narrow line-widths below
a few 100~mHz \cite{Jiang:2011,Kessler:2012,Amairi:2013} for a new generation of frequency standards. However new composite pulse protocols represent an alternative approach to this issue. Relaxation processes for the clock populations are also included into the presented analysis. Notice also that the MHR and GHR($\varphi_{\textup{1}}$) protocols  are not fully equivalent in the presence of decoherence produced by the finite laser linewidth~\cite{Hobson:2016,Zanon-Willette:2016b}.

\subsection{Matrix solution to optical Bloch equations}
\label{sec:IVA-Bloch-Equations}

\indent The relevant analysis has to be performed within a formalism based on the two-level $\rho$ density matrix where the atomic decoherence can be treated
properly \cite{Tabatchikova:2013,Tabatchikova:2015}. The direct numerical integration of density matrix  equations includes dephasing of the off-diagonal elements \cite{Allen:1975,Berman:2011} and relaxation terms of populations, in order to describe dissipative processes such as spontaneous emission, dephasing and decoherence within a closed two-level configuration.
The atomic evolution includes a decoherence term $\gamma_{c}$, a spontaneous emission rate denoted $\Gamma$ and a population difference relaxation $\zeta$
induced by  collisions. The Bloch variables $\textup{U}_{l}\equiv\rho_{\textup{ge}}^{*}+\rho_{\textup{ge}}$,
$\textup{V}_{l}\equiv i(\rho_{\textup{ge}}^{*}-\rho_{\textup{ge}})$ and $\textup{W}_{l}\equiv\rho_{\textup{ee}}-\rho_{\textup{gg}}$ are used to describe the atomic excitation after the $l$-th optical pulse of the
composite pulse sequence. The general set of time-dependent optical Bloch equations
is given by \cite{Zanon-Willette:2017}:
\begin{equation}
	\left\lbrace \begin{split}
		\dot{\textup{U}}_{l} &= -\gamma_{c} \textup{U}_{l} + \delta_{l} \textup{V}_{l} - \Omega_{l} \sin\varphi_{l} \textup{W}_{l}, \\
		\dot{\textup{V}}_{l} &= -\delta_{l} \textup{U}_{l} - \gamma_{c} \textup{V}_{l} + \Omega_{l} \cos\varphi_{l} \textup{W}_{l}, \\
		\dot{\textup{W}}_{l} &= \Omega_{l} \sin\varphi_{l} \textup{U}_{l} - \Omega_{l} \cos\varphi_{l} \textup{V}_{l}
			- (\Gamma+\zeta) \textup{W}_{l} - \Gamma, \\
	\end{split} \right.
	\label{eq:IVA-Bloch-System}
\end{equation}
The three components vector $\textup{M}(\theta_{l})\equiv( \textup{U}(\theta_{l}),\textup{V}(\theta_{l}),\textup{W}(\theta_{l}))$ solution to the previous set
of equations is \cite{Jaynes:1955,Schoemaker:1978}
\begin{equation}
	\textup{M}(\theta_{l}) = \textup{R}(\theta_{l}) \left[ \textup{M}_{l}(0) - \textup{M}_{l}(\infty) \right] + \textup{M}_{l}(\infty),
	\label{eq:IVA-Bloch-Solution}
\end{equation}
characterized by the pulse area $\theta_{l}$. $\textup{M}_{l}(0)\equiv(\textup{U}_{l}(0),\textup{V}_{l}(0),\textup{W}_{l}(0))$
stands for the system's state before the $l$-th pulse. The steady-state solution matrix
$\textup{M}_{l}(\infty)\equiv(\textup{U}_{l}(\infty),\textup{V}_{l}(\infty),\textup{W}_{l}(\infty))$ is obtained by
switching off time-dependent derivatives in Eq.~(\ref{eq:IVA-Bloch-System}) for its three components.
See Appendix~\ref{sec:VIIC-Rotation-Matrix} for its definition.\\
\indent The rotation matrix $\textup{R}(\theta_{l})$, taking
decoherence and relaxation terms into account, is written as follows:
\begin{equation}
		\textup{R}(\theta_{l}) = e^{-\gamma_{c}\tau_{l}} e^{-\beta_{l}\tau_{l}},
	\label{eq:IVA-Bloch-Rotation-Matrix}
\end{equation}
requiring the exponentiation of the following $\beta_{l}$ matrix:
\begin{equation}
			\beta_{l} = \left( \begin{array}{ccc}
			0 & \delta & -\Omega_{l} \sin\varphi_{l} \\
			-\delta & 0 & \Omega_{l} \cos\varphi_{l} \\
			\Omega_{l} \sin\varphi_{l} & -\Omega_{l} \cos\varphi_{l} & \Delta\gamma
		\end{array} \right),
	\label{eq:IVA-Bloch-Rotation-Matrix-B}
\end{equation}
where we have defined $\Delta\gamma=\gamma_{c}-(\Gamma+\zeta)$. The matrix $\textup{R}(\theta_{l})$ can be exactly expressed
as a square matrix of time-dependent matrix elements $\textup{R}_{\textup{mn}}(\theta_{l})$ ($\textup{m,n} \in \{1,2,3\}$), presented in detail in Appendix~\ref{sec:VIIC-Rotation-Matrix}.\\
\indent Consider now a sequence of pulses separated by a free evolution T, as in the HR scheme for example. The free evolution matrix $\textup{R}(\theta_{k}=\delta\textup{T})$ without
laser field is given by
\begin{equation}
	\begin{split}
		\textup{R}(\theta_{k}=\delta\textup{T}) = e^{-\gamma_{c}\textup{T}} \left( \begin{array}{ccc}
			\cos\delta\textup{T} & \sin\delta\textup{T} & 0 \\
			-\sin\delta\textup{T} & \cos\delta\textup{T} & 0 \\
			0 & 0 & e^{\Delta\gamma\textup{T}}
		\end{array} \right).
	\end{split}
\end{equation}
The corresponding stationary solution $\textup{M}_{\textup{T}}(\infty)\equiv(\textup{U}_{\textup{T}}(\infty), \textup{V}_{\textup{T}}(\infty),\textup{W}_{\textup{T}}(\infty))$
is also found by switching off the $\Omega_{l}$  laser field in Eq.~(\ref{eq:IVA-Bloch-System}) during free evolution time.

\subsection{Composite interrogation protocol with arbitrary sequence of pulses}
\label{sec:IVB-Composite-Pulses}

\begin{figure*}[t!!]
	\centering
	\subfloat[\textup{HR}-$\pi$ protocol: ideal case]{
		\resizebox{0.4\linewidth}{!}{
			\includegraphics[angle=0]{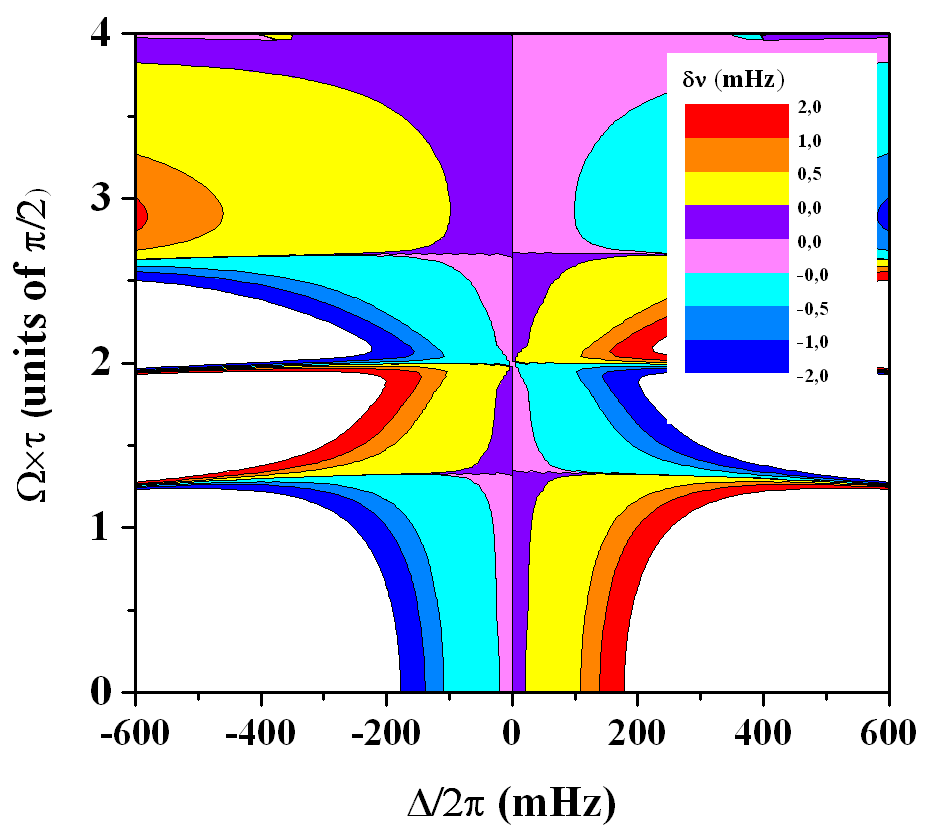}
	    }
		\resizebox{0.46\linewidth}{!}{
	        \includegraphics[angle=0]{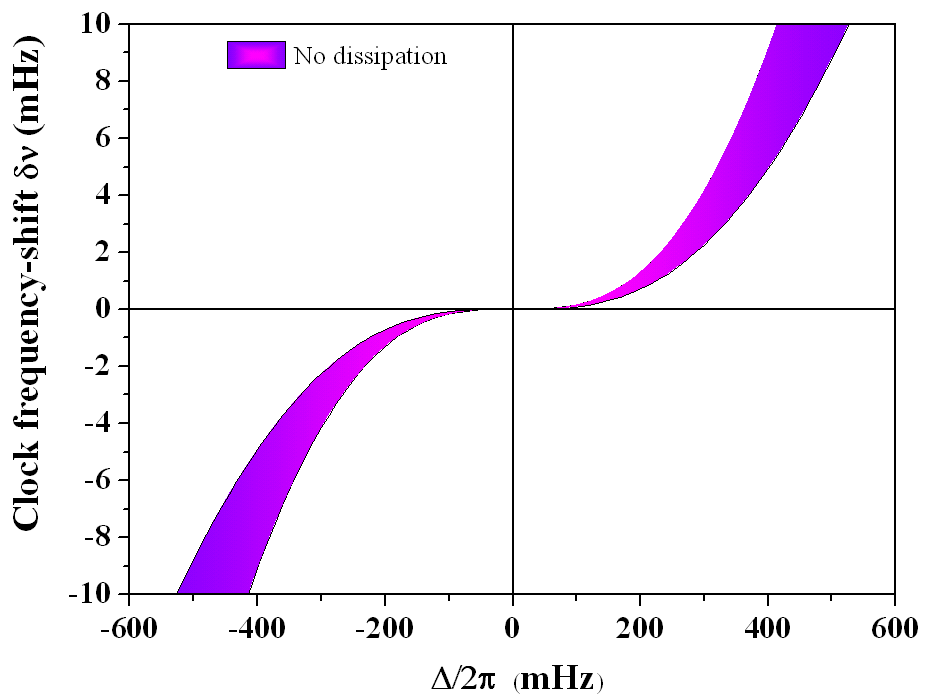}
		}
		\label{fig:IVC-Map2D-Shift-HR-a}
	} \\
	\subfloat[\textup{HR}-$\pi$ protocol: decoherence only]{
		\resizebox{0.4\linewidth}{!}{
			\includegraphics[angle=0]{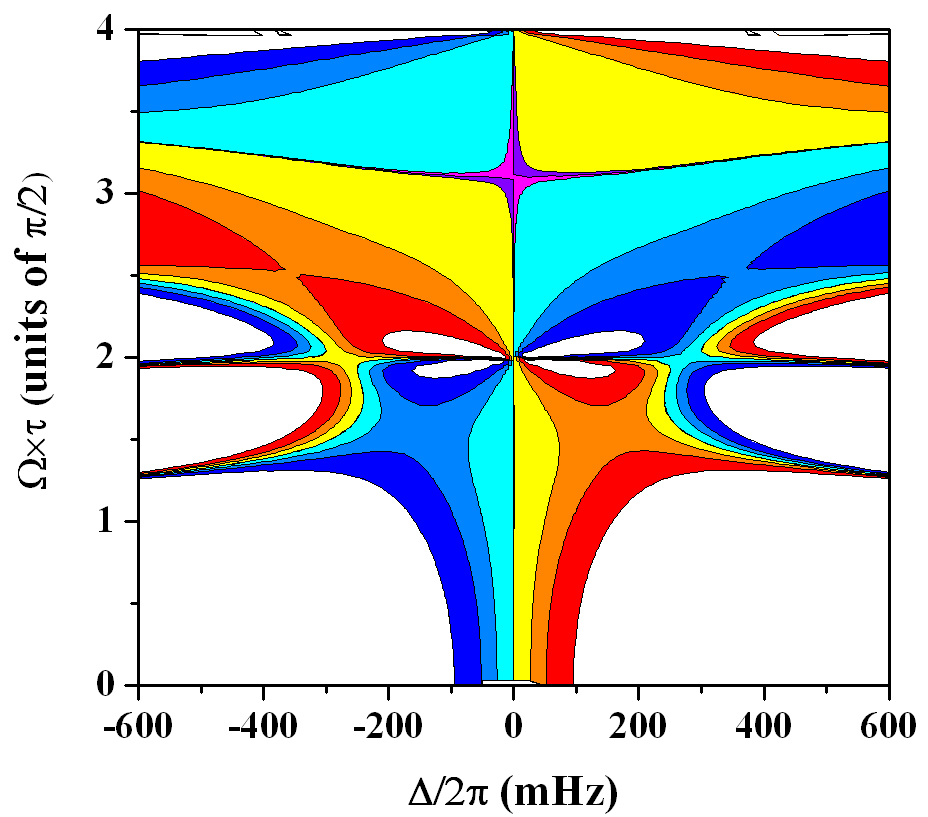}
	    }
		\resizebox{0.46\linewidth}{!}{
	        \includegraphics[angle=0]{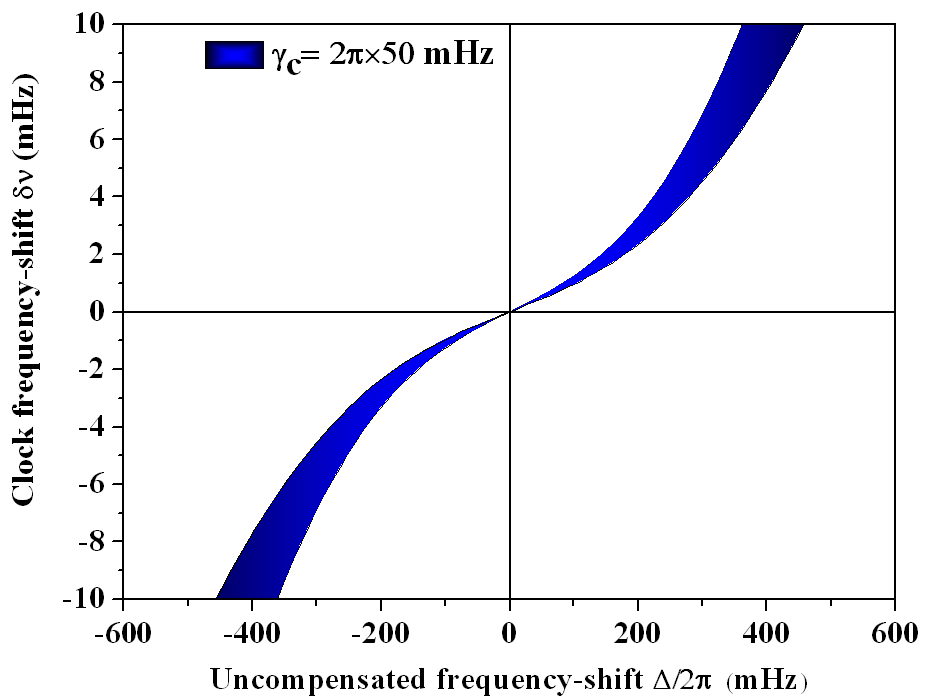}
		}
		\label{fig:IVC-Map2D-Shift-HR-b}
	} \\
	\subfloat[\textup{HR}-$\pi$ protocol: decoherence and relaxation]{
		\resizebox{0.4\linewidth}{!}{
			\includegraphics[angle=0]{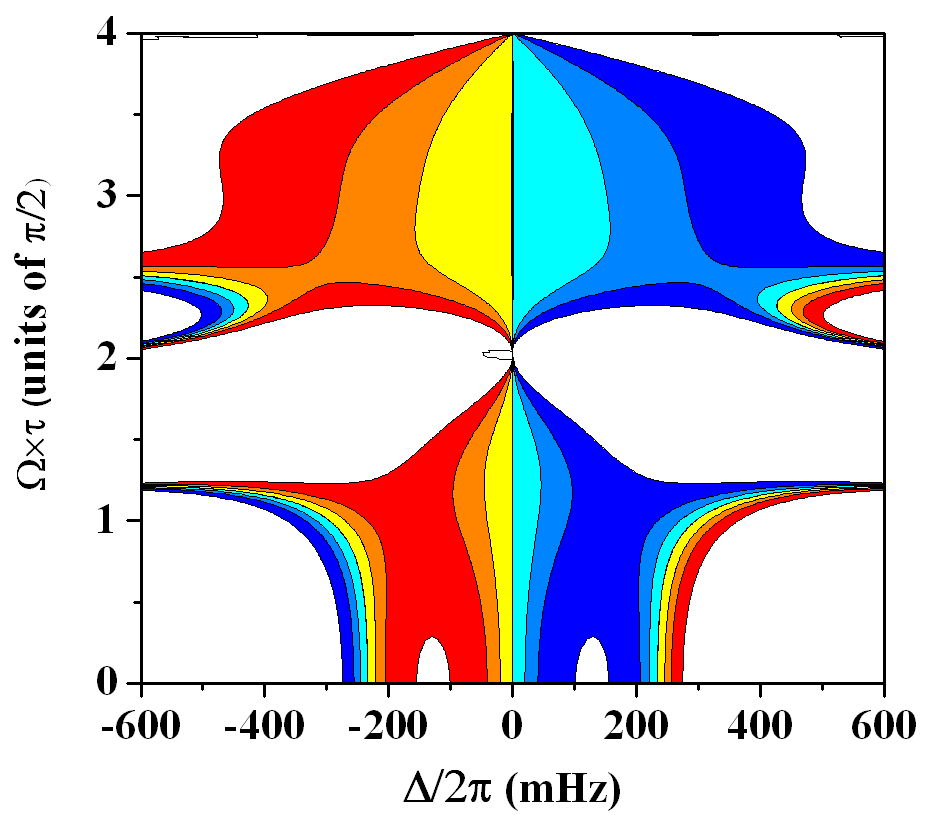}
	    }
		\resizebox{0.46\linewidth}{!}{
			\includegraphics[angle=0]{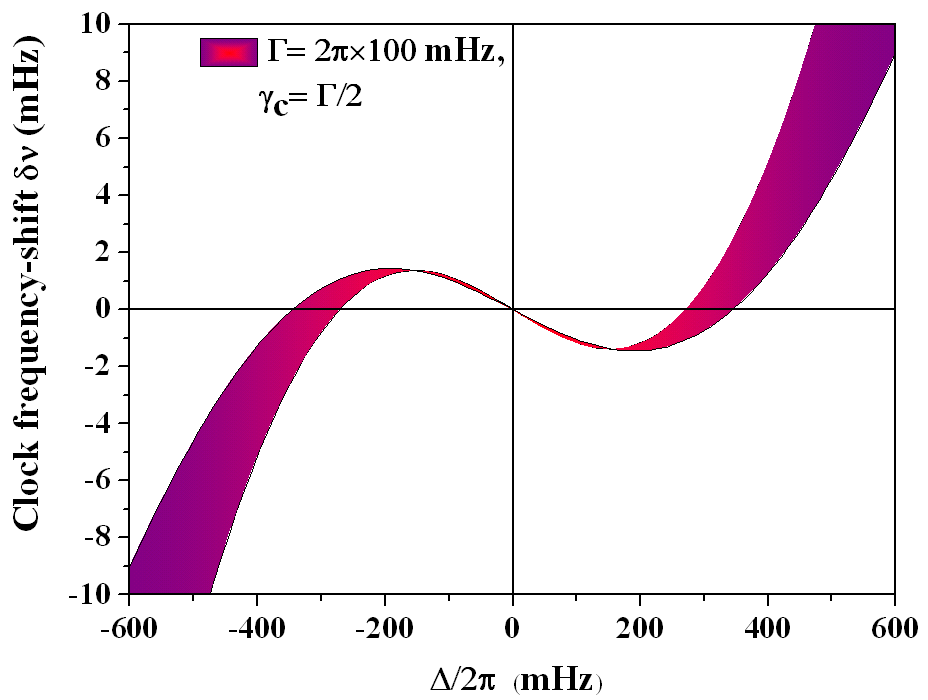}
		}
		\label{fig:IVC-Map2D-Shift-HR-c}
	}
	\caption{
		(color online) On the left 2D contour and density plot diagrams of the  $\delta\widetilde{\nu}$ frequency-shift lock point defined by Eq.~(\ref{eq:IIIA-Shift-DeltaNu}) using the new phase-shift of Eq.~(\ref{eq:IVB-Pulse-Sequence-Phase-Shift-GHR}) against uncompensated frequency-shifts
		$\Delta/2\pi$ (horizontal axis) and pulse area  $\Omega\tau$ (vertical axis). Right column shows a planar cut of density
		plots at fixed pulse area $\Omega\tau=\pi/2$ with a $\pm 10$\% pulse area variation (shadow area). For $\gamma_{c}/2\pi=50$~mHz and $\Gamma/2\pi=100$~mHz, the (b) case corresponds to decoherence produced by laser linewidth contribution and the (c) case to decoherence and relaxation by spontaneous emission.}
	\label{fig:IVC-Map2D-Shift-HR}
\end{figure*}

\indent When we consider an arbitrary sequence of composite pulses where each laser interaction zone is associated with different
areas tailored in phase, frequency and duration, each laser pulse interaction introduced by $\textup{M}(\theta_{l})$ is described by an equation identical to
Eq.~(\ref{eq:IVA-Bloch-Solution}).
An arbitrary sequence of pulses can be constructed by iteration to compute the final response
$\textup{M}(\theta_{1},...,\theta_{\textup{n}})$.

\indent  Using exact analytic expressions to solve the Bloch equations for a single given Rabi pulse, the expression for a full sequence of
$\textup{n}$ pulses can be generalized to:
\begin{equation}
	\begin{split}
		\textup{M}(\theta_{1},...,\theta_{\textup{n}}) & =\sum_{\textup{p}=1}^{\textup{n}} \left[
			\left( \overleftarrow{\prod_{l=\textup{p}}^{\textup{n}}} \textup{R}(\theta_{l}) \right)
			\left( \textup{M}_{\textup{p}-1}(\infty) - \textup{M}_{\textup{p}}(\infty) \right)
		\right] \\
		&+ \textup{M}_{\textup{n}}(\infty),
	\end{split}
	\label{eq:IVB-Pulse-Sequence-Canonical}
\end{equation}
where backward arrows indicate a matrix product from right to left with growing indices and state initialization denoted $\textup{M}_{0}(\infty)\equiv\textup{M}_{1}(0)$ by convention.

\indent Ref.~\cite{Zanon-Willette:2017} has derived an exact analytic solution for an arbitrary sequence of pulses  including a generalized canonical form for the associated phase shift.
Such a solution with a single free evolution time can always be expressed in a reduced canonical
form as
\begin{equation}
	\textup{M}(\theta_{1},...,\theta_{\textup{n}}) \equiv \mathcal{A}+ \mathcal{B} \cos( \delta\textup{T} + \phi),
	\label{eq:IVB-Pulse-Sequence-Canonical-Reduced}
\end{equation}
where  the new offset $\mathcal{A}$, amplitude $\mathcal{B}$ and new phase-shift $\phi$ are analytical functions including dissipation explicitly computed in \cite{Zanon-Willette:2017}. Notice the equivalence between the present canonical form and those of Eqs.~\eqref{eq:IIB-Prob-Transition-Canonical} and \eqref{eq:IIIA-Error-Signal-Canonical_2}.

\indent Using Eq.~(\ref{eq:IVB-Pulse-Sequence-Canonical-Reduced}), we can study the three-pulse interrogation schemes. By looking
at the third component $\textup{W}(\theta_{1},\theta_{2}=\delta\textup{T},\theta_{\textup{3}},\theta_{\textup{4}})$, we can retrieve
population transitions to construct an error signal.
The derived generalized expression for the $\rho_{ee}$ population of the upper level is formally identical to the transition
probability given by Eq.~\eqref{eq:IIIA-Prob-Transition-GHR} when dissipation is switched-off. The phase-shift $\phi$ including
dissipation accumulated over the entire sequence of pulses is now expressed as:
\begin{equation}
	\phi = \arctan \left[
		\frac{
			\textup{R}_{32}(\theta_{\textup{3}},\theta_{\textup{4}}) \textup{M}_{1}(\theta_{\textup{1}})
			+ \textup{R}_{31}(\theta_{\textup{3}},\theta_{\textup{4}}) \textup{M}_{2}(\theta_{\textup{1}})
		}{
			\textup{R}_{31}(\theta_{\textup{3}},\theta_{\textup{4}}) \textup{M}_{1}(\theta_{\textup{1}})
			+ \textup{R}_{32}(\theta_{\textup{3}},\theta_{\textup{4}}) \textup{M}_{2}(\theta_{\textup{1}})
		}
	\right],
	\label{eq:IVB-Pulse-Sequence-Phase-Shift-GHR}
\end{equation}
with
\begin{equation}
	\begin{split}
		\textup{M}_{1}(\theta_{\textup{1}}) &=
			\left[ 1 - \textup{R}_{11}(\theta_{\textup{1}}) \right] \textup{U}_{\textup{1}}(\infty)
			- \textup{R}_{12}(\theta_{\textup{1}}) \textup{V}_{\textup{1}}(\infty) \\
		&+ \textup{R}_{13}(\theta_{\textup{1}}) \left[ \textup{W}_{\textup{1}}(0) - \textup{W}_{\textup{1}}(\infty) \right],\\
		\textup{M}_{2}(\theta_{\textup{1}}) &=
			\left[ 1 - \textup{R}_{22}(\theta_{\textup{1}}) \right] \textup{V}_{\textup{1}}(\infty)
			- \textup{R}_{21}(\theta_{\textup{1}}) \textup{U}_{\textup{1}}(\infty) \\
		&+ \textup{R}_{23}(\theta_{\textup{1}}) \left[ \textup{W}_{\textup{1}}(0) - \textup{W}_{\textup{1}}(\infty) \right],
	\end{split}
\end{equation}
and for the components of the product matrix $\textup{R}(\theta_3,\theta_4) = \textup{R}(\theta_4)\textup{R}(\theta_3)$.
\begin{equation}
	\begin{split}
		\textup{R}_{31}(\theta_{\textup{3}},\theta_{\textup{4}}) &=
			\textup{R}_{31}(\theta_{\textup{4}}) \textup{R}_{11}(\theta_{\textup{3}})
			+ \textup{R}_{32}(\theta_{\textup{4}}) \textup{R}_{21}(\theta_{\textup{3}}) \\
			&+ \textup{R}_{33}(\theta_{\textup{4}}) \textup{R}_{31}(\theta_{\textup{3}}), \\
		\textup{R}_{32}(\theta_{\textup{3}},\theta_{\textup{4}}) &=
			\textup{R}_{31}(\theta_{\textup{4}}) \textup{R}_{12}(\theta_{\textup{3}})
			+ \textup{R}_{32}(\theta_{\textup{4}}) \textup{R}_{22}(\theta_{\textup{3}}) \\
			&+ \textup{R}_{33}(\theta_{\textup{4}}) \textup{R}_{32}(\theta_{\textup{3}}),
	\end{split}
\end{equation}
The above
equations may be used to evaluate the dependencies of clock frequency-shifts to residual light-shifts in presence of decoherence and
relaxation processes for any composite pulse protocols, in particular those presented in Tab.~\ref{tab:III-Composite-Pulses-Protocols}.

\subsection{Elimination of dissipation effects on clock frequency-shift}
\label{sec:IVC-Dissipation-Effects}

\begin{figure*}[t!!]
	\center
	\subfloat[\textup{MHR} protocol]{
		\resizebox{0.4\linewidth}{!}{
			\includegraphics[angle=0]{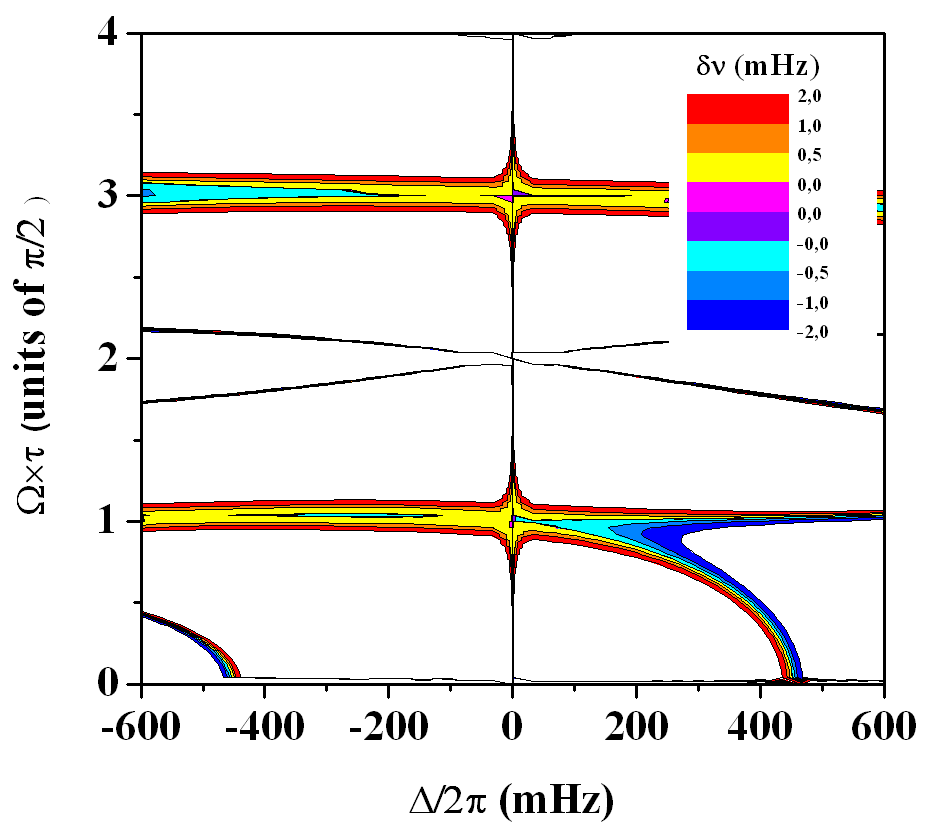}
	     }
		\resizebox{0.46\linewidth}{!}{
	        \includegraphics[angle=0]{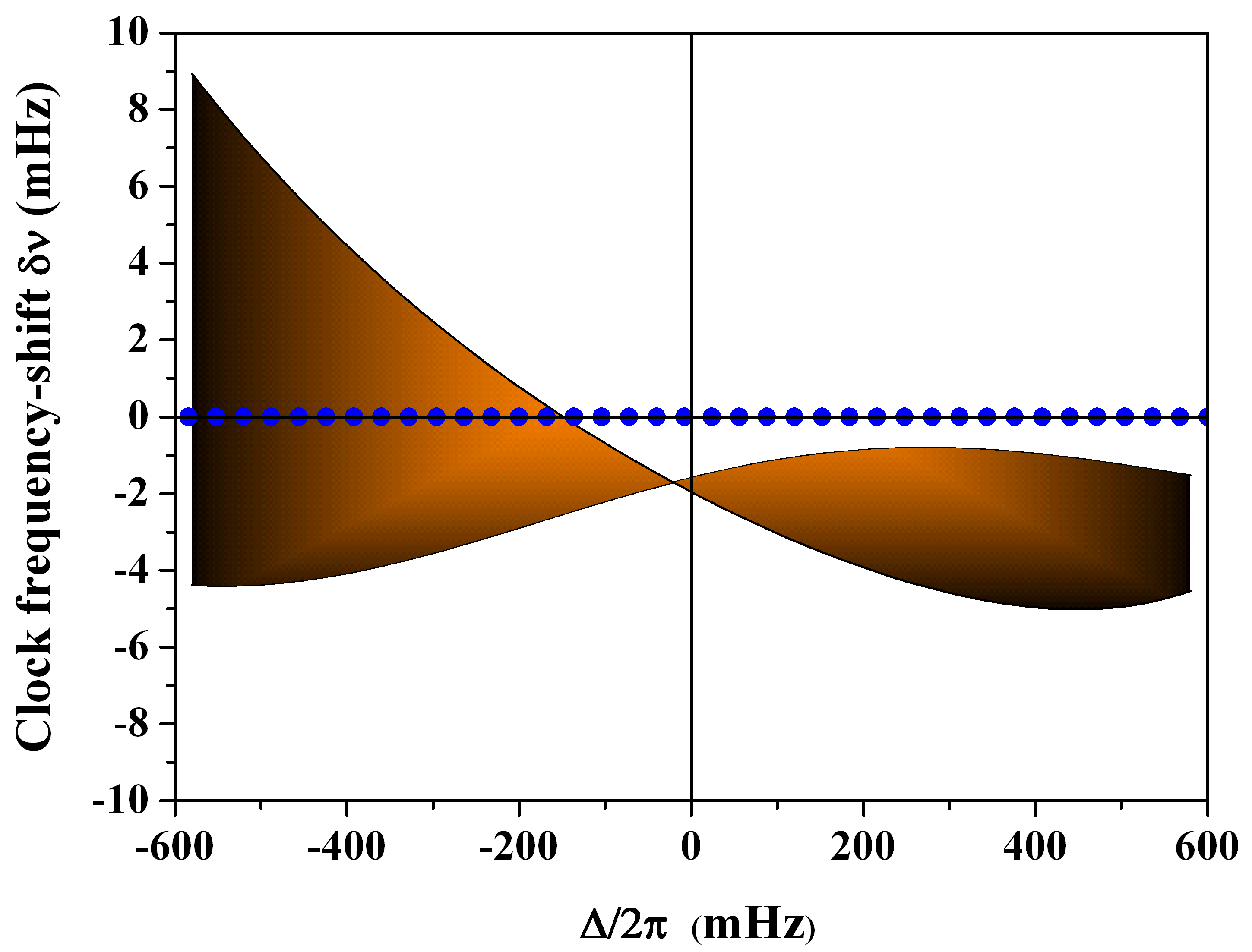}
		}
		\label{fig:IVC-Map2D-Shift-MHR-GHR-a}
	} \\
	\subfloat[\textup{GHR}($\pi/4$) protocol]{
		\resizebox{0.4\linewidth}{!}{
			\includegraphics[angle=0]{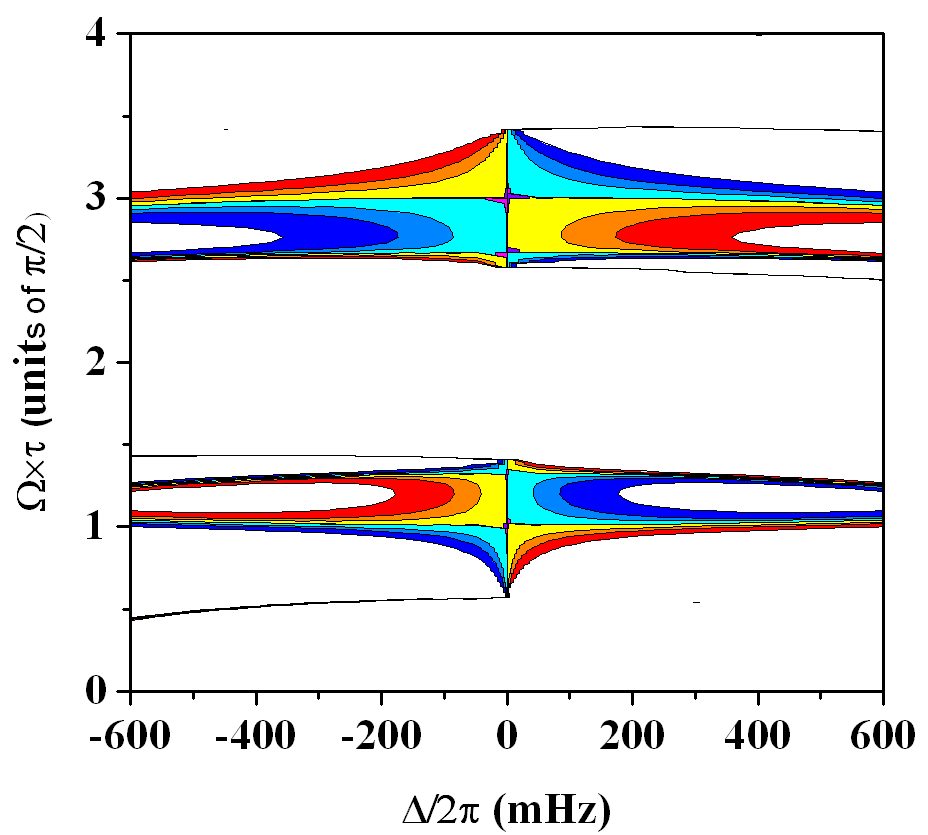}
	    }
		\resizebox{0.46\linewidth}{!}{
			\includegraphics[angle=0]{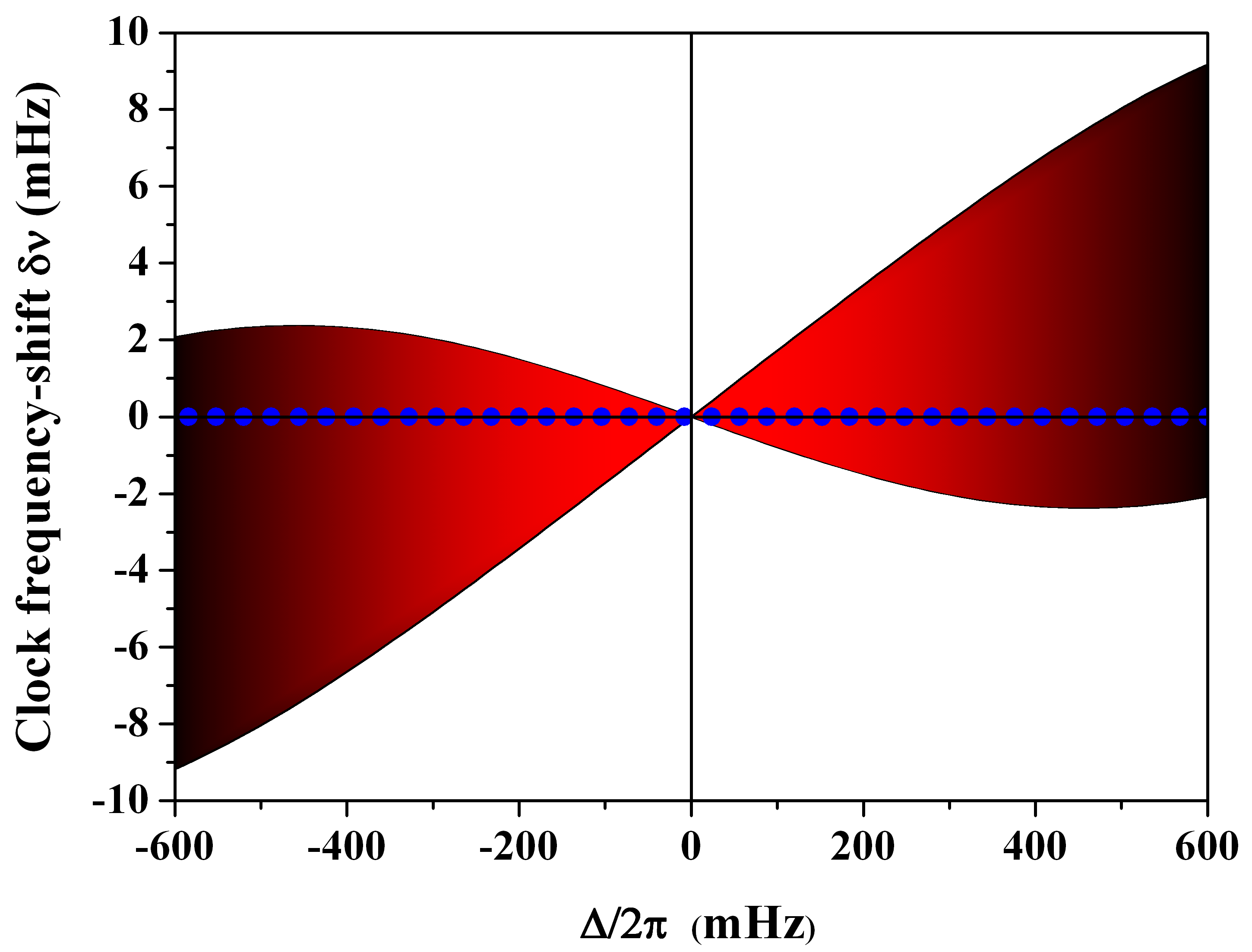}
		}
		\label{fig:IVC-Map2D-Shift-MHR-GHR-b}
	} \\
	\subfloat[\textup{GHR}($3\pi/4$) protocol ]{
		\resizebox{0.4\linewidth}{!}{
			\includegraphics[angle=0]{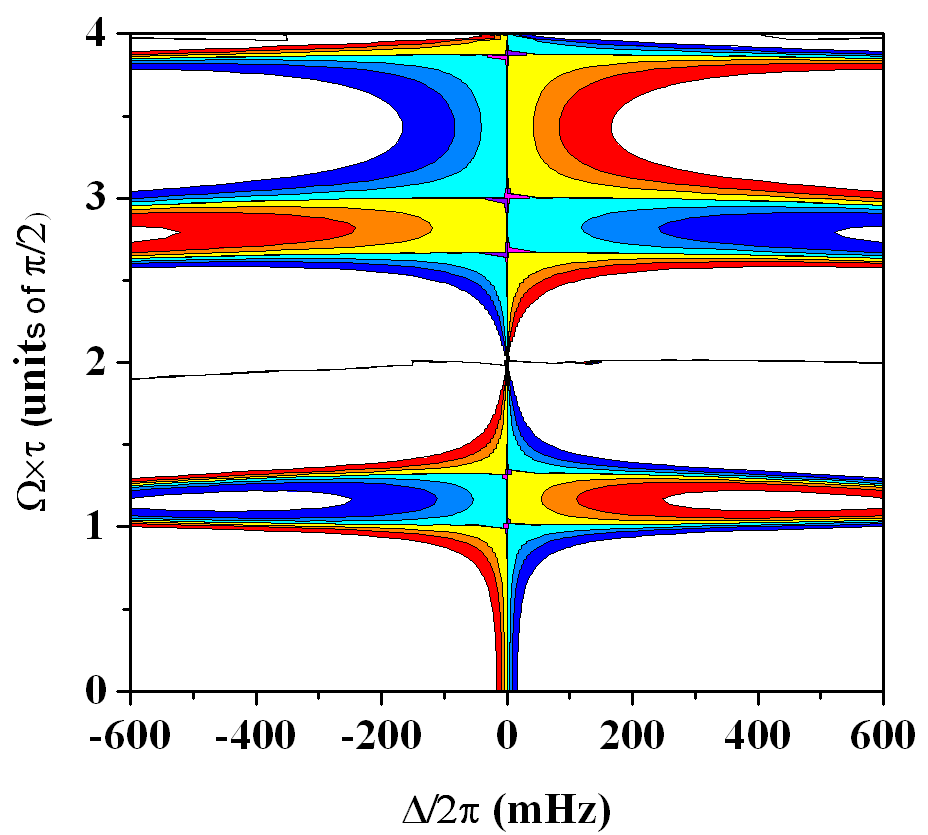}
	    }
		\resizebox{0.46\linewidth}{!}{
			\includegraphics[angle=0]{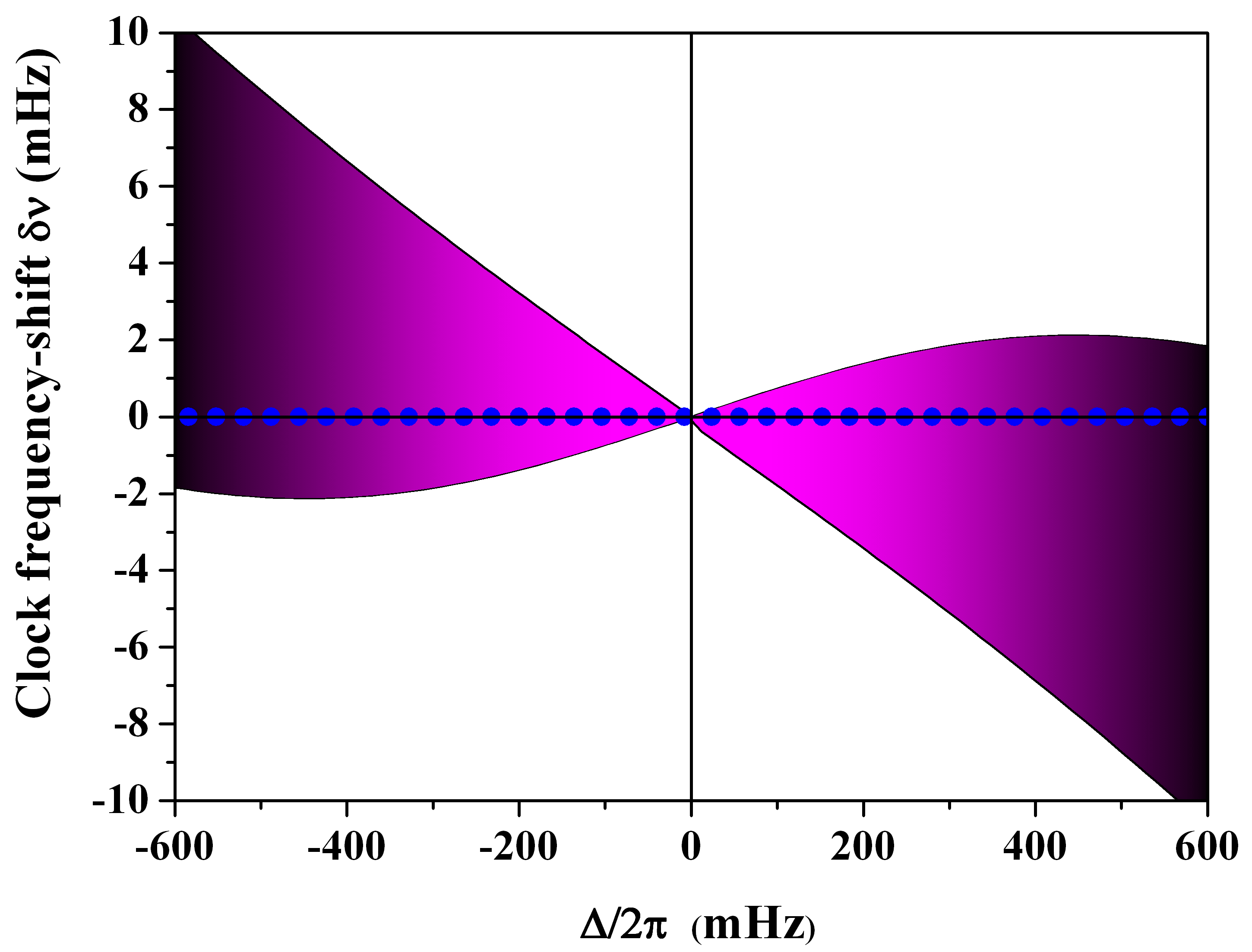}
		}
		\label{fig:IVC-Map2D-Shift-MHR-GHR-c}
	}
	\caption{
		(color online) 2D contour and density plot diagrams of the $\delta\widetilde{\nu}$ frequency-shift lock point derived using the phase of  Eq.~(\ref{eq:IVB-Pulse-Sequence-Phase-Shift-GHR}) in presence of $\gamma_{c}/2\pi=50$~mHz decoherence, and no relaxation, against uncompensated frequency-shifts $\Delta/2\pi$ (horizontal axis) and pulse area $\Omega\tau$ (vertical axis).
		Right column shows planar cuts of density plots at the given pulse area $\Omega\tau=\pi/2$ with a $\pm 10$\% pulse area variation (shadow area). The dotted line is the clock frequency shift robustness against uncompensated frequency-shifts when dissipation is ignored}.
	\label{fig:IVC-Map2D-Shift-MHR-GHR}
\end{figure*}

\indent The influence of decoherence or relaxation by spontaneous emission on the HR-$\pi$ probing scheme is analyzed using the 2D
contour and density plot diagrams shown on the left in Figs.~\ref{fig:IVC-Map2D-Shift-HR}. All clock-frequency shifts
$\delta\widetilde{\nu}$, introduced by  Eq.~(\ref{eq:IIIA-Shift-DeltaNu}) in Sec. III for the dispersive clock lock, and determined using the new phase-shift of Eq.~\eqref{eq:IVB-Pulse-Sequence-Phase-Shift-GHR}, are plotted against uncompensated frequency-shifts and large pulse area variations. Because ac Stark-shifts
increase quadratically with pulse area, the diagrams explore also regions of several $\pi/2$ laser pulse area units corresponding to the application of a large laser frequency-step for pre-compensation of the central fringe frequency-shift
\cite{Taichenachev:2009}. Within contour plots, the colored values of clock-frequency shifts have been deliberately limited between -2~mHz and +2~mHz (see
layout on graphs) for constraining the clock relative accuracy below $10^{-18}$. For white surrounding regions, the relative accuracy of the residual
shift exceeds a few $10^{-18}$ level.
Clock frequency shifts affecting the lock point are reported on the right of Figs.~\ref{fig:IVC-Map2D-Shift-HR}. One effect of
dissipation is to restore a weak linear dependence to uncompensated residual light-shifts while slope rotation depends on the
specific dissipation process.

\indent Fig.~\ref{fig:IVC-Map2D-Shift-MHR-GHR} reports the clock response to decoherence only for three different protocols. For the MHR protocol,
the right plot of Fig.~\ref{fig:IVC-Map2D-Shift-MHR-GHR-a} shows a significative frequency-shift at zero residual uncompensated light-shift while its pulse area stability lock point is off axis from $\Delta=0$. Thus the MHR parasite shift produces a large sensitivity to variations of the Rabi frequency and reduces optimal performances of the stabilization scheme~\cite{Yudin:2016}. The GHR($\varphi_3$) protocols do not suffer from these parasite shifts but the immunity to residual uncompensated shifts is lost when pulse area is not constant as shown in \cite{Yudin:2016}. For the GHR$(\pi/4)$ and GHR$(3\pi/4)$ protocols in Fig.~\ref{fig:IVC-Map2D-Shift-MHR-GHR-b} and Fig.~\ref{fig:IVC-Map2D-Shift-MHR-GHR-c}, respectively, stability islands exist even when the pulse area is not a perfect $\pi/2$.

\subsection{Universal interrogation protocols combining $\pi/4$ and $3\pi/4$ phase-steps}
\label{sec:IVD-Universal-Protocols}

\begin{figure}[t!!]
	\center
	\subfloat[\textup{GHR}($\pi/4,3\pi/4$) protocol: state preparation method]{
		\resizebox{\linewidth}{!}{
			\includegraphics[angle=0]{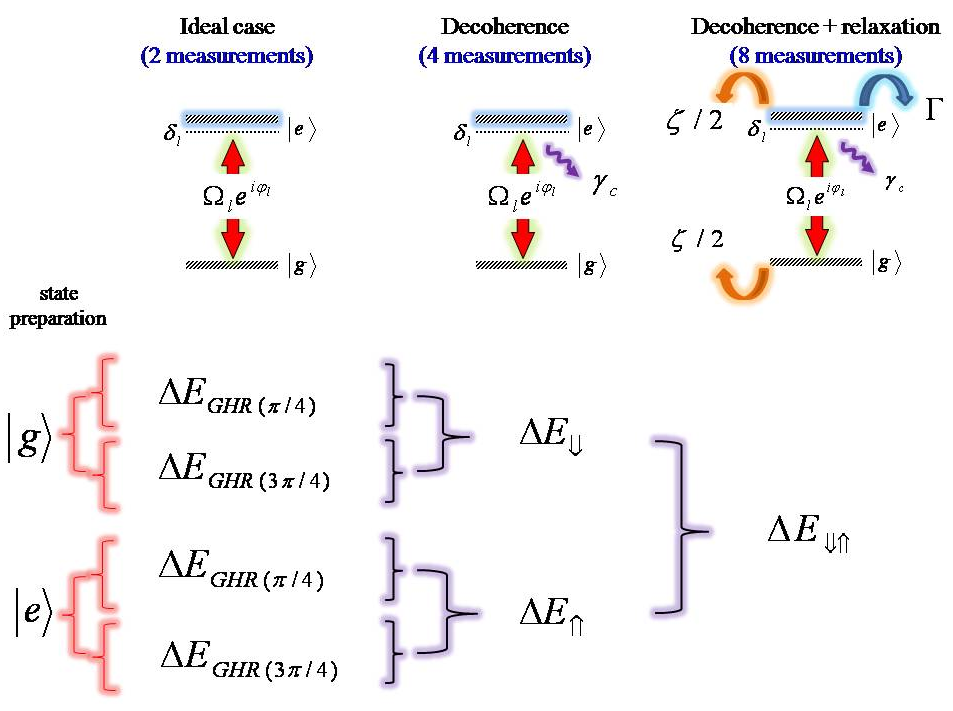}
		}
		\label{fig:IVE-Universal-Protocols-a}
	} \\
	\subfloat[\textup{GHR}($\pi/4,3\pi/4$) protocol: time-reversal method]{
		\resizebox{\linewidth}{!}{
			\includegraphics[angle=0]{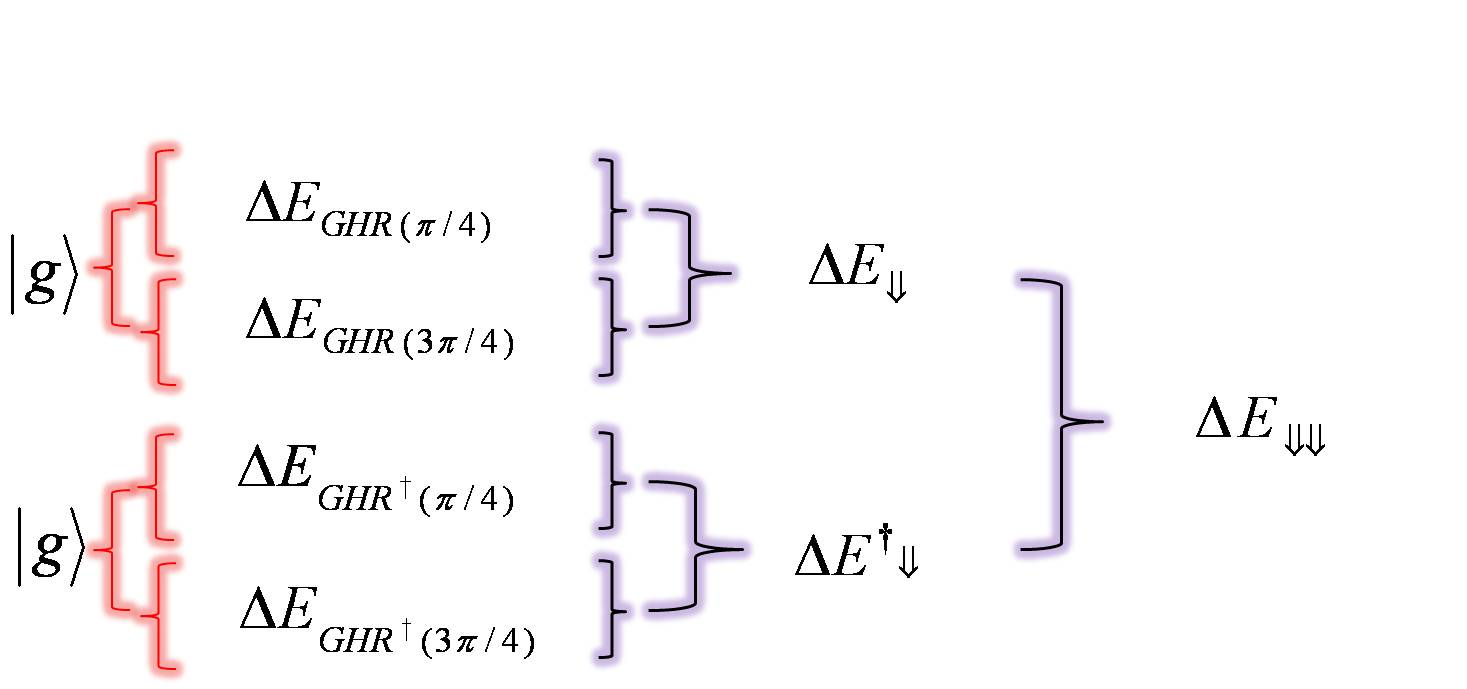}
		}
		\label{fig:IVE-Universal-Protocols-b}
	}
	\caption{
		(color online) Composite pulse sequences for two universal protocols. The top one (a) is the hybrid method
		combining two GHR($\pi/4,3\pi/4$) sequences applied to two different initialization states before being combined together
		by error signal differentiation. The bottom one (b) combines a GHR($\pi/4,3\pi/4$) sequence with its time-reversal partner GHR$^{\dagger}(\pi/4,3\pi/4)$ applied to the same initialization state, and combined together by error signal summation.
	}
	\label{fig:IVE-Universal-Protocols}
\end{figure}

\begin{figure}[t!!]
	\center
	\subfloat[\textup{HR}$-\pi$ error signal for three values of $\Delta$]{
		\resizebox{0.8\linewidth}{!}{
			\includegraphics[angle=0]{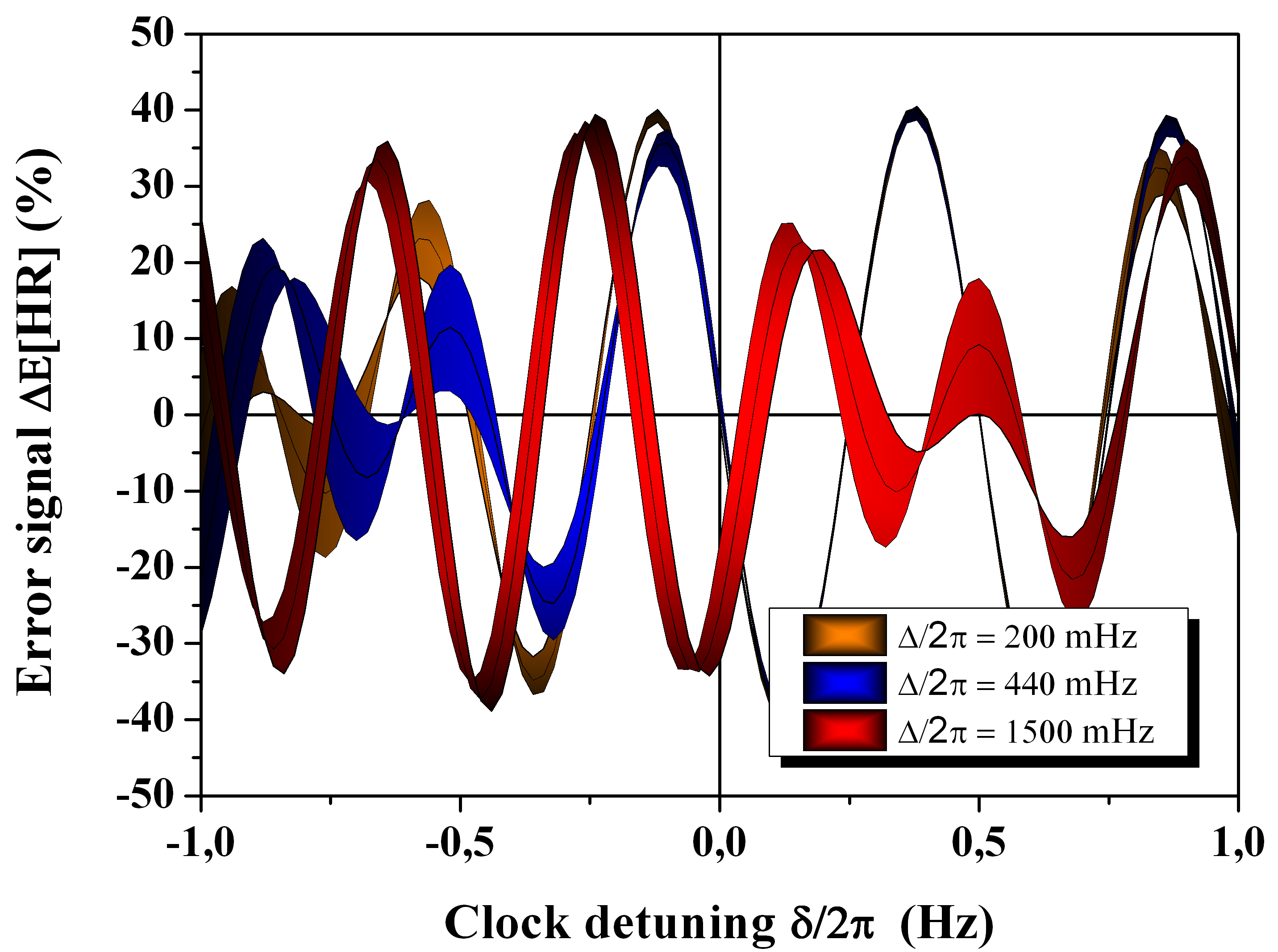}
		}
		\label{fig:IVE-Phase-Modulation-HR-relaxation-decoherence}
	}\\
	\subfloat[$\Delta\textup{E}_{\Downarrow\Uparrow}$ ($\Delta\textup{E}_{\Downarrow\Downarrow}$) error signals for same values of $\Delta$]{
		\resizebox{0.8\linewidth}{!}{
			\includegraphics[angle=0]{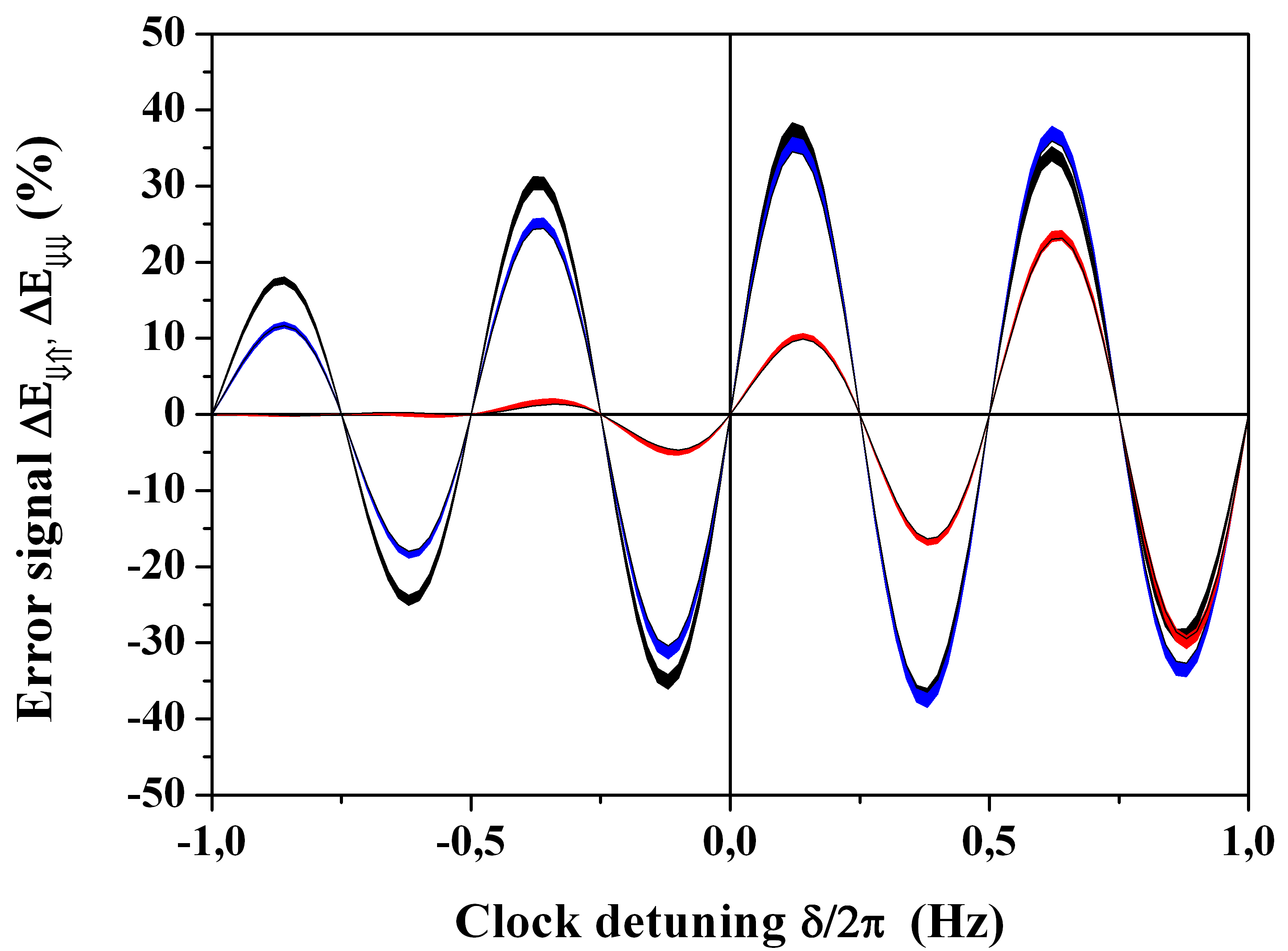}
		}
		\label{fig:IVE-Phase-Modulation-GHR-relaxation-decoherence}
	}\\
	\subfloat[Lock point frequency shift of \textup{HR}$-\pi$ and $\Delta\textup{E}_{\Downarrow\Uparrow}$ ($\Delta\textup{E}_{\Downarrow\Downarrow}$) error signals]{
		\resizebox{0.8\linewidth}{!}{
			\includegraphics[angle=0]{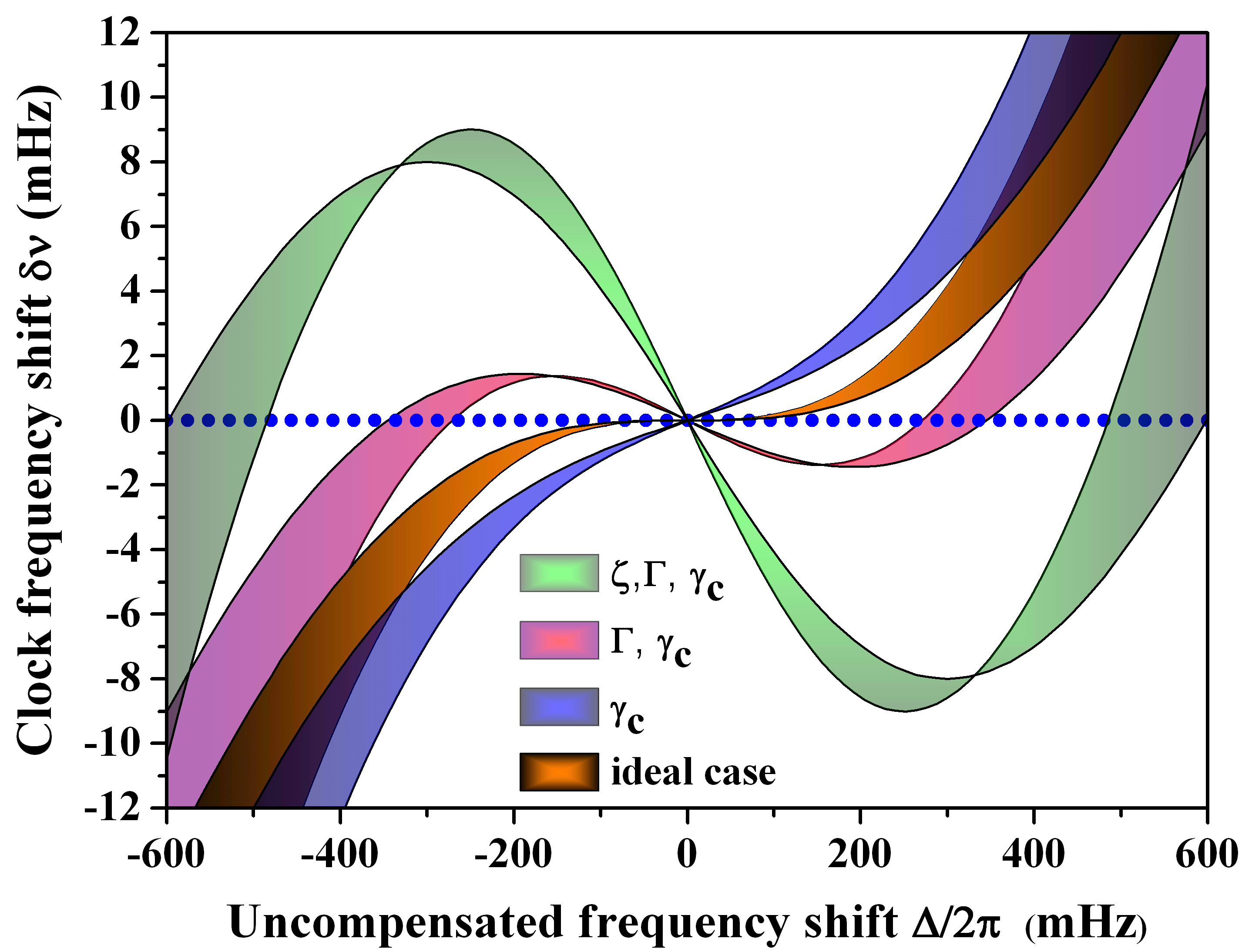}
		}
		\label{fig:IVE-Phase-Modulation-GHR-Combo-Sensitivity}
	}
	\caption{(color online) Error signal line shape distortion versus the clock detuning for HR$-\pi$ and GHR$(\pi/4,3\pi/4)$ (GHR$^{\dagger}(\pi/4,3\pi/4)$) protocols and associated frequency lock point sensitivities under action of decoherence $\gamma_{c}=(\Gamma+\zeta)/2$, relaxation $\Gamma/2\pi=100$~mHz and collisions $\zeta/2\pi=100$~mHz versus uncompensated residual light-shifts. Solid lines are related to the HR-$\pi$ frequency lock point under various dissipative processes that are activated. The dotted line is related to $\Delta\textup{E}_{\Downarrow\Uparrow}$ ($\Delta\textup{E}_{\Downarrow\Downarrow}$) frequency lock points in presence of decoherence and relaxation. Same parameters as in Fig.~\ref{fig:IIB-Composite-Pulses-R-HR-Pi}}
	\label{fig:IVE-Phase-Modulation-GHR-Combo}
\end{figure}

\indent Another stabilization scheme can be generated from the combination of error signals from both GHR($\pi/4$) and GHR($3\pi/4$)
protocols, as suggested by the noticeable antisymmetry of their slopes observed on the right side of Fig.~\ref{fig:IVC-Map2D-Shift-MHR-GHR-b}
and Fig.~\ref{fig:IVC-Map2D-Shift-MHR-GHR-c} in presence of decoherence. This hybrid scheme denoted GHR($\pi/4,3\pi/4$)
was defined in Eq.~\eqref{eq:GHR-Pi-3Pi}.
For this protocol, the error signal and its frequency-shift lock-point condition are written, as of Eq.~\eqref{eq:IIIA-Error-Signal-Lock-Condition}:
\begin{equation}
	\Delta\textup{E}(\pi/4,3\pi/4)|_{\delta=\delta\widetilde{\nu}} = 0,
\end{equation}
with the $\delta\widetilde{\nu}$ frequency-shift derived from Eq.~\eqref{eq:IIIA-Shift-DeltaNu}.

\indent In presence of a large residual offset, this scheme reduces the distortion, associated to the
GHR($\pi/4$) and GHR($3\pi/4$) protocols. In addition the GHR($\pi/4,3\pi/4$) protocol leads to a potential suppression of the decoherence effect as proposed in \cite{Yudin:2016}. However it is still sensitive to the presence of relaxation.

\indent A more recent development on composite pulse protocols in ref.~\cite{Zanon-Willette:2017} has  allowed a
further improvement of the GHR($\pi/4,3\pi/4$) protocol. The protocol can be greatly improved by applying it twice while varying the initial state, which is
now considered as a parameter to the protocol itself. A new error signal, detailed in Fig.~\ref{fig:IVE-Universal-Protocols-a}, is
built in the following way:
\begin{equation}
	\begin{split}
		\Delta\textup{E}_{\Downarrow} &= \Delta\textup{E}(\pi/4,3\pi/4)_{\Downarrow}, \\
		\Delta\textup{E}_{\Uparrow} &= \Delta\textup{E}(\pi/4,3\pi/4)_{\Uparrow}, \\
		\Delta\textup{E}_{\Downarrow\Uparrow} &=
			\frac{1}{2} \left( \Delta\textup{E}_{\Downarrow} - \Delta\textup{E}_{\Uparrow} \right),
	\end{split}
\end{equation}
where the index $\Downarrow$ ($\Uparrow$) means the protocol is applied with population initialization in ground state $|g\rangle\equiv\Downarrow$ (excited state $|e\rangle\equiv\Uparrow$), respectively.

\indent This new combination of pulse sequences produces a completely anti-symmetric signal on the population difference Bloch variable
$\textup{W}$, which directly translates to the transition probability signal. By comparing under identical but now stringent variations of atomic parameters, HR$-\pi$ protocol shown in Fig.~\ref{fig:IVE-Phase-Modulation-GHR-Combo}(a) and this protocol shown in Fig.~\ref{fig:IVE-Phase-Modulation-GHR-Combo}(b), it allows for an exact correction of probe-induced frequency shifts in Fig.~\ref{fig:IVE-Phase-Modulation-GHR-Combo}(c) under arbitrarily large light shift effects or laser intensity variations. It yields an exact frequency lock-point of the central fringe while being immune to both decoherence and relaxation. As remarkable feature, this approach is also independent of the initial atomic states as long as they are distinct from each other. This property removes the need for a high precision quantum state initial preparation.

\indent Another approach is to introduce a time reversal of the GHR($\pi/4,3\pi/4$) protocol. A reciprocal sequence is built
by inverting the pulse order in the composite pulse sequence while also inverting all laser phase signs, the so-called
GHR$^{\dagger}$($\pi/4,3\pi/4$) protocol. It is possible to construct its error signal, detailed in  Fig.~\ref{fig:IVE-Universal-Protocols-b}, as follows:
\begin{equation}
	\Delta\textup{E}_{\Downarrow\Downarrow} =
		\frac{1}{2} \left( \Delta\textup{E}_{\Downarrow} + \Delta\textup{E}_{\Downarrow}^{\dagger} \right),
\end{equation}
where $\Delta\textup{E}_{\Downarrow}^{\dagger}$ stands for the reciprocal sequence's error signal. This protocol offers the same
properties as the previous initial state variation approach with the additional effect of producing an exact anti-symmetric signal for the detection of any Bloch variables. This new feature allows for further detection techniques, for example it is now possible to derive the uncompensated
shift from either $\textup{U}$ or $\textup{V}$ Bloch variables and predict how the central fringe is shifted
from its lock-point before being compensated by applying the error signal.
Those last two protocols were introduced and discussed in heavy details in \cite{Zanon-Willette:2017}.

\section{Implementation in quantum metrology}
\label{sec:VI-Clock-Implementations}

\indent In this last section, we present two different optical clocks based on excitation of a single ion and interrogation of multiple neutral atoms,
respectively, which have used composite pulse sequences to eliminate with a very high efficiency the probe induced
frequency-shift on clock resonances. Composite pulse sequences have been recently implemented experimentally in the single
$^{171}Yb^{+}$ ion clock developed at PTB (Germany)~\cite{Huntemann:2012b,Huntemann:2016} and in the optical 1D lattice clock with
$^{88}Sr$ bosonic atoms at NPL (UK)~\cite{Hobson:2016}.
The single ion clock achieved a reduction of the probe induced frequency-shift by more than four orders of
magnitude using the HR protocol~\cite{Huntemann:2016}. Using a MHR scheme, the optical lattice clock has successfully demonstrated
suppression of a sizable $2\times10^{-13}$ probe Stark shift to below $1\times10^{-16}$ even with very large errors
in shift compensation~\cite{Hobson:2016}.
Both clocks benefit from a nearly field free environment and a strong localization of one single particle within an RF
trap or several atoms within an optical lattice, allowing for a Doppler-recoil free spectroscopy of ultra-narrow
atomic transitions with high accuracy. They are based on very narrow optical transitions, the quadrupole E2 or the octupole E3
transition for the single ion clocks or the $^{1}S_{0}-^{3}P_{0}$ forbidden transition in alkaline-earth atomic systems.
In both systems, systematic, and problematic, clock-frequency shifts originate from the interaction of the quantum two-level
system with the applied laser probe field, and the light-shift produced by off-resonant states not directly addressed by the probe
limits the achievable accuracy to the 10$^{-18}$ relative level.

\subsection{HR protocol with the single trapped $^{171}$Yb$^{+}$ clock}
\label{sec:VA-Clock-Yb}

\begin{figure}[t!!]
	\center
	\resizebox{0.95\linewidth}{!}{
		\includegraphics[angle=0]{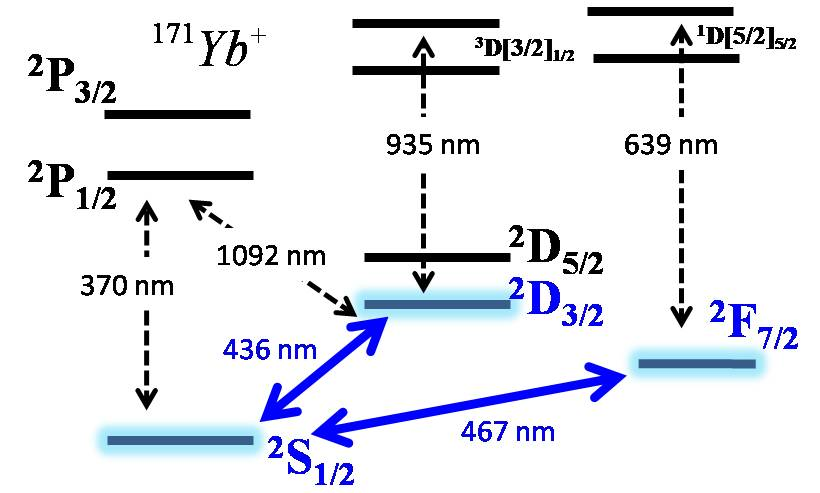}
	}
	\caption{
		Energy level scheme of $^{171}$Yb$^{+}$ ion. wo clock transitions, a weakly allowed electric quadrupole (E2) transition at 436~nm with a natural line-width of a few Hz and a strongly forbidden electric octupole (E3) transition at 467~nm with a natural nHz line-width are accessible.
	}
	\label{fig:VI-levels_Yb_171}
\end{figure}

\indent The ytterbium-ion clock transition is unique among optical frequency standards in that the lowest-lying excited state
is the $^{2}F_{7/2}$ state, which decays to the $^{2}S_{1/2}$ ground state via an electric octupole transition at 467 nm,
see Fig.~\ref{fig:VI-levels_Yb_171}. The
$^{2}F_{7/2}$ state is extremely long-lived, with an estimated lifetime of around 6~years. Hence the natural
line-width of this transition is of order of the nHz, and is not a limit to the performance of the standard. Instead
the stability limit is determined by the probe laser line-width that can be achieved.

\indent The experimental setup is described in Ref.~\citep{Huntemann:2012a, Huntemann:2012b}. A single ion is confined in a radio
frequency Paul trap. During a first period the ion is successively laser-cooled and pumped to the $^{2}S_{1/2}(F=0)$ ground state. The
$^{2}S_{1/2}(F=0) \leftrightarrow \,^2F_{7/2}(F=3, m_F=0)$ clock transition is then probed by applying an HR
sequence, \textit{i. e.}: time
pulse sequence $[\tau,\,T ,\, 2\tau,\, \tau]$, detuning steps
$[\delta+\Delta, \, \delta, \, \delta+\Delta, \, \delta+\Delta]$, phase steps $[\pm\pi/2, \, 0, \, \pi , \, 0]$. The
intensity, detuning and phase of the probe laser beam are precisely shaped by an acousto-optic modulator (AOM). Up to 10 mW of light
power can be focused to a beam waist diameter of 40 $\mu$m at the trap center.
The population of the excited state is known from the decrease of fluorescence at the beginning of the cooling period. Multiple
repetitions of the sequence allow computation of the excitation probability. After the detection, a 760~nm laser
pumps the ion remaining in $^2F_{7/2}$ level again toward an excited level with a lifetime of 29~ns which
predominantly decays to the ground state ~\citep{Huntemann:2012a}.
The pre-compensated light shift value $\Delta_{\mathrm{step}}$ is obtained by measuring the resonance frequency of a Rabi interrogation compared to
an unperturbed transition frequency value previously measured~\citep{Huntemann:2012a}, $\Delta_{\mathrm{step}} \approx 1$ kHz.

\begin{figure}[t!!]
	\center
	\resizebox{\linewidth}{!}{
		\includegraphics[angle=0]{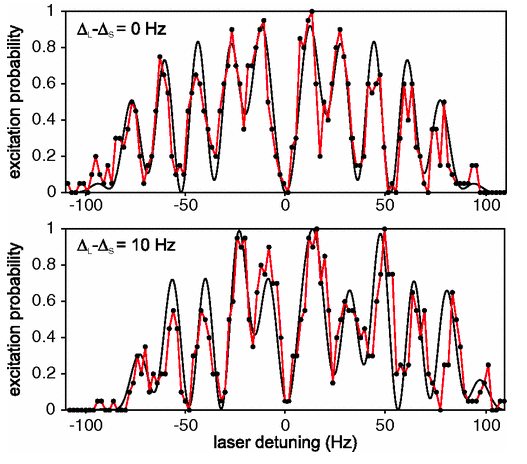}
	}
	\caption{
		Hyper-Ramsey fringe pattern in the single $^{171}$Yb$^{+}$ ion optical clock. Excitation spectra using the HR protocol with
		$\tau= 9$ ms and $T=36$ ms. Top : fully compensated light shift,
		$\Delta_{\textup{step}}=\Delta_\textup{ls}, \, \Delta_\textup{ls}=1090$ Hz. Bottom: partially compensated
		light shift, $\Delta_l=\Delta_\textup{ls}-\Delta_{\textup{step}}= 10$ Hz. The data points are the result of 20
		interrogations at each frequency step. Solid black lines show the calculated line shapes. Reprinted with permission from
		ref.~\citep{Huntemann:2012b}, $\Delta_L,\, \Delta_S$ are the original notations for
		$\Delta_{\textup{ls}},\, \Delta_{\textup{step}}$.
	}
	\label{fig:VA-Optical-Clock-Yb-Signal}
\end{figure}

\indent Experimental records of hyper-Ramsey resonances obtained by Huntemann \textit{et al.}~\citep{Huntemann:2012b} are shown in
Fig.~\ref{fig:VA-Optical-Clock-Yb-Signal} in both cases: fully compensated light shift, $\Delta_{\mathrm{step}}=\Delta_\textup{ls}$,
and partially compensated shift $\Delta_{\mathrm{step}}-\Delta_\textup{ls}= 10$ Hz, for a light shift $\Delta_\textup{ls}=1090$ Hz. Although the
resonance shape is altered when the shift is not fully compensated, it is worthwhile to note that the central minimum is unaffected.

\begin{figure}[t!!]
	\center
	\resizebox{\linewidth}{!}{
		\includegraphics[angle=0]{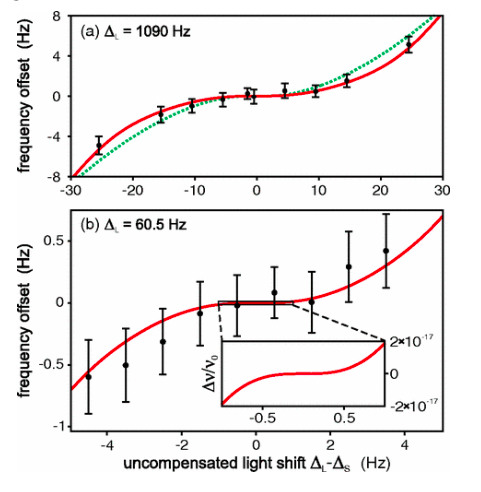}
	}
	\caption{
		Single $^{171}$Yb$^{+}$ ion optical clock. Frequency offset of the probe laser frequency locked on the clock
		transition as a function of the uncompensated shift $\Delta_\textup{l}=\Delta_\textup{ls}-\Delta_\textup{step}$. Solid red line:
		predicted offset when the error signal is obtained by a $\pm \pi/2$ modulation of the first pulse phase. (a) $\tau = 9$ ms,
		$T=36$ ms, $\Delta_\textup{ls}=1090$ Hz. The dashed line shows the position of the central fringe minimum. (b) $\tau = 36$ ms,
		$T=144$ ms, $\Delta_\textup{ls}=60.5$ Hz (linewidth of the central fringe 3.2 Hz). The inset is a zoom showing the offset in
		fractional value. Reprinted with permission from ref.~\citep{Huntemann:2012b}.
	}
	\label{fig:VA-Optical-Clock-Yb}
\end{figure}

\indent Fig.~\ref{fig:VA-Optical-Clock-Yb} shows the clock frequency shift as a function of the error on the frequency step
$\Delta_{\mathrm{step}}-\Delta_\textup{ls}$. A phase modulation $\pm \pi/2$ is applied on the first light pulse to generate an error signal in
order to lock the probe laser frequency. Measurements are performed with two different sets of laser intensity
(light shift), pulse duration, and free evolution time $T$. Experimental results are in very good agreement with the calculated
dependence (solid red line). Note that the shift is smaller for the phase-modulated laser lock than for the minimum of the central
fringe (dashed line in Fig.~\ref{fig:VA-Optical-Clock-Yb}(a)). Authors of Ref.~\citep{Huntemann:2012b}
highlight that the light shift is reduced by four orders of magnitude in the case of Fig.~\ref{fig:VA-Optical-Clock-Yb}(b).
However, the cubic dependence of the shift to the uncompensated light shift is still visible. This hyper-Ramsey
technique has been applied by the same team to cancel the probe-induced light shift with a factional uncertainty of
$1.1\times 10^{-18}$ leading to the outstanding result of a relative systematic uncertainty on the clock frequency of
$3.2 \times 10^{-18}$~\citep{Huntemann:2016}.

\subsection{MHR protocol with the optical lattice $^{88}$Sr clock}
\label{sec:VB-Clock-Sr}

\begin{figure}[t!!]
	\center
	\resizebox{0.95\linewidth}{!}{
		\includegraphics[angle=0]{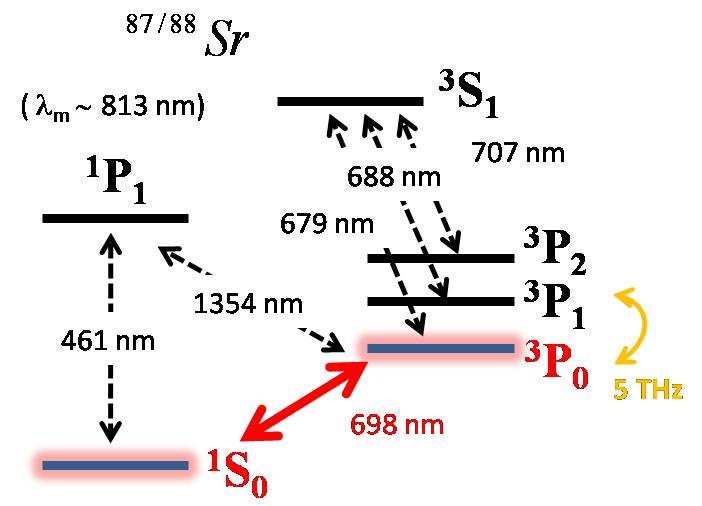}
	}
	\caption{
		Energy level scheme of neutral $^{88}Sr$ atoms. The ultra-narrow forbidden 698~nm bosonic clock transition is activated by a small admixture of the $^{3}$P$_{1}$ with the metastable state $^{3}$P$_{0}$ using a weak static B field (see ref.~\cite{Taichenachev:2006}).
	}
	\label{fig:VI-levels_Sr_88}
\end{figure}

\indent The $^1S_0 \leftrightarrow\, ^3P_0$ transition in alkaline-earth-like atoms, see Fig.~\ref{fig:VI-levels_Sr_88}, is very attractive for optical-lattice based
neutral atom clocks due to its long-lived atomic states. It is a doubly forbidden (spin and angular momentum) transition. The even
isotopes (bosons) have no nuclear spin, thus no hyperfine structure which weakly allows the transition in odd
isotopes by hyperfine level mixing. Nevertheless the strongly forbidden transition can be magnetically induced by
adding a small constant magnetic field weakly mixing the nearby $^3P_1$ state into the $^3P_0$ state~\citep{Taichenachev:2006}. The
transition can then be probed by a single photon excitation.

\indent A modified hyper-Ramsey scheme was implemented on a $^{88}$Sr lattice clock by the team of Gill at NPL~\citep{Hobson:2016}.
$^{88}$Sr atoms, emitted from an atomic beam, are slowed, captured and cooled in two successive magneto-optical
traps before loading a one-dimensional vertically oriented optical lattice operating near the magic wavelength~\cite{Katori:2003}.
The $^1S_0 \leftrightarrow$ $^3P_0$ transition at 698 nm is probed by magnetically induced spectroscopy, using a
2.5 mT mixing magnetic field. Up to 2.7 mW of probe laser can be focused to a waist of about 250 $\mu$m. For this
experiment, the probe light-shift of about 80 Hz is monitored by two Rabi interrogations operating at two different
laser intensity, and interleaved with MHR interrogations. We recall that in the MHR scheme the phase modulation $\pm \pi/2$ in the
first pulse of the HR scheme is replaced by an alternating phase step $\pi/2$ in the first pulse with $-\pi/2$ in
the last pulse (or the opposite) in order to generate the error signal, see Fig.~\ref{fig:IIIC-Phase-Modulation-MHR-GHR-a} and Table~II.

\begin{figure}[t!!]
	\center
	\resizebox{\linewidth}{!}{
		\includegraphics[angle=0]{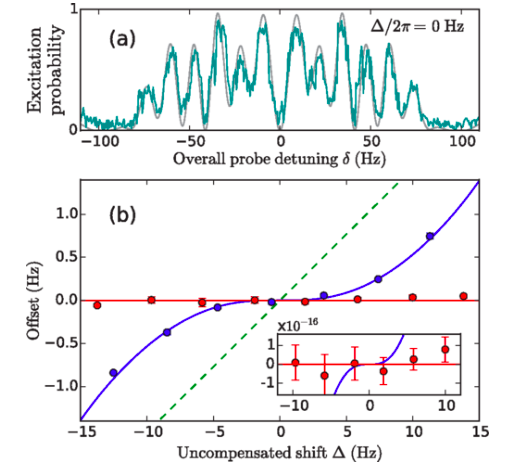}
	}
	\caption{
		The optical lattice clock with neutral $^{88}Sr$ atoms. Top: frequency scan over the HR resonant feature
		for $\tau=10$ ms, $T=50$ ms, $\theta_l=\pi/2$, and $\varphi_l=0$ in the first and last pulse. The theoretical model is
		overlaid in gray with no fitting on the green experimental record. Bottom: modeled and measured residual light-shifts for
		the different HR and MHR protocols. The HR lock (blue dots and line) shows good suppression compared with modified Ramsey
		(dashed green), but MHR is better (red dots and line). Inset: Enlarged view showing the residual lock offset in fractional
		frequency units. Reprinted with permission from ref.~\cite{Hobson:2016}.
	}
	\label{fig:VB-Optical-Clock-Sr}
\end{figure}

\indent The HR signal recorded with no phase step in the first and last pulse is shown in Fig.~\ref{fig:VB-Optical-Clock-Sr}(a).
The line shape is then the same as in Fig.~\ref{fig:VA-Optical-Clock-Yb-Signal}(a), in good agreement with the theoretical model. A
comparison of measured frequency shifts against uncompensated shifts is shown in
Fig.~\ref{fig:VB-Optical-Clock-Sr}(b) for the MHR and HR protocols. Note the very good agreement
 with modeled shifts. The computed shift for a modified Ramsey (MR) scheme is also shown. The MR
technique is the usual two-pulse Ramsey interrogation with a compensating frequency step in each
pulse~\cite{Taichenachev:2009}.
The resulting frequency shift is canceled for a perfect compensated shift, but a linear dependence to the uncompensated shift
remains. The cubic dependence is observed with the HR interrogation, as in Fig.~\ref{fig:VA-Optical-Clock-Yb}.
On the contrary, the shift is fully compensated by the MHR interrogation. The $2 \times 10^{-13}$ light shift does
not bias the locked frequency anymore at the uncertainty level of measurements
($1 \times 10^{-16}$).
 
\section{Conclusion and perspectives}
\label{sec:VI-Conclusion}

\indent We reviewed various spectroscopic probe techniques based on composite pulses, as the Rabi and composite-adjacent NMR-like pulses, and the R, HR, MHR, and GHR  optical clock protocols. These composite having different merits for the clock operation,  are based on an appropriate manipulation of the probe
laser parameters with the aim of generating error signals which offer a very high degree of immunity to light-shift perturbations and laser intensity variations.   Among the different laser parameters a special role is played by the probe laser phase, well controlled in the present optical clock investigations. A useful approach is based on the introduction of phase steps into the laser pulses probing the atomic evolution. The theoretical analysis predicts that error signals including a combination of $\pi/4$ and $3\pi/4$ laser phase-steps generated by using phase-steps modulation and leading to the GHR resonances are able  to suppress the sensitivity to both pulse length error and residual light-shifts induced by the probe laser.  As a key element for the target pulse sequences, the clock frequency-shift associated to a composite pulse interrogation scheme is modeled in order to obtain a strongly non-linear response of the quantum system to probe induced frequency-shifts. \\
\indent The numerical simulation here presented, and the experimental investigations performed so far, demonstrate the very high degree of clock stability reached by implementing these protocols.   For the probe induced perturbations the clock uncertainty  perturbations is reduced below the highest relative accuracy reached at the present, within the $10^{-19}$  range required for the next generation of optical frequency standards. An alternative composite protocol approach  is  based on the exploration of the clock response to pulse sequence whose free evolution times are scaled by integers, and the successive  manipulation of the clock measurements.   The  synthesization of a clock frequency-shift is derived from those measurements with the different free evolution times.\\
\indent The above success of the composite pulse protocols was extended to another question connected to the fast progress in the precision and accuracy of the optical clocks, the limit imposed by a decoherence in the atomic response. This represents an important issue because for experimentalists the decoherence is considered as a hard limit for the final precision. The probe laser frequency instability representing an atomic decoherence has been included into the composite pulse analysis.  In presence of laser induced decoherence, the laser stabilization
schemes belonging to the first class presented in Sec. III do not perform a perfect probe-light-shift suppression anymore. Instead the synthetic protocol provides a very robust error signal against light-shift variation and pulse amplitude
error in presence of decoherence. The control of the decoherence role on the clock precision, i.e., the measurement reliability, is predicted by the Heisenberg principle. Instead the control on the clock accuracy, i.e., the realization of an unperturbed measurement, a surprising result, is connected to the main feature of Ramsey-type protocols: the atomic free-evolution determines the accuracy, if the perturbations of the interrogation phases are eliminated or corrected.\\
 \indent Decoherence, relaxation of the atomic population by spontaneous emission, and weak collisions, are considered also
in the context of universal protocols,
 based on the combined use of GHR($\pi/4,3\pi/4$) schemes, leading to cancellation of the probe induced shift.
It will become important to
evaluate its effects for some clock transitions in the near future. Further analysis will be required to achieve experimental implementation of a protocol immune to both decoherence and
relaxation.\\
\indent Besides the accuracy of an atomic clock, its frequency stability is an important parameter with a strong dependence on the interrogation time. Some phase step protocols presented here are based on a double interrogation of the clock atoms, followed by a manipulation of the double-interrogation results. A straight application of this approach will greatly increase the required
interrogation time. As an alternative, ultracold atom optics techniques may be applied where a single atomic cloud is split  into two (or more components), each of them probed by a different pulse sequence, with an on-line elaboration of the global error signal.\\
\indent The basic ideas of composite pulse approach can be applied to any other measurement based on the interferometric response of an interrogated quantum system. The final accuracy of the interference reading can be greatly increased by inserting a free evolution time as in the Ramsey original scheme, or by using the more elaborated pulse sequence presented in this work. Therefore  new  composite pulse sequences might be considered to develop very sensitive space interferometers for
gravitational waves detection \cite{Vutha:2015,Loeb:2015,Kolkowitz:2016}, or  be extended, for a better control of some systematics, to mass
spectrometry  where Ramsey-type interrogation schemes have been already introduced~\cite{Bollen:1992,Kretzschmar:2007,Eibach:2011}. In quantum computation where the qubit performances are limited by decoherence/relaxation, their control  by proper composite pulse sequences represents an important issue to be explored. Also the qubits error correction schemes could represent an area where the construction of ad-hoc interrogation Hamiltonians may have important applications.

 \section*{Acknowledgments}
This paper is dedicated to the memory of N.F. Ramsey and A. Clairon. The contributions by N.F. Ramsey  to the
whole field of frequency standards for almost 70 years remain the key elements contributing to new evolutions, as those discussed in this work. The work by A. Clairon on
microwave-to-visible frequency standards and measurements, his design and construction of the first Cs fountain (the model of present primary frequency standards) also represents outstanding contribution to the present flourishing of the optical clocks.\\
\indent T. Zanon-Willette deeply acknowledges C. Janssen, B. Darquié, S. Blatt, Y. Té and M. Glass-Maujean for suggestions and a
careful reading of the manuscript. 
\indent V.I. Yudin and A.V. Taichenachev were supported by the Russian
Scientific Foundation (No. 16-12-00052). V.I. Yudin was also
supported by the Ministry of Education and Science of the Russian
Federation (No. 3.1326.2017/4.6), and Russian Foundation for Basic
Research (No. 17-02-00570).

\section*{Appendix}
\label{sec:VII-Appendix}

\subsection{Envelopes and phase for the HR protocol}
\label{sec:VIIA-Envelopes-HR}

Introducing for simplicity of notation $\tilde{\theta}_l=\theta_l/2$, the envelopes and phase are given by
\begin{subequations}
	\begin{align}
		\alpha^{2} &= \left( 1 + \frac{\delta_{1}^{2}}{\omega_{1}^{2}} \tan^2\tilde{\theta}_1 \right)
			\left( 1 + \frac{\delta_{34}^{2}}{\omega_{34}^{2}} \tan^{2}\tilde{\theta}_{34} \right) \notag \\
		&\times \left( \frac{\Omega_3}{\omega_3} \tan\tilde{\theta}_3 + \frac{\Omega_4}{\omega_4} \tan\tilde{\theta}_4 \right)^2
			\cos^2\tilde{\theta}_1 \cos^2\tilde{\theta}_3 \cos^2\tilde{\theta}_4, \\
		\beta &= \frac{
				\frac{\Omega_1}{\omega_1} \tan\tilde{\theta}_1 \left(
					1 - \frac{\delta_3 \delta_4 + \Omega_3 \Omega_4}{\omega_3 \omega_4} \tan\tilde{\theta}_3 \tan\tilde{\theta}_4
				\right)
			}{
				\left( 1 - \frac{\delta_1 \delta_{34}}{\omega_1 \omega_{34}} \tan\tilde{\theta}_1 \tan\tilde{\theta}_{34} \right)
				\left( \frac{\Omega_3}{\omega_3} \tan\tilde{\theta}_3 + \frac{\Omega_4}{\omega_4} \tan\tilde{\theta}_4 \right)
			} \notag \\
		&\times \frac{
				1 - \frac{
					\frac{\delta_3}{\omega_3} \tan\tilde{\theta}_3 + \frac{\delta_4}{\omega_4} \tan\tilde{\theta}_4
				}{
					1 - \frac{\delta_3 \delta_4 + \Omega_3 \Omega_4}{\omega_3 \omega_4} \tan\tilde{\theta}_3 \tan\tilde{\theta}_4
				}
				\frac{
					\frac{\delta_1}{\omega_1} \tan\tilde{\theta}_1 + \frac{\delta_{34}}{\omega_{34}} \tan\tilde{\theta}_{34}
				}{
					1 - \frac{\delta_1 \delta_{34}}{\omega_1 \omega_{34}} \tan\tilde{\theta}_1 \tan\tilde{\theta}_{34}
				}
			}{
				1 + \left(
					\frac{
						\frac{\delta_1}{\omega_1} \tan\tilde{\theta}_1 + \frac{\delta_{34}}{\omega_{34}} \tan\tilde{\theta}_{34}
					}{
						1 - \frac{\delta_1 \delta_{34}}{\omega_1 \omega_{34}} \tan\tilde{\theta}_1 \tan\tilde{\theta}_{34}
					}
				\right)^{2}
			}, \\
	\tan\Phi &= \frac{
		\frac{
			\frac{\delta_3}{\omega_3} \tan\tilde{\theta}_3 + \frac{\delta_4}{\omega_4} \tan\tilde{\theta}_4
		}{
			1 - \frac{\delta_3\delta_4 + \Omega_3\Omega_4}{\omega_3\omega_4} \tan\tilde{\theta}_3 \tan\tilde{\theta}_4
		}
		+ \frac{
			\frac{\delta_1}{\omega_1} \tan\tilde{\theta}_1 + \frac{\delta_{34}}{\omega_{34}} \tan\tilde{\theta}_{34}
		}{
			1 - \frac{\delta_1\delta_{34}}{\omega_1\omega_{34}} \tan\tilde{\theta}_1 \tan\tilde{\theta}_{34}
		}
	}{
		1 - \frac{
			\frac{\delta_3}{\omega_3} \tan\tilde{\theta}_3 + \frac{\delta_4}{\omega_4} \tan\tilde{\theta}_4
		}{
			1 - \frac{\delta_3\delta_4 + \Omega_3\Omega_4}{\omega_3\omega_4} \tan\tilde{\theta}_3 \tan\tilde{\theta}_4
		}
		\frac{
			\frac{\delta_1}{\omega_1} \tan\tilde{\theta}_1 + \frac{\delta_{34}}{\omega_{34}} \tan\tilde{\theta}_{34}
		}{
			1 - \frac{\delta_1\delta_{34}}{\omega_1\omega_{34}} \tan\tilde{\theta}_1 \tan\tilde{\theta}_{34}
		}
	},
	\label{eq:A-tan-phi}
	\end{align}
\end{subequations}
with the reduced notation
\begin{equation}
	\frac{\delta_{34}}{\omega_{34}} \tan\tilde{\theta}_{34} =
		\frac{
			\left( \delta_3\Omega_4 - \Omega_3\delta_4 \right) \tan\tilde{\theta}_{3} \tan\tilde{\theta}_{4}
		}{
			\Omega_{3}\omega_{4} \tan\tilde{\theta}_{3} + \Omega_{4}\omega_{3} \tan\tilde{\theta}_{4}
		}.
\label{eq:reducednotation}
\end{equation}

\subsection{Envelopes and reduced elements for the GHR protocol}
\label{sec:VIIB-Envelopes-GHR}

\indent The $\alpha_{\textup{l}}$ and $\beta_{\textup{l}}$ envelopes of Eq.(\ref{eq:IIIA-Prob-Transition-GHR}) are expressed for
two different pulse sequences as follows:

\indent $\bullet$ For the GHR$(\varphi_{l})$ sequence:
\begin{subequations}
	\begin{align}
		\alpha_{\{\textup{l}\}} &= \left[
				\textup{M}_{+}(\theta_1) c_{\textup{g}}(0) +
				\textup{M}_{\dagger}(\theta_1) e^{i\varphi_1} c_{\textup{e}}(0)
			\right]
			\textup{M}_{\dagger}(\theta_4,\theta_3) \nonumber \\
			&\times \chi(\theta_1,\theta_3,\theta_4), \\
		\beta_{\{\textup{l}\}} e^{i\Phi_{\{\textup{l}\}}}&= \left[
				\frac{
					\textup{M}_{\dagger}(\theta_1) e^{-i\varphi_1} c_{\textup{g}}(0) +
					\textup{M}_{-}(\theta_1) c_{\textup{e}}(0)
				}{
					\textup{M}_{+}(\theta_1) c_{\textup{g}}(0) +
					\textup{M}_{\dagger}(\theta_1) e^{i\varphi_1} c_{\textup{e}}(0)
				}
			\right]
			\cdot \frac{
				\textup{M}_{-}(\theta_3,\theta_4)
			}{
				\textup{M}_{\dagger}(\theta_4,\theta_3)
			},
	\end{align}
\end{subequations}
where we introduce the phase factor
\begin{equation}
	\chi(\theta_1,\theta_3,\theta_4)=\displaystyle\prod_{\substack{l=\textup{1,3,4}}} \exp\left[ -i \delta_{l} \frac{\tau_{l}}{2} \right].
\end{equation}
The reduced matrix components for the GHR$(\varphi_{l})$ scheme are
\begin{subequations}
	\begin{align}
		\textup{M}_{-}(\theta_3,\theta_4)&=
			\textup{M}_{-}(\theta_3) \textup{M}_{-}(\theta_4) +
			\textup{M}_{\dagger}(\theta_3) \textup{M}_{\dagger}(\theta_4)
			e^{i(\varphi_3-\varphi_4)}, \\
		\textup{M}_{\dagger}(\theta_4,\theta_3)&=
			\textup{M}_{\dagger}(\theta_4) \textup{M}_{+}(\theta_3) e^{-i\varphi_4} +
			\textup{M}_{\dagger}(\theta_3) \textup{M}_{-}(\theta_4) e^{-i\varphi_3}.
	\end{align}
\end{subequations}
\indent $\bullet$ For the GHR$^{\dagger}(\varphi_{l})$ sequence:
\begin{subequations}
	\begin{align}
		\alpha_{\{\textup{l}\}}&= \left[
				\textup{M}_{+}(\theta_1,\theta_2) c_{g}(0) +
				\textup{M}_{\dagger}(\theta_2,\theta_1) c_{e}(0)
			\right]
			\textup{M}_{\dagger}(\theta_4) e^{-i\varphi_4} \nonumber \\
			& \times \chi(\theta_1,\theta_2,\theta_4), \\
		\beta_{\{\textup{l}\}} e^{i\Phi_{\{\textup{l}\}}}&= \left[
				\frac{
					\textup{M}_{\dagger}(\theta_1,\theta_2) c_{g}(0) + \textup{M}_{-}(\theta_1,\theta_2) c_{e}(0)
				}{
					\textup{M}_{+}(\theta_1,\theta_2) c_{g}(0) + \textup{M}_{\dagger}(\theta_2,\theta_1) c_{e}(0)
				}
			\right]
			\cdot \frac{
				\textup{M}_{-}(\theta_4)
			}{
				\textup{M}_{\dagger}(\theta_4)
			},
	\end{align}
\end{subequations}
Reduced matrix components for the GHR$^{\dagger}(\varphi_{l})$ scheme are written as:
\begin{subequations}
	\begin{align}
		M_{+}(\theta_1,\theta_2)&=
			M_{+}(\theta_1) M_{+}(\theta_2) +
			M_{\dagger}(\theta_1) M_{\dagger}(\theta_2) e^{-i(\varphi_1-\varphi_2)}, \\
		M_{-}(\theta_1,\theta_2)&=
			M_{-}(\theta_1) M_{-}(\theta_2) +
			M_{\dagger}(\theta_1) M_{\dagger}(\theta_2) e^{i(\varphi_1-\varphi_2)}, \\
		M_{\dagger}(\theta_1,\theta_2)&=
			M_{\dagger}(\theta_2) M_{+}(\theta_1) e^{-i\varphi_2} +
			M_{\dagger}(\theta_1) M_{-}(\theta_2) e^{-i\varphi_1}, \\
		M_{\dagger}(\theta_2,\theta_1)&=
			M_{\dagger}(\theta_1) M_{+}(\theta_2) e^{i\varphi_1} +
			M_{\dagger}(\theta_2) M_{-}(\theta_1) e^{i\varphi_2},
	\end{align}
\end{subequations}

\subsection{Time-dependent components of the rotation matrix}
\label{sec:VIIC-Rotation-Matrix}

\indent The optical Bloch equations of Eq.~\eqref{eq:IVA-Bloch-System} describe the laser field interaction with a two-state quantum system in order to examine the GHR resonance including decoherence and relaxation. The general solution $\textup{M}(\theta_{l})$ in a matrix form including the $\textup{M}_{l}(\infty)$ steady-state is written as \cite{Jaynes:1955,Schoemaker:1978}
\begin{equation}
	\begin{split}
		\textup{M}(\theta_{l}) &= \textup{R}(\theta_{l}) \left[
			\textup{M}_{l}(0) - \textup{M}_{l}(\infty)
		\right] + \textup{M}_{l}(\infty), \\
		\textup{M}_{l}(\infty) &= -\frac{\Gamma}{\mathcal{D}} \left( \begin{array}{ccc}
			\delta_{l} \Omega_{l} \cos\varphi_{l} - \gamma_{c} \Omega_{l} \sin\varphi_{l} \\
			\gamma_{c} \Omega_{l} \cos\varphi_{l} + \delta_{l} \Omega_{l} \sin\varphi_{l} \\
			\gamma_{c}^{2} + \delta_{l}^{2}
		\end{array} \right), \\
		\mathcal{D} &= \gamma_{c} \Omega_{l}^{2} + ( \Gamma + \zeta )( \gamma_{c}^{2} + \delta_{l}^{2} ), \\
	\end{split}
\end{equation}
 the generalized pulse area is $\theta_{l}=\omega_{l}\tau_{l}$.  The evolution
square matrix elements $\textup{R}_{\textup{mn}}(\theta_{l})$ following refs.~\cite{Schoemaker:1978,Berman:2011} are given by:
\begin{equation}
	\begin{split}
		\textup{R}_{11}(\theta_{l}) = e^{-\gamma_{c}\tau_{l}} & \left(
			a_{0} - a_{2} \left[ \delta_{l}^{2} + \Omega_{l}^{2} \sin^{2}\varphi_{l} \right]
		\right), \\
		\textup{R}_{12}(\theta_{l}) = e^{-\gamma_{c}\tau_{l}} & \left(
			a_{1} \delta_{l} + a_{2} \Omega_{l}^{2} \sin\varphi_{l} \cos\varphi_{l}
		\right), \\
		\textup{R}_{13}(\theta_{l}) = e^{-\gamma_{c}\tau_{l}} & \left(
				a_{2} \left[ \delta_{l} \Omega_{l} \cos\varphi_{l} - \Delta\gamma \Omega_{l} \sin\varphi_{l} \right]
			\right. \\
			& \left. - a_{1} \Omega_{l} \sin\varphi_{l} \right), \\
		\textup{R}_{21}(\theta_{l}) = e^{-\gamma_{c}\tau_{l}} & \left(
			-a_{1} \delta_{l} + a_{2} \Omega_{l}^{2} \sin\varphi_{l} \cos\varphi_{l}
		\right), \\
		\textup{R}_{22}(\theta_{l}) = e^{-\gamma_{c}\tau_{l}} & \left(
			a_{0} - a_{2} \left[ \delta_{l}^{2} + \Omega_{l}^{2} \cos^{2}\varphi_{l} \right]
		\right), \\
		\textup{R}_{23}(\theta_{l}) = e^{-\gamma_{c}\tau_{l}} & \left(
				a_{2} \left[ \delta_{l} \Omega_{l} \sin\varphi_{l} + \Delta\gamma \Omega_{l} \cos\varphi_{l} \right]
			\right. \\
			& \left. + a_{1} \Omega_{l} \cos\varphi_{l} \right), \\
		\textup{R}_{31}(\theta_{l}) = e^{-\gamma_{c}\tau_{l}} & \left(
				a_{2} \left[ \delta_{l} \Omega_{l} \cos\varphi_{l} + \Delta\gamma \Omega_{l} \sin\varphi_{l} \right]
			\right. \\
			& \left. + a_{1} \Omega_{l} \sin\varphi_{l} \right), \\
		\textup{R}_{32}(\theta_{l}) = e^{-\gamma_{c}\tau_{l}} & \left(
				a_{2} \left[ \delta_{l} \Omega_{l} \sin\varphi_{l} - \Delta\gamma \Omega_{l} \cos\varphi_{l} \right]
			\right. \\
			& \left. - a_{1} \Omega_{l} \cos\varphi_{l} \right), \\
		\textup{R}_{33}(\theta_{l}) = e^{-\gamma_{c}\tau_{l}} & \left(
			a_{0} + a_{1} \Delta\gamma - a_{2} \left[ \Omega_{l}^{2} - \Delta\gamma^{2} \right]
		\right).
	\end{split}
\end{equation}
Auxiliary time-dependent functions $\textup{a}_{0}\equiv\textup{a}_{0}(\theta_{l}),\textup{a}_{1}\equiv\textup{a}_{1}(\theta_{l}),
\textup{a}_{2}\equiv\textup{a}_{2}(\theta_{l})$ are given by \cite{Torrey:1949,Schoemaker:1978}:
\begin{equation}
	\begin{split}
		a_{0}(\theta_{l}) &= \left[
				\left( \textup{SD}_{3} - \textup{TD}_{2} \right) \sin\theta_{l} +
				\left( \textup{SD}_{2} + \textup{TD}_{3} \right) \cos\theta_{l}
			\right] e^{\rho_{l}\tau_{l}} \\
			&+ \left( \textup{D}_{0} \eta_{l} + \textup{g}_{l}^{2} \right) \textup{R} e^{\eta_{l}\tau_{l}}, \\
		a_{1}(\theta_{l}) &= \left[
				\left( \textup{SD}_{1} - \textup{T} \omega_{l} \right) \sin\theta_{l} +
				\left( \textup{S} \omega_{l} + \textup{TD}_{1} \right) \cos\theta_{l}
			\right] e^{\rho_{l}\tau_{l}} \\
			&+ \textup{D}_{0} \textup{R} e^{\eta_{l}\tau_{l}}, \\
		a_{2}(\theta_{l}) &= \left[	\textup{S} \sin\theta_{l} + \textup{T} \cos\theta_{l} \right] e^{\rho_{l}\tau_{l}} +
			\textup{R} e^{\eta_{l}\tau_{l}},
	\end{split}
\end{equation}
and relations between derivatives as \cite{Schoemaker:1978}:
\begin{equation}
	\begin{split}
		\dot{a}_{0}(\theta_{l}) &= \delta_{l}^{2} \Delta\gamma a_{2}(\theta_{l}), \\
		\dot{a}_{1}(\theta_{l}) &= a_{0}(\theta_{l}) - g_{l}^{2} a_{2}(\theta_{l}), \\
		\dot{a}_{2}(\theta_{l}) &= a_{1}(\theta_{l}) + \Delta\gamma a_{2}(\theta_{l}),
	\end{split}
\end{equation}
Where we introduced the following notations:
\begin{equation}
	\begin{split}
		g_{l}^{2} &= \Omega_{l}^{2} + \delta_{l}^{2}, \\
		\textup{D}_{0} &= \eta_{l} - \Delta\gamma, \\
		\textup{D}_{1} &= \rho_{l} - \Delta\gamma, \\
		\textup{D}_{2} &= \omega_{l} \left( 2\rho_{l} - \Delta\gamma \right), \\
		\textup{D}_{3} &= \rho_{l}^{2} - \omega_{l}^{2} - \rho_{l} \Delta\gamma + g_{l}^{2}, \\
	\end{split}
\end{equation}
and
\begin{equation}
	\begin{split}
		\textup{R} &= \frac{1}{ \left( \rho_{l} - \eta_{l} \right)^{2} + \omega_{l}^{2} }, \\
		\textup{S} &= \frac{
			\left( \rho_{l} - \eta_{l} \right)
		}{
			\omega_{l} \left( \left( \rho_{l} - \eta_{l} \right)^{2} + \omega_{l}^{2} \right)
		}, \\
		\textup{T} &= \frac{-1}{ \left( \rho_{l} - \eta_{l} \right)^{2} + \omega_{l}^{2} }.
	\end{split}
\end{equation}
The three roots of the matrix (one real root $\eta_{l}$ and two complex roots $\rho_{l}\pm i\omega_{l}$) are given by Cardan's
cubic solutions leading to damped oscillations into the atomic response to the clock interrogation sequence as $\eta_{l},\rho_{l}$ and a generalized angular frequency $\omega_{l}$ written as:
\begin{equation}
	\begin{split}
		\eta_{l} &= \frac{1}{3} \left( \Delta\gamma - \textup{C} - \frac{\Delta_{0}}{\textup{C}} \right), \\
		\rho_{l} &= \frac{1}{3} \left( \Delta\gamma + \frac{\textup{C}}{2} + \frac{\Delta_{0}}{2\textup{C}} \right), \\
		\omega_{l} &= \frac{\sqrt{3}}{6} \left( -\textup{C} + \frac{\Delta_{0}}{\textup{C}} \right), \\
		\Delta_{0} &= \Delta\gamma^{2} - 3 g_{l}^{2}, \\
		\Delta_{1} &= -2 \Delta\gamma^{3} + 9 g_{l}^{2} \Delta\gamma - 27 \delta_{l}^{2} \Delta\gamma, \\
		\textup{C} &= \sqrt[3]{ \frac{ \Delta_{1} + \sqrt{ \Delta_{1}^{2} - 4 \Delta_{0}^{3} } }{2} }.
	\end{split}
\end{equation}



\newpage


\begin{thebibliography}{96}
\expandafter\ifx\csname natexlab\endcsname\relax\def\natexlab#1{#1}\fi
\expandafter\ifx\csname bibnamefont\endcsname\relax
  \def\bibnamefont#1{#1}\fi
\expandafter\ifx\csname bibfnamefont\endcsname\relax
  \def\bibfnamefont#1{#1}\fi
\expandafter\ifx\csname citenamefont\endcsname\relax
  \def\citenamefont#1{#1}\fi
\expandafter\ifx\csname url\endcsname\relax
  \def\url#1{\texttt{#1}}\fi
\expandafter\ifx\csname urlprefix\endcsname\relax\def\urlprefix{URL }\fi
\providecommand{\bibinfo}[2]{#2}
\providecommand{\eprint}[2][]{\url{#2}}

\bibitem[{\citenamefont{Rabi et~al.}(1938)\citenamefont{Rabi, Zacharias,
  Millman, and Kusch}}]{Rabi:1938}
\bibinfo{author}{\bibfnamefont{I.~I.} \bibnamefont{Rabi}},
  \bibinfo{author}{\bibfnamefont{J.~R.} \bibnamefont{Zacharias}},
  \bibinfo{author}{\bibfnamefont{S.}~\bibnamefont{Millman}}, \bibnamefont{and}
  \bibinfo{author}{\bibfnamefont{P.}~\bibnamefont{Kusch}},
  \bibinfo{journal}{Phys. Rev.} \textbf{\bibinfo{volume}{53}},
  \bibinfo{pages}{318} (\bibinfo{year}{1938}).

\bibitem[{\citenamefont{Rabi et~al.}(1939)\citenamefont{Rabi, Millman, Kusch,
  and Zacharias}}]{Rabi:1939}
\bibinfo{author}{\bibfnamefont{I.~I.} \bibnamefont{Rabi}},
  \bibinfo{author}{\bibfnamefont{S.}~\bibnamefont{Millman}},
  \bibinfo{author}{\bibfnamefont{P.}~\bibnamefont{Kusch}}, \bibnamefont{and}
  \bibinfo{author}{\bibfnamefont{J.~R.} \bibnamefont{Zacharias}},
  \bibinfo{journal}{Phys. Rev.} \textbf{\bibinfo{volume}{55}},
  \bibinfo{pages}{526} (\bibinfo{year}{1939}).

\bibitem[{\citenamefont{Ramsey}(1950)}]{Ramsey:1950}
\bibinfo{author}{\bibfnamefont{N.}~\bibnamefont{Ramsey}},
  \bibinfo{journal}{Phys. Rev.} \textbf{\bibinfo{volume}{78}},
  \bibinfo{pages}{695} (\bibinfo{year}{1950}).

\bibitem[{\citenamefont{Ramsey}(1956)}]{Ramsey:1956}
\bibinfo{author}{\bibfnamefont{N.~F.} \bibnamefont{Ramsey}},
  \emph{\bibinfo{title}{Molecular beams}} (\bibinfo{publisher}{Clarendon
  Press}, \bibinfo{address}{Oxford}, \bibinfo{year}{1956}).

\bibitem[{\citenamefont{Essen and Parry}(1955)}]{Essen:1955}
\bibinfo{author}{\bibfnamefont{L.}~\bibnamefont{Essen}} \bibnamefont{and}
  \bibinfo{author}{\bibfnamefont{J.~V.~L.} \bibnamefont{Parry}},
  \bibinfo{journal}{Nature} \textbf{\bibinfo{volume}{176}},
  \bibinfo{pages}{280} (\bibinfo{year}{1955}).

\bibitem[{\citenamefont{Ramsey}(1990)}]{Ramsey:1990}
\bibinfo{author}{\bibfnamefont{N.~F.} \bibnamefont{Ramsey}},
  \bibinfo{journal}{Rev. Mod. Phys.} \textbf{\bibinfo{volume}{62}},
  \bibinfo{pages}{541} (\bibinfo{year}{1990}).

\bibitem[{\citenamefont{Vanier and Audoin}(1989)}]{Vanier:1989}
\bibinfo{author}{\bibfnamefont{J.}~\bibnamefont{Vanier}} \bibnamefont{and}
  \bibinfo{author}{\bibfnamefont{C.}~\bibnamefont{Audoin}},
  \emph{\bibinfo{title}{The quantum physics of atomic frequency standards}}
  (\bibinfo{publisher}{Adam Hilger IOP}, \bibinfo{address}{Bristol},
  \bibinfo{year}{1989}).

\bibitem[{\citenamefont{Ludlow et~al.}(2015)\citenamefont{Ludlow, Boyd, Ye,
  Peik, and Schmidt}}]{Ludlow:2015}
\bibinfo{author}{\bibfnamefont{A.~D.} \bibnamefont{Ludlow}},
  \bibinfo{author}{\bibfnamefont{M.~M.} \bibnamefont{Boyd}},
  \bibinfo{author}{\bibfnamefont{J.}~\bibnamefont{Ye}},
  \bibinfo{author}{\bibfnamefont{E.}~\bibnamefont{Peik}}, \bibnamefont{and}
  \bibinfo{author}{\bibfnamefont{P.~O.} \bibnamefont{Schmidt}},
  \bibinfo{journal}{Rev. Mod. Phys.} \textbf{\bibinfo{volume}{87}},
  \bibinfo{pages}{637} (\bibinfo{year}{2015}).

\bibitem[{\citenamefont{Rosenband et~al.}(2008)\citenamefont{Rosenband, Hume,
  Schmidt, Chou, Brusch, Lorini, Oskay, Drullinger, Fortier, Stalnaker
  et~al.}}]{Rosenband:2008}
\bibinfo{author}{\bibfnamefont{T.}~\bibnamefont{Rosenband}},
  \bibinfo{author}{\bibfnamefont{D.~B.} \bibnamefont{Hume}},
  \bibinfo{author}{\bibfnamefont{P.~O.} \bibnamefont{Schmidt}},
  \bibinfo{author}{\bibfnamefont{C.~W.} \bibnamefont{Chou}},
  \bibinfo{author}{\bibfnamefont{A.}~\bibnamefont{Brusch}},
  \bibinfo{author}{\bibfnamefont{L.}~\bibnamefont{Lorini}},
  \bibinfo{author}{\bibfnamefont{W.~H.} \bibnamefont{Oskay}},
  \bibinfo{author}{\bibfnamefont{R.~E.} \bibnamefont{Drullinger}},
  \bibinfo{author}{\bibfnamefont{T.~M.} \bibnamefont{Fortier}},
  \bibinfo{author}{\bibfnamefont{J.~E.} \bibnamefont{Stalnaker}},
  \bibnamefont{et~al.}, \bibinfo{journal}{Science}
  \textbf{\bibinfo{volume}{319}}, \bibinfo{pages}{1808} (\bibinfo{year}{2008}).

\bibitem[{\citenamefont{Margolis}(2009)}]{Margolis:2009}
\bibinfo{author}{\bibfnamefont{H.~S.} \bibnamefont{Margolis}},
  \bibinfo{journal}{Eur. Phys. J. Special Topics}
  \textbf{\bibinfo{volume}{172}}, \bibinfo{pages}{97} (\bibinfo{year}{2009}).

\bibitem[{\citenamefont{Chou et~al.}(2010)\citenamefont{Chou, Hume, Koelemeij,
  Wineland, and Rosenband}}]{Chou:2010-2}
\bibinfo{author}{\bibfnamefont{C.~W.} \bibnamefont{Chou}},
  \bibinfo{author}{\bibfnamefont{D.~B.} \bibnamefont{Hume}},
  \bibinfo{author}{\bibfnamefont{J.~C.~J.} \bibnamefont{Koelemeij}},
  \bibinfo{author}{\bibfnamefont{D.~J.} \bibnamefont{Wineland}},
  \bibnamefont{and}
  \bibinfo{author}{\bibfnamefont{T.}~\bibnamefont{Rosenband}},
  \bibinfo{journal}{Phys. Rev. Lett.} \textbf{\bibinfo{volume}{104}},
  \bibinfo{pages}{070802} (\bibinfo{year}{2010}).

\bibitem[{\citenamefont{Ye et~al.}(2008)\citenamefont{Ye, Kimble, and
  Hidetoshi}}]{Ye:2008}
\bibinfo{author}{\bibfnamefont{J.}~\bibnamefont{Ye}},
  \bibinfo{author}{\bibfnamefont{H.~J.} \bibnamefont{Kimble}},
  \bibnamefont{and}
  \bibinfo{author}{\bibfnamefont{K.}~\bibnamefont{Hidetoshi}},
  \bibinfo{journal}{Science} \textbf{\bibinfo{volume}{27}},
  \bibinfo{pages}{1734} (\bibinfo{year}{2008}).

\bibitem[{\citenamefont{Derevianko and Hidetoshi}(2011)}]{Derevianko:2011}
\bibinfo{author}{\bibfnamefont{A.}~\bibnamefont{Derevianko}} \bibnamefont{and}
  \bibinfo{author}{\bibfnamefont{K.}~\bibnamefont{Hidetoshi}},
  \bibinfo{journal}{Rev. Mod. Phys.} \textbf{\bibinfo{volume}{83}},
  \bibinfo{pages}{331} (\bibinfo{year}{2011}).

\bibitem[{\citenamefont{Katori}(2011)}]{Katori:2011}
\bibinfo{author}{\bibfnamefont{H.}~\bibnamefont{Katori}},
  \bibinfo{journal}{Nat. Photon.} \textbf{\bibinfo{volume}{5}},
  \bibinfo{pages}{203} (\bibinfo{year}{2011}).

\bibitem[{\citenamefont{Levitt}(1982)}]{Levitt:1982-1}
\bibinfo{author}{\bibfnamefont{M.}~\bibnamefont{Levitt}}, \bibinfo{journal}{J.
  Mag. Res.} \textbf{\bibinfo{volume}{48}}, \bibinfo{pages}{234}
  (\bibinfo{year}{1982}).

\bibitem[{\citenamefont{Vandersypen and Chuang}(2005)}]{Vandersypen:2005}
\bibinfo{author}{\bibfnamefont{L.}~\bibnamefont{Vandersypen}} \bibnamefont{and}
  \bibinfo{author}{\bibfnamefont{I.}~\bibnamefont{Chuang}},
  \bibinfo{journal}{Rev. Mod. Phys.} \textbf{\bibinfo{volume}{76}},
  \bibinfo{pages}{1037} (\bibinfo{year}{2005}).

\bibitem[{\citenamefont{Braun and Glaser}(2014)}]{Braun:2014}
\bibinfo{author}{\bibfnamefont{M.}~\bibnamefont{Braun}} \bibnamefont{and}
  \bibinfo{author}{\bibfnamefont{S.~J.} \bibnamefont{Glaser}},
  \bibinfo{journal}{New J. Phys.} \textbf{\bibinfo{volume}{16}},
  \bibinfo{pages}{115002} (\bibinfo{year}{2014}).

\bibitem[{\citenamefont{Taichenachev et~al.}(2009)\citenamefont{Taichenachev,
  Yudin, Oates, Barber, Lemke, Ludlow, Sterr, Lisdat, and
  Riehle}}]{Taichenachev:2009}
\bibinfo{author}{\bibfnamefont{A.}~\bibnamefont{Taichenachev}},
  \bibinfo{author}{\bibfnamefont{V.}~\bibnamefont{Yudin}},
  \bibinfo{author}{\bibfnamefont{C.}~\bibnamefont{Oates}},
  \bibinfo{author}{\bibfnamefont{Z.}~\bibnamefont{Barber}},
  \bibinfo{author}{\bibfnamefont{N.}~\bibnamefont{Lemke}},
  \bibinfo{author}{\bibfnamefont{A.}~\bibnamefont{Ludlow}},
  \bibinfo{author}{\bibfnamefont{U.}~\bibnamefont{Sterr}},
  \bibinfo{author}{\bibfnamefont{C.}~\bibnamefont{Lisdat}}, \bibnamefont{and}
  \bibinfo{author}{\bibfnamefont{F.}~\bibnamefont{Riehle}},
  \bibinfo{journal}{JETP Lett.} \textbf{\bibinfo{volume}{90}},
  \bibinfo{pages}{713} (\bibinfo{year}{2009}).

\bibitem[{\citenamefont{Yudin et~al.}(2010)\citenamefont{Yudin, Taichenachev,
  Oates, Barber, Lemke, Ludlow, Sterr, Lisdat, and Riehle}}]{Yudin:2010}
\bibinfo{author}{\bibfnamefont{V.~I.} \bibnamefont{Yudin}},
  \bibinfo{author}{\bibfnamefont{A.~V.} \bibnamefont{Taichenachev}},
  \bibinfo{author}{\bibfnamefont{C.~W.} \bibnamefont{Oates}},
  \bibinfo{author}{\bibfnamefont{Z.~W.} \bibnamefont{Barber}},
  \bibinfo{author}{\bibfnamefont{N.~D.} \bibnamefont{Lemke}},
  \bibinfo{author}{\bibfnamefont{A.~D.} \bibnamefont{Ludlow}},
  \bibinfo{author}{\bibfnamefont{U.}~\bibnamefont{Sterr}},
  \bibinfo{author}{\bibfnamefont{C.}~\bibnamefont{Lisdat}}, \bibnamefont{and}
  \bibinfo{author}{\bibfnamefont{F.}~\bibnamefont{Riehle}},
  \bibinfo{journal}{Phys. Rev. A} \textbf{\bibinfo{volume}{82}},
  \bibinfo{pages}{011804(R)} (\bibinfo{year}{2010}).

\bibitem[{\citenamefont{Huntemann
  et~al.}(2012{\natexlab{a}})\citenamefont{Huntemann, Lipphardt, Okhapkin,
  Tamm, Peik, Taichenachev, and Yudin}}]{Huntemann:2012b}
\bibinfo{author}{\bibfnamefont{N.}~\bibnamefont{Huntemann}},
  \bibinfo{author}{\bibfnamefont{B.}~\bibnamefont{Lipphardt}},
  \bibinfo{author}{\bibfnamefont{M.}~\bibnamefont{Okhapkin}},
  \bibinfo{author}{\bibfnamefont{C.}~\bibnamefont{Tamm}},
  \bibinfo{author}{\bibfnamefont{E.}~\bibnamefont{Peik}},
  \bibinfo{author}{\bibfnamefont{A.~V.} \bibnamefont{Taichenachev}},
  \bibnamefont{and} \bibinfo{author}{\bibfnamefont{V.}~\bibnamefont{Yudin}},
  \bibinfo{journal}{Phys. Rev. Lett.} \textbf{\bibinfo{volume}{109}},
  \bibinfo{pages}{213002} (\bibinfo{year}{2012}{\natexlab{a}}).

\bibitem[{\citenamefont{Huntemann et~al.}(2016)\citenamefont{Huntemann, Sanner,
  Lipphardt, Tamm, and Peik}}]{Huntemann:2016}
\bibinfo{author}{\bibfnamefont{N.}~\bibnamefont{Huntemann}},
  \bibinfo{author}{\bibfnamefont{C.}~\bibnamefont{Sanner}},
  \bibinfo{author}{\bibfnamefont{B.}~\bibnamefont{Lipphardt}},
  \bibinfo{author}{\bibfnamefont{C.}~\bibnamefont{Tamm}}, \bibnamefont{and}
  \bibinfo{author}{\bibfnamefont{E.}~\bibnamefont{Peik}},
  \bibinfo{journal}{Phys. Rev. Lett.} \textbf{\bibinfo{volume}{116}},
  \bibinfo{pages}{063001} (\bibinfo{year}{2016}).

\bibitem[{\citenamefont{Hobson et~al.}(2016)\citenamefont{Hobson, Bowden, King,
  Baird, Hill, and Gill}}]{Hobson:2016}
\bibinfo{author}{\bibfnamefont{R.}~\bibnamefont{Hobson}},
  \bibinfo{author}{\bibfnamefont{W.}~\bibnamefont{Bowden}},
  \bibinfo{author}{\bibfnamefont{S.~A.} \bibnamefont{King}},
  \bibinfo{author}{\bibfnamefont{P.~E.~G.} \bibnamefont{Baird}},
  \bibinfo{author}{\bibfnamefont{I.~R.} \bibnamefont{Hill}}, \bibnamefont{and}
  \bibinfo{author}{\bibfnamefont{P.}~\bibnamefont{Gill}},
  \bibinfo{journal}{Phys. Rev. A} \textbf{\bibinfo{volume}{93}},
  \bibinfo{pages}{010501(R)} (\bibinfo{year}{2016}).

\bibitem[{\citenamefont{Hinkley et~al.}(2013)\citenamefont{Hinkley, Sherman,
  Phillips, Schioppo, Lemke, Beloy, Pizzocaro, Oates, and
  Ludlow}}]{Hinkley:2013}
\bibinfo{author}{\bibfnamefont{N.}~\bibnamefont{Hinkley}},
  \bibinfo{author}{\bibfnamefont{J.~A.} \bibnamefont{Sherman}},
  \bibinfo{author}{\bibfnamefont{N.~B.} \bibnamefont{Phillips}},
  \bibinfo{author}{\bibfnamefont{M.}~\bibnamefont{Schioppo}},
  \bibinfo{author}{\bibfnamefont{N.~D.} \bibnamefont{Lemke}},
  \bibinfo{author}{\bibfnamefont{K.}~\bibnamefont{Beloy}},
  \bibinfo{author}{\bibfnamefont{M.}~\bibnamefont{Pizzocaro}},
  \bibinfo{author}{\bibfnamefont{C.~W.} \bibnamefont{Oates}}, \bibnamefont{and}
  \bibinfo{author}{\bibfnamefont{A.~D.} \bibnamefont{Ludlow}},
  \bibinfo{journal}{Science} \textbf{\bibinfo{volume}{341}},
  \bibinfo{pages}{1215} (\bibinfo{year}{2013}).

\bibitem[{\citenamefont{Nicholson et~al.}(2015)\citenamefont{Nicholson,
  Campbell, Hutson, Marti, Bloom, McNally, Zhang, Barrett, Safronova, Strouse
  et~al.}}]{Nicholson:2015}
\bibinfo{author}{\bibfnamefont{T.~L.} \bibnamefont{Nicholson}},
  \bibinfo{author}{\bibfnamefont{S.~L.} \bibnamefont{Campbell}},
  \bibinfo{author}{\bibfnamefont{R.~B.} \bibnamefont{Hutson}},
  \bibinfo{author}{\bibfnamefont{G.~E.} \bibnamefont{Marti}},
  \bibinfo{author}{\bibfnamefont{B.~J.} \bibnamefont{Bloom}},
  \bibinfo{author}{\bibfnamefont{R.~L.} \bibnamefont{McNally}},
  \bibinfo{author}{\bibfnamefont{W.}~\bibnamefont{Zhang}},
  \bibinfo{author}{\bibfnamefont{M.~D.} \bibnamefont{Barrett}},
  \bibinfo{author}{\bibfnamefont{M.~S.} \bibnamefont{Safronova}},
  \bibinfo{author}{\bibfnamefont{G.~F.} \bibnamefont{Strouse}},
  \bibnamefont{et~al.}, \bibinfo{journal}{Nature Comm.}
  \textbf{\bibinfo{volume}{6}}, \bibinfo{pages}{7896} (\bibinfo{year}{2015}).

\bibitem[{\citenamefont{Taichenachev et~al.}(2006)\citenamefont{Taichenachev,
  Yudin, Oates, Hoyt, Barber, and Hollberg}}]{Taichenachev:2006}
\bibinfo{author}{\bibfnamefont{A.~V.} \bibnamefont{Taichenachev}},
  \bibinfo{author}{\bibfnamefont{V.~I.} \bibnamefont{Yudin}},
  \bibinfo{author}{\bibfnamefont{C.~W.} \bibnamefont{Oates}},
  \bibinfo{author}{\bibfnamefont{C.~W.} \bibnamefont{Hoyt}},
  \bibinfo{author}{\bibfnamefont{Z.~W.} \bibnamefont{Barber}},
  \bibnamefont{and} \bibinfo{author}{\bibfnamefont{L.}~\bibnamefont{Hollberg}},
  \bibinfo{journal}{Phys. Rev. Lett.} \textbf{\bibinfo{volume}{96}},
  \bibinfo{pages}{083001} (\bibinfo{year}{2006}).

\bibitem[{\citenamefont{Barber et~al.}(2006)\citenamefont{Barber, Hoyt, Oates,
  Hollberg, Taichenachev, and Yudin}}]{Barber:2006}
\bibinfo{author}{\bibfnamefont{Z.~W.} \bibnamefont{Barber}},
  \bibinfo{author}{\bibfnamefont{C.~W.} \bibnamefont{Hoyt}},
  \bibinfo{author}{\bibfnamefont{C.~W.} \bibnamefont{Oates}},
  \bibinfo{author}{\bibfnamefont{L.}~\bibnamefont{Hollberg}},
  \bibinfo{author}{\bibfnamefont{A.~V.} \bibnamefont{Taichenachev}},
  \bibnamefont{and} \bibinfo{author}{\bibfnamefont{V.~I.} \bibnamefont{Yudin}},
  \bibinfo{journal}{Phys. Rev. Lett.} \textbf{\bibinfo{volume}{96}},
  \bibinfo{pages}{083002} (\bibinfo{year}{2006}).

\bibitem[{\citenamefont{Ovsiannikov et~al.}(2007)\citenamefont{Ovsiannikov,
  Pal'chikov, Taichenachev, Yudin, Katori, and Takamoto}}]{Ovsiannikov:2007}
\bibinfo{author}{\bibfnamefont{V.~D.} \bibnamefont{Ovsiannikov}},
  \bibinfo{author}{\bibfnamefont{V.~G.} \bibnamefont{Pal'chikov}},
  \bibinfo{author}{\bibfnamefont{A.~V.} \bibnamefont{Taichenachev}},
  \bibinfo{author}{\bibfnamefont{V.~I.} \bibnamefont{Yudin}},
  \bibinfo{author}{\bibfnamefont{H.}~\bibnamefont{Katori}}, \bibnamefont{and}
  \bibinfo{author}{\bibfnamefont{M.}~\bibnamefont{Takamoto}},
  \bibinfo{journal}{Phys. Rev. A.} \textbf{\bibinfo{volume}{75}},
  \bibinfo{pages}{020501(R)} (\bibinfo{year}{2007}).

\bibitem[{\citenamefont{Santra et~al.}(2005)\citenamefont{Santra, Arimondo,
  Ido, Greene, and Ye}}]{Santra:2005}
\bibinfo{author}{\bibfnamefont{R.}~\bibnamefont{Santra}},
  \bibinfo{author}{\bibfnamefont{E.}~\bibnamefont{Arimondo}},
  \bibinfo{author}{\bibfnamefont{T.}~\bibnamefont{Ido}},
  \bibinfo{author}{\bibfnamefont{C.}~\bibnamefont{Greene}}, \bibnamefont{and}
  \bibinfo{author}{\bibfnamefont{J.}~\bibnamefont{Ye}}, \bibinfo{journal}{Phys.
  Rev. Lett.} \textbf{\bibinfo{volume}{94}}, \bibinfo{pages}{173002}
  (\bibinfo{year}{2005}).

\bibitem[{\citenamefont{Zanon-Willette
  et~al.}(2006)\citenamefont{Zanon-Willette, Ludlow, Blatt, Boyd, Arimondo, and
  Ye}}]{Zanon-Willette:2006}
\bibinfo{author}{\bibfnamefont{T.}~\bibnamefont{Zanon-Willette}},
  \bibinfo{author}{\bibfnamefont{A.~D.} \bibnamefont{Ludlow}},
  \bibinfo{author}{\bibfnamefont{S.}~\bibnamefont{Blatt}},
  \bibinfo{author}{\bibfnamefont{M.~M.} \bibnamefont{Boyd}},
  \bibinfo{author}{\bibfnamefont{E.}~\bibnamefont{Arimondo}}, \bibnamefont{and}
  \bibinfo{author}{\bibfnamefont{J.}~\bibnamefont{Ye}}, \bibinfo{journal}{Phys.
  Rev. Lett.} \textbf{\bibinfo{volume}{97}}, \bibinfo{pages}{233001}
  (\bibinfo{year}{2006}).

\bibitem[{\citenamefont{Zanon-Willette
  et~al.}(2014)\citenamefont{Zanon-Willette, Almonacil, de~Clercq, Ludlow, and
  Arimondo}}]{Zanon-Willette:2014}
\bibinfo{author}{\bibfnamefont{T.}~\bibnamefont{Zanon-Willette}},
  \bibinfo{author}{\bibfnamefont{S.}~\bibnamefont{Almonacil}},
  \bibinfo{author}{\bibfnamefont{E.}~\bibnamefont{de~Clercq}},
  \bibinfo{author}{\bibfnamefont{A.~D.} \bibnamefont{Ludlow}},
  \bibnamefont{and} \bibinfo{author}{\bibfnamefont{E.}~\bibnamefont{Arimondo}},
  \bibinfo{journal}{Phys. Rev. A} \textbf{\bibinfo{volume}{90}},
  \bibinfo{pages}{053427} (\bibinfo{year}{2014}).

\bibitem[{\citenamefont{Stortini and Marzoli}(2005)}]{Stortini:2005}
\bibinfo{author}{\bibfnamefont{S.}~\bibnamefont{Stortini}} \bibnamefont{and}
  \bibinfo{author}{\bibfnamefont{I.}~\bibnamefont{Marzoli}},
  \bibinfo{journal}{Eur. Phys. J. D} \textbf{\bibinfo{volume}{32}},
  \bibinfo{pages}{209} (\bibinfo{year}{2005}).

\bibitem[{\citenamefont{Dunning et~al.}(2014)\citenamefont{Dunning, Gregory,
  Bateman, Cooper, Himsworth, Jones, and Freegarde}}]{Dunning:2014}
\bibinfo{author}{\bibfnamefont{A.}~\bibnamefont{Dunning}},
  \bibinfo{author}{\bibfnamefont{R.}~\bibnamefont{Gregory}},
  \bibinfo{author}{\bibfnamefont{J.}~\bibnamefont{Bateman}},
  \bibinfo{author}{\bibfnamefont{N.}~\bibnamefont{Cooper}},
  \bibinfo{author}{\bibfnamefont{M.}~\bibnamefont{Himsworth}},
  \bibinfo{author}{\bibfnamefont{J.~A.} \bibnamefont{Jones}}, \bibnamefont{and}
  \bibinfo{author}{\bibfnamefont{T.}~\bibnamefont{Freegarde}},
  \bibinfo{journal}{Phys. Rev. A} \textbf{\bibinfo{volume}{90}},
  \bibinfo{pages}{033608} (\bibinfo{year}{2014}).

\bibitem[{\citenamefont{Lin et~al.}(2016)\citenamefont{Lin, Gaebler, Reiter,
  Tan, Bowler, Wan, Keith, Knill, Glancy, Coakley et~al.}}]{LIn:2016}
\bibinfo{author}{\bibfnamefont{Y.}~\bibnamefont{Lin}},
  \bibinfo{author}{\bibfnamefont{J.~P.} \bibnamefont{Gaebler}},
  \bibinfo{author}{\bibfnamefont{F.}~\bibnamefont{Reiter}},
  \bibinfo{author}{\bibfnamefont{T.~R.} \bibnamefont{Tan}},
  \bibinfo{author}{\bibfnamefont{R.}~\bibnamefont{Bowler}},
  \bibinfo{author}{\bibfnamefont{Y.}~\bibnamefont{Wan}},
  \bibinfo{author}{\bibfnamefont{A.}~\bibnamefont{Keith}},
  \bibinfo{author}{\bibfnamefont{E.}~\bibnamefont{Knill}},
  \bibinfo{author}{\bibfnamefont{S.}~\bibnamefont{Glancy}},
  \bibinfo{author}{\bibfnamefont{K.}~\bibnamefont{Coakley}},
  \bibnamefont{et~al.}, \bibinfo{journal}{Phys. Rev. Lett.}
  \textbf{\bibinfo{volume}{117}}, \bibinfo{pages}{140502}
  (\bibinfo{year}{2016}).

\bibitem[{\citenamefont{Vitanov et~al.}(2015)\citenamefont{Vitanov, Gloger,
  Kaufmann, Kaufmann, Collath, Tanveer~Baig, Johanning, and
  Wunderlich}}]{Vitanov:2015}
\bibinfo{author}{\bibfnamefont{N.~V.} \bibnamefont{Vitanov}},
  \bibinfo{author}{\bibfnamefont{T.~F.} \bibnamefont{Gloger}},
  \bibinfo{author}{\bibfnamefont{P.}~\bibnamefont{Kaufmann}},
  \bibinfo{author}{\bibfnamefont{D.}~\bibnamefont{Kaufmann}},
  \bibinfo{author}{\bibfnamefont{T.}~\bibnamefont{Collath}},
  \bibinfo{author}{\bibfnamefont{M.}~\bibnamefont{Tanveer~Baig}},
  \bibinfo{author}{\bibfnamefont{M.}~\bibnamefont{Johanning}},
  \bibnamefont{and}
  \bibinfo{author}{\bibfnamefont{C.}~\bibnamefont{Wunderlich}},
  \bibinfo{journal}{Phys. Rev. A} \textbf{\bibinfo{volume}{91}},
  \bibinfo{pages}{033406} (\bibinfo{year}{2015}).

\bibitem[{\citenamefont{Zanon-Willette
  et~al.}(2015)\citenamefont{Zanon-Willette, Yudin, and
  Taichenachev}}]{Zanon:2015}
\bibinfo{author}{\bibfnamefont{T.}~\bibnamefont{Zanon-Willette}},
  \bibinfo{author}{\bibfnamefont{V.~I.} \bibnamefont{Yudin}}, \bibnamefont{and}
  \bibinfo{author}{\bibfnamefont{A.~V.} \bibnamefont{Taichenachev}},
  \bibinfo{journal}{Phys. Rev. A} \textbf{\bibinfo{volume}{92}},
  \bibinfo{pages}{023416} (\bibinfo{year}{2015}).

\bibitem[{\citenamefont{Zanon-Willette
  et~al.}(2016{\natexlab{a}})\citenamefont{Zanon-Willette, Minissale, Yudin,
  and Taichenachev}}]{Zanon-Willette:2016a}
\bibinfo{author}{\bibfnamefont{T.}~\bibnamefont{Zanon-Willette}},
  \bibinfo{author}{\bibfnamefont{M.}~\bibnamefont{Minissale}},
  \bibinfo{author}{\bibfnamefont{V.~I.} \bibnamefont{Yudin}}, \bibnamefont{and}
  \bibinfo{author}{\bibfnamefont{A.~V.} \bibnamefont{Taichenachev}},
  \bibinfo{journal}{J. Phys.: Conf. Ser.} \textbf{\bibinfo{volume}{723}},
  \bibinfo{pages}{012057} (\bibinfo{year}{2016}{\natexlab{a}}).

\bibitem[{\citenamefont{Zanon-Willette
  et~al.}(2016{\natexlab{b}})\citenamefont{Zanon-Willette, de~Clercq, and
  Arimondo}}]{Zanon-Willette:2016b}
\bibinfo{author}{\bibfnamefont{T.}~\bibnamefont{Zanon-Willette}},
  \bibinfo{author}{\bibfnamefont{E.}~\bibnamefont{de~Clercq}},
  \bibnamefont{and} \bibinfo{author}{\bibfnamefont{E.}~\bibnamefont{Arimondo}},
  \bibinfo{journal}{Phys. Rev. A} \textbf{\bibinfo{volume}{93}},
  \bibinfo{pages}{042506} (\bibinfo{year}{2016}{\natexlab{b}}).

\bibitem[{\citenamefont{Yudin et~al.}(2016)\citenamefont{Yudin, Taichenachev,
  Basalaev, and Zanon-Willette}}]{Yudin:2016}
\bibinfo{author}{\bibfnamefont{V.~I.} \bibnamefont{Yudin}},
  \bibinfo{author}{\bibfnamefont{A.~V.} \bibnamefont{Taichenachev}},
  \bibinfo{author}{\bibfnamefont{M.~Y.} \bibnamefont{Basalaev}},
  \bibnamefont{and}
  \bibinfo{author}{\bibfnamefont{T.}~\bibnamefont{Zanon-Willette}},
  \bibinfo{journal}{Phys. Rev. A} \textbf{\bibinfo{volume}{94}},
  \bibinfo{pages}{052505} (\bibinfo{year}{2016}).

\bibitem[{\citenamefont{Dalibard et~al.}(2011)\citenamefont{Dalibard, Gerbier,
  Juzeli\ifmmode~\bar{u}\else \={u}\fi{}nas, and \"Ohberg}}]{Dalibard:2011}
\bibinfo{author}{\bibfnamefont{J.}~\bibnamefont{Dalibard}},
  \bibinfo{author}{\bibfnamefont{F.}~\bibnamefont{Gerbier}},
  \bibinfo{author}{\bibfnamefont{G.}~\bibnamefont{Juzeli\ifmmode~\bar{u}\else
  \={u}\fi{}nas}}, \bibnamefont{and}
  \bibinfo{author}{\bibfnamefont{P.}~\bibnamefont{\"Ohberg}},
  \bibinfo{journal}{Rev. Mod. Phys.} \textbf{\bibinfo{volume}{83}},
  \bibinfo{pages}{1523} (\bibinfo{year}{2011}).

\bibitem[{\citenamefont{Ramsey and Silsbee}(1951)}]{Ramsey:1951}
\bibinfo{author}{\bibfnamefont{N.~F.} \bibnamefont{Ramsey}} \bibnamefont{and}
  \bibinfo{author}{\bibfnamefont{H.~B.} \bibnamefont{Silsbee}},
  \bibinfo{journal}{Phys. Rev.} \textbf{\bibinfo{volume}{84}},
  \bibinfo{pages}{506} (\bibinfo{year}{1951}).

\bibitem[{\citenamefont{Morinaga et~al.}(1989)\citenamefont{Morinaga, Riehle,
  Ishikawa, and Helmcke}}]{Morinaga:1989}
\bibinfo{author}{\bibfnamefont{A.}~\bibnamefont{Morinaga}},
  \bibinfo{author}{\bibfnamefont{F.}~\bibnamefont{Riehle}},
  \bibinfo{author}{\bibfnamefont{J.}~\bibnamefont{Ishikawa}}, \bibnamefont{and}
  \bibinfo{author}{\bibfnamefont{J.}~\bibnamefont{Helmcke}},
  \bibinfo{journal}{Appl. Phys. B} \textbf{\bibinfo{volume}{48}},
  \bibinfo{pages}{165} (\bibinfo{year}{1989}).

\bibitem[{\citenamefont{Klipstein et~al.}(2001)\citenamefont{Klipstein, Dick,
  Jefferts, and Walls}}]{Klipstein:2001}
\bibinfo{author}{\bibfnamefont{W.~M.} \bibnamefont{Klipstein}},
  \bibinfo{author}{\bibfnamefont{G.~J.} \bibnamefont{Dick}},
  \bibinfo{author}{\bibfnamefont{S.~R.} \bibnamefont{Jefferts}},
  \bibnamefont{and} \bibinfo{author}{\bibfnamefont{F.~L.} \bibnamefont{Walls}},
  \bibinfo{journal}{Proc. 2001 IEEE Int. Freq. Contr. Symp.} pp.
  \bibinfo{pages}{25--32} (\bibinfo{year}{2001}).

\bibitem[{\citenamefont{Letchumanan et~al.}(2004)\citenamefont{Letchumanan,
  Gill, Riis, and Sinclair}}]{Letchumanan:2004}
\bibinfo{author}{\bibfnamefont{V.}~\bibnamefont{Letchumanan}},
  \bibinfo{author}{\bibfnamefont{P.}~\bibnamefont{Gill}},
  \bibinfo{author}{\bibfnamefont{E.}~\bibnamefont{Riis}}, \bibnamefont{and}
  \bibinfo{author}{\bibfnamefont{A.~G.} \bibnamefont{Sinclair}},
  \bibinfo{journal}{Phys. Rev. A} \textbf{\bibinfo{volume}{70}},
  \bibinfo{pages}{033419} (\bibinfo{year}{2004}).

\bibitem[{\citenamefont{Letchumanan et~al.}(2006)\citenamefont{Letchumanan,
  Gill, Sinclair, and Riis}}]{Letchumanan:2006}
\bibinfo{author}{\bibfnamefont{V.}~\bibnamefont{Letchumanan}},
  \bibinfo{author}{\bibfnamefont{P.}~\bibnamefont{Gill}},
  \bibinfo{author}{\bibfnamefont{A.~G.} \bibnamefont{Sinclair}},
  \bibnamefont{and} \bibinfo{author}{\bibfnamefont{E.}~\bibnamefont{Riis}},
  \bibinfo{journal}{J. Opt. Soc. Am. B} \textbf{\bibinfo{volume}{23}},
  \bibinfo{pages}{714} (\bibinfo{year}{2006}).

\bibitem[{\citenamefont{Numata et~al.}(2004)\citenamefont{Numata, Kemery, and
  Camp}}]{Numata:2004}
\bibinfo{author}{\bibfnamefont{K.}~\bibnamefont{Numata}},
  \bibinfo{author}{\bibfnamefont{A.}~\bibnamefont{Kemery}}, \bibnamefont{and}
  \bibinfo{author}{\bibfnamefont{J.}~\bibnamefont{Camp}},
  \bibinfo{journal}{Phys. Rev. Lett.} \textbf{\bibinfo{volume}{93}},
  \bibinfo{pages}{250602} (\bibinfo{year}{2004}).

\bibitem[{\citenamefont{Ludlow et~al.}(2007)\citenamefont{Ludlow, Huang,
  Notcutt, Zanon-Willette, Foreman, Boyd, Blatt, and Ye}}]{Ludlow:2007}
\bibinfo{author}{\bibfnamefont{A.~D.} \bibnamefont{Ludlow}},
  \bibinfo{author}{\bibfnamefont{X.}~\bibnamefont{Huang}},
  \bibinfo{author}{\bibfnamefont{M.}~\bibnamefont{Notcutt}},
  \bibinfo{author}{\bibfnamefont{T.}~\bibnamefont{Zanon-Willette}},
  \bibinfo{author}{\bibfnamefont{S.~M.} \bibnamefont{Foreman}},
  \bibinfo{author}{\bibfnamefont{M.~M.} \bibnamefont{Boyd}},
  \bibinfo{author}{\bibfnamefont{S.}~\bibnamefont{Blatt}}, \bibnamefont{and}
  \bibinfo{author}{\bibfnamefont{J.}~\bibnamefont{Ye}}, \bibinfo{journal}{Opt.
  Lett.} \textbf{\bibinfo{volume}{32}}, \bibinfo{pages}{641}
  (\bibinfo{year}{2007}).

\bibitem[{\citenamefont{Jiang et~al.}(2011)\citenamefont{Jiang, Ludlow, Lemke,
  Fox, Sherman, Ma, and Oates}}]{Jiang:2011}
\bibinfo{author}{\bibfnamefont{Y.~Y.} \bibnamefont{Jiang}},
  \bibinfo{author}{\bibfnamefont{A.~D.} \bibnamefont{Ludlow}},
  \bibinfo{author}{\bibfnamefont{N.~D.} \bibnamefont{Lemke}},
  \bibinfo{author}{\bibfnamefont{R.~W.} \bibnamefont{Fox}},
  \bibinfo{author}{\bibfnamefont{J.~A.} \bibnamefont{Sherman}},
  \bibinfo{author}{\bibfnamefont{L.-S.} \bibnamefont{Ma}}, \bibnamefont{and}
  \bibinfo{author}{\bibfnamefont{C.~W.} \bibnamefont{Oates}},
  \bibinfo{journal}{Nature Photon.} \textbf{\bibinfo{volume}{5}},
  \bibinfo{pages}{158} (\bibinfo{year}{2011}).

\bibitem[{\citenamefont{Kessler et~al.}(2012)\citenamefont{Kessler, Hagemann,
  Grebing, Legero, Sterr, F.Riehle, Martin, Chen, and Ye}}]{Kessler:2012}
\bibinfo{author}{\bibfnamefont{T.}~\bibnamefont{Kessler}},
  \bibinfo{author}{\bibfnamefont{C.}~\bibnamefont{Hagemann}},
  \bibinfo{author}{\bibfnamefont{C.}~\bibnamefont{Grebing}},
  \bibinfo{author}{\bibfnamefont{T.}~\bibnamefont{Legero}},
  \bibinfo{author}{\bibfnamefont{U.}~\bibnamefont{Sterr}},
  \bibinfo{author}{\bibnamefont{F.Riehle}},
  \bibinfo{author}{\bibfnamefont{M.~J.} \bibnamefont{Martin}},
  \bibinfo{author}{\bibfnamefont{L.}~\bibnamefont{Chen}}, \bibnamefont{and}
  \bibinfo{author}{\bibfnamefont{J.}~\bibnamefont{Ye}},
  \bibinfo{journal}{Nature Photon.} \textbf{\bibinfo{volume}{6}},
  \bibinfo{pages}{687} (\bibinfo{year}{2012}).

\bibitem[{\citenamefont{Amairi et~al.}(2013)\citenamefont{Amairi, Legero,
  Kessler, Sterr, W\"ubbena, Mandel, and Schmidt}}]{Amairi:2013}
\bibinfo{author}{\bibfnamefont{S.}~\bibnamefont{Amairi}},
  \bibinfo{author}{\bibfnamefont{T.}~\bibnamefont{Legero}},
  \bibinfo{author}{\bibfnamefont{T.}~\bibnamefont{Kessler}},
  \bibinfo{author}{\bibfnamefont{U.}~\bibnamefont{Sterr}},
  \bibinfo{author}{\bibfnamefont{J.~B.} \bibnamefont{W\"ubbena}},
  \bibinfo{author}{\bibfnamefont{O.}~\bibnamefont{Mandel}}, \bibnamefont{and}
  \bibinfo{author}{\bibfnamefont{P.~O.} \bibnamefont{Schmidt}},
  \bibinfo{journal}{Appl. Phys. B} \textbf{\bibinfo{volume}{113}},
  \bibinfo{pages}{233} (\bibinfo{year}{2013}).

\bibitem[{\citenamefont{Breuer}(2007)}]{Breuer:2007}
\bibinfo{author}{\bibfnamefont{H.}~\bibnamefont{Breuer}},
  \emph{\bibinfo{title}{The Theory of Open Quantum Systems}}
  (\bibinfo{publisher}{Oxford University Press}, \bibinfo{address}{Oxford},
  \bibinfo{year}{2007}).

\bibitem[{\citenamefont{M\"uller et~al.}(2012)\citenamefont{M\"uller, Diehl,
  Pupillo, and Zoller}}]{Mueller:2012}
\bibinfo{author}{\bibfnamefont{M.}~\bibnamefont{M\"uller}},
  \bibinfo{author}{\bibfnamefont{S.}~\bibnamefont{Diehl}},
  \bibinfo{author}{\bibfnamefont{G.}~\bibnamefont{Pupillo}}, \bibnamefont{and}
  \bibinfo{author}{\bibfnamefont{P.}~\bibnamefont{Zoller}},
  \bibinfo{journal}{Adv. Atom. Mol. Opt. Phys.} \textbf{\bibinfo{volume}{61}},
  \bibinfo{pages}{1} (\bibinfo{year}{2012}).

\bibitem[{\citenamefont{Katori et~al.}(2003)\citenamefont{Katori, Takamoto,
  Pal'chikov, and Ovsiannikov}}]{Katori:2003}
\bibinfo{author}{\bibfnamefont{H.}~\bibnamefont{Katori}},
  \bibinfo{author}{\bibfnamefont{M.}~\bibnamefont{Takamoto}},
  \bibinfo{author}{\bibfnamefont{V.~G.} \bibnamefont{Pal'chikov}},
  \bibnamefont{and} \bibinfo{author}{\bibfnamefont{V.~D.}
  \bibnamefont{Ovsiannikov}}, \bibinfo{journal}{Phys. Rev. Lett.}
  \textbf{\bibinfo{volume}{91}}, \bibinfo{pages}{173005}
  (\bibinfo{year}{2003}).

\bibitem[{\citenamefont{Akatsuka et~al.}(2008)\citenamefont{Akatsuka, Takamoto,
  and Katori}}]{Akatsuka:2008}
\bibinfo{author}{\bibfnamefont{T.}~\bibnamefont{Akatsuka}},
  \bibinfo{author}{\bibfnamefont{M.}~\bibnamefont{Takamoto}}, \bibnamefont{and}
  \bibinfo{author}{\bibfnamefont{H.}~\bibnamefont{Katori}},
  \bibinfo{journal}{Nat. Phys.} \textbf{\bibinfo{volume}{4}},
  \bibinfo{pages}{954} (\bibinfo{year}{2008}).

\bibitem[{\citenamefont{Baillard et~al.}(2007)\citenamefont{Baillard, Fouch\'e,
  Targat, Westergaard, Lecallier, Coq, Rovera, Bize, and
  Lemonde}}]{Baillard:2007}
\bibinfo{author}{\bibfnamefont{X.}~\bibnamefont{Baillard}},
  \bibinfo{author}{\bibfnamefont{M.}~\bibnamefont{Fouch\'e}},
  \bibinfo{author}{\bibfnamefont{R.~L.} \bibnamefont{Targat}},
  \bibinfo{author}{\bibfnamefont{P.~G.} \bibnamefont{Westergaard}},
  \bibinfo{author}{\bibfnamefont{A.}~\bibnamefont{Lecallier}},
  \bibinfo{author}{\bibfnamefont{Y.~L.} \bibnamefont{Coq}},
  \bibinfo{author}{\bibfnamefont{G.~D.} \bibnamefont{Rovera}},
  \bibinfo{author}{\bibfnamefont{S.}~\bibnamefont{Bize}}, \bibnamefont{and}
  \bibinfo{author}{\bibfnamefont{P.}~\bibnamefont{Lemonde}},
  \bibinfo{journal}{Opt. Lett.} \textbf{\bibinfo{volume}{32}},
  \bibinfo{pages}{1812} (\bibinfo{year}{2007}).

\bibitem[{\citenamefont{Kulosa et~al.}(2015)\citenamefont{Kulosa, Fim, Zipfel,
  R\"uhmann, Sauer, Jha, Gibble, Ertmer, Rasel, Safronova
  et~al.}}]{Kulosa:2015}
\bibinfo{author}{\bibfnamefont{A.~P.} \bibnamefont{Kulosa}},
  \bibinfo{author}{\bibfnamefont{D.}~\bibnamefont{Fim}},
  \bibinfo{author}{\bibfnamefont{K.~H.} \bibnamefont{Zipfel}},
  \bibinfo{author}{\bibfnamefont{S.}~\bibnamefont{R\"uhmann}},
  \bibinfo{author}{\bibfnamefont{S.}~\bibnamefont{Sauer}},
  \bibinfo{author}{\bibfnamefont{N.}~\bibnamefont{Jha}},
  \bibinfo{author}{\bibfnamefont{K.}~\bibnamefont{Gibble}},
  \bibinfo{author}{\bibfnamefont{W.}~\bibnamefont{Ertmer}},
  \bibinfo{author}{\bibfnamefont{E.~M.} \bibnamefont{Rasel}},
  \bibinfo{author}{\bibfnamefont{M.~S.} \bibnamefont{Safronova}},
  \bibnamefont{et~al.}, \bibinfo{journal}{Phys. Rev. Lett.}
  \textbf{\bibinfo{volume}{115}}, \bibinfo{pages}{240801}
  (\bibinfo{year}{2015}).

\bibitem[{\citenamefont{Cohen-Tannoudji and
  Gu\'ery-Odelin}(2011)}]{Cohen-Tannoudji:2011}
\bibinfo{author}{\bibfnamefont{C.}~\bibnamefont{Cohen-Tannoudji}}
  \bibnamefont{and}
  \bibinfo{author}{\bibfnamefont{D.}~\bibnamefont{Gu\'ery-Odelin}},
  \emph{\bibinfo{title}{Advances in Atomic Physics: an Overview}}
  (\bibinfo{publisher}{World Scientific}, \bibinfo{address}{Singapore},
  \bibinfo{year}{2011}).

\bibitem[{\citenamefont{Bloch and Rabi}(1945)}]{Rabi:1945}
\bibinfo{author}{\bibfnamefont{F.}~\bibnamefont{Bloch}} \bibnamefont{and}
  \bibinfo{author}{\bibfnamefont{I.~I.} \bibnamefont{Rabi}},
  \bibinfo{journal}{Rev. Mod. Phys.} \textbf{\bibinfo{volume}{17}},
  \bibinfo{pages}{237} (\bibinfo{year}{1945}).

\bibitem[{\citenamefont{Rabi et~al.}(1954)\citenamefont{Rabi, Ramsey, and
  Schwinger}}]{Rabi:1954}
\bibinfo{author}{\bibfnamefont{I.~I.} \bibnamefont{Rabi}},
  \bibinfo{author}{\bibfnamefont{N.~F.} \bibnamefont{Ramsey}},
  \bibnamefont{and}
  \bibinfo{author}{\bibfnamefont{J.}~\bibnamefont{Schwinger}},
  \bibinfo{journal}{Rev. Mod. Phys.} \textbf{\bibinfo{volume}{26}},
  \bibinfo{pages}{167} (\bibinfo{year}{1954}).

\bibitem[{\citenamefont{Jaynes}(1955)}]{Jaynes:1955}
\bibinfo{author}{\bibfnamefont{E.~T.} \bibnamefont{Jaynes}},
  \bibinfo{journal}{Phys. Rev.} \textbf{\bibinfo{volume}{98}},
  \bibinfo{pages}{1999} (\bibinfo{year}{1955}).

\bibitem[{\citenamefont{Levitt}(1986)}]{Levitt:1986}
\bibinfo{author}{\bibfnamefont{M.}~\bibnamefont{Levitt}},
  \bibinfo{journal}{Prog. Nucl. Mag. Res. Spect.}
  \textbf{\bibinfo{volume}{18}}, \bibinfo{pages}{61} (\bibinfo{year}{1986}).

\bibitem[{\citenamefont{Kasevich et~al.}(1989)\citenamefont{Kasevich, Erling,
  Chu, and DeVoe}}]{Kasevich:1989}
\bibinfo{author}{\bibfnamefont{M.}~\bibnamefont{Kasevich}},
  \bibinfo{author}{\bibfnamefont{R.}~\bibnamefont{Erling}},
  \bibinfo{author}{\bibfnamefont{S.}~\bibnamefont{Chu}}, \bibnamefont{and}
  \bibinfo{author}{\bibfnamefont{R.~G.} \bibnamefont{DeVoe}},
  \bibinfo{journal}{Phys. Rev. Lett.} \textbf{\bibinfo{volume}{63}},
  \bibinfo{pages}{612} (\bibinfo{year}{1989}).

\bibitem[{\citenamefont{Clairon et~al.}(1991)\citenamefont{Clairon, Salomon,
  Guellati, and Phillips}}]{Clairon:1991}
\bibinfo{author}{\bibfnamefont{A.}~\bibnamefont{Clairon}},
  \bibinfo{author}{\bibfnamefont{C.}~\bibnamefont{Salomon}},
  \bibinfo{author}{\bibfnamefont{S.}~\bibnamefont{Guellati}}, \bibnamefont{and}
  \bibinfo{author}{\bibfnamefont{W.}~\bibnamefont{Phillips}},
  \bibinfo{journal}{Europhys. Lett.} \textbf{\bibinfo{volume}{16}},
  \bibinfo{pages}{165} (\bibinfo{year}{1991}).

\bibitem[{\citenamefont{Campbell and Phillips}(2011)}]{Campbell:2011}
\bibinfo{author}{\bibfnamefont{G.~K.} \bibnamefont{Campbell}} \bibnamefont{and}
  \bibinfo{author}{\bibfnamefont{W.~D.} \bibnamefont{Phillips}},
  \bibinfo{journal}{Phil. Trans. R. Soc. A} \textbf{\bibinfo{volume}{369}},
  \bibinfo{pages}{4078} (\bibinfo{year}{2011}).

\bibitem[{\citenamefont{Shirley}(1963)}]{Shirley:1963}
\bibinfo{author}{\bibfnamefont{J.~H.} \bibnamefont{Shirley}},
  \bibinfo{journal}{J. Appl. Phys.} \textbf{\bibinfo{volume}{34}},
  \bibinfo{pages}{783} (\bibinfo{year}{1963}).

\bibitem[{\citenamefont{Fabjan and Pipkin}(1972)}]{Fabjan:1972}
\bibinfo{author}{\bibfnamefont{C.}~\bibnamefont{Fabjan}} \bibnamefont{and}
  \bibinfo{author}{\bibfnamefont{F.}~\bibnamefont{Pipkin}},
  \bibinfo{journal}{Phys. Rev. A} \textbf{\bibinfo{volume}{6}},
  \bibinfo{pages}{556} (\bibinfo{year}{1972}).

\bibitem[{\citenamefont{Code and Ramsey}(1971)}]{Code:1971}
\bibinfo{author}{\bibfnamefont{R.~F.} \bibnamefont{Code}} \bibnamefont{and}
  \bibinfo{author}{\bibfnamefont{N.~F.} \bibnamefont{Ramsey}},
  \bibinfo{journal}{Phys. Rev. A} \textbf{\bibinfo{volume}{4}},
  \bibinfo{pages}{1945} (\bibinfo{year}{1971}).

\bibitem[{\citenamefont{Greene}(1978)}]{Greene:1978}
\bibinfo{author}{\bibfnamefont{G.~L.} \bibnamefont{Greene}},
  \bibinfo{journal}{Phys. Rev. A} \textbf{\bibinfo{volume}{18}},
  \bibinfo{pages}{1057} (\bibinfo{year}{1978}).

\bibitem[{\citenamefont{Bord\'e}(1983)}]{Borde:1983}
\bibinfo{author}{\bibfnamefont{C.}~\bibnamefont{Bord\'e}},
  \bibinfo{journal}{Advances in Laser Spectroscopy, Edited by F.T. Arecchi, F.
  Strumia and H. Walther, Plenum Publishing Corporation}
  (\bibinfo{year}{1983}).

\bibitem[{\citenamefont{Marrocco et~al.}(1998)\citenamefont{Marrocco,
  Weidinger, Sang, and Walthertitle}}]{Marrocco:1998}
\bibinfo{author}{\bibfnamefont{M.}~\bibnamefont{Marrocco}},
  \bibinfo{author}{\bibfnamefont{M.}~\bibnamefont{Weidinger}},
  \bibinfo{author}{\bibfnamefont{R.~T.} \bibnamefont{Sang}}, \bibnamefont{and}
  \bibinfo{author}{\bibfnamefont{H.}~\bibnamefont{Walthertitle}},
  \bibinfo{journal}{Phys. Rev. Lett.} \textbf{\bibinfo{volume}{81}},
  \bibinfo{pages}{5784} (\bibinfo{year}{1998}).

\bibitem[{\citenamefont{Abramowitz and Stegun}(1968)}]{Abramowitz:1968}
\bibinfo{author}{\bibfnamefont{M.}~\bibnamefont{Abramowitz}} \bibnamefont{and}
  \bibinfo{author}{\bibfnamefont{I.~A.} \bibnamefont{Stegun}},
  \emph{\bibinfo{title}{Handbook of mathematical functions}}
  (\bibinfo{publisher}{National Bureau of Standards Applied Mathematics Series
  - 55}, \bibinfo{address}{Washington}, \bibinfo{year}{1968}).

\bibitem[{\citenamefont{Bando et~al.}(2013)\citenamefont{Bando, Ichikawa,
  Kondo, and Nakahara}}]{Bando:2013}
\bibinfo{author}{\bibfnamefont{M.}~\bibnamefont{Bando}},
  \bibinfo{author}{\bibfnamefont{T.}~\bibnamefont{Ichikawa}},
  \bibinfo{author}{\bibfnamefont{Y.}~\bibnamefont{Kondo}}, \bibnamefont{and}
  \bibinfo{author}{\bibfnamefont{M.}~\bibnamefont{Nakahara}},
  \bibinfo{journal}{J. Phys. Soc. of Japan} \textbf{\bibinfo{volume}{82}},
  \bibinfo{pages}{014004} (\bibinfo{year}{2013}).

\bibitem[{\citenamefont{Shaka and Freeman}(1983)}]{Shaka:1983}
\bibinfo{author}{\bibfnamefont{A.~J.} \bibnamefont{Shaka}} \bibnamefont{and}
  \bibinfo{author}{\bibfnamefont{R.}~\bibnamefont{Freeman}},
  \bibinfo{journal}{J. Mag. Res.} \textbf{\bibinfo{volume}{55}},
  \bibinfo{pages}{487} (\bibinfo{year}{1983}).

\bibitem[{\citenamefont{Riis and Sinclair}(2004)}]{Riis:2004}
\bibinfo{author}{\bibfnamefont{E.}~\bibnamefont{Riis}} \bibnamefont{and}
  \bibinfo{author}{\bibfnamefont{A.}~\bibnamefont{Sinclair}},
  \bibinfo{journal}{J. Phys. B: At. Mol. Opt. Phys.}
  \textbf{\bibinfo{volume}{37}}, \bibinfo{pages}{4719} (\bibinfo{year}{2004}).

\bibitem[{\citenamefont{Kabytayev et~al.}(2014)\citenamefont{Kabytayev, Green,
  Khodjasteh, Biercuk, Viola, and Brown}}]{Kabytayev:2014}
\bibinfo{author}{\bibfnamefont{C.}~\bibnamefont{Kabytayev}},
  \bibinfo{author}{\bibfnamefont{T.~J.} \bibnamefont{Green}},
  \bibinfo{author}{\bibfnamefont{K.}~\bibnamefont{Khodjasteh}},
  \bibinfo{author}{\bibfnamefont{M.~J.} \bibnamefont{Biercuk}},
  \bibinfo{author}{\bibfnamefont{L.}~\bibnamefont{Viola}}, \bibnamefont{and}
  \bibinfo{author}{\bibfnamefont{K.~R.} \bibnamefont{Brown}},
  \bibinfo{journal}{Phys. Rev. A} \textbf{\bibinfo{volume}{90}},
  \bibinfo{pages}{012316} (\bibinfo{year}{2014}).

\bibitem[{\citenamefont{Chen et~al.}(2012)\citenamefont{Chen, Bohnet, Weiner,
  and Thompson}}]{Chen:2012}
\bibinfo{author}{\bibfnamefont{Z.}~\bibnamefont{Chen}},
  \bibinfo{author}{\bibfnamefont{J.~G.} \bibnamefont{Bohnet}},
  \bibinfo{author}{\bibfnamefont{J.~M.} \bibnamefont{Weiner}},
  \bibnamefont{and} \bibinfo{author}{\bibfnamefont{J.~K.}
  \bibnamefont{Thompson}}, \bibinfo{journal}{Phys. Rev. A}
  \textbf{\bibinfo{volume}{86}}, \bibinfo{pages}{032313}
  (\bibinfo{year}{2012}).

\bibitem[{\citenamefont{Yudin et~al.}(2011)\citenamefont{Yudin, Taichenachev,
  Okhapkin, Bagayev, Tamm, Peik, Huntemann, Mehlst$\ddot{\text{a}}$ubler, and
  Riehle}}]{Yudin:2011}
\bibinfo{author}{\bibfnamefont{V.~I.} \bibnamefont{Yudin}},
  \bibinfo{author}{\bibfnamefont{A.~V.} \bibnamefont{Taichenachev}},
  \bibinfo{author}{\bibfnamefont{M.~V.} \bibnamefont{Okhapkin}},
  \bibinfo{author}{\bibfnamefont{S.~N.} \bibnamefont{Bagayev}},
  \bibinfo{author}{\bibfnamefont{C.}~\bibnamefont{Tamm}},
  \bibinfo{author}{\bibfnamefont{E.}~\bibnamefont{Peik}},
  \bibinfo{author}{\bibfnamefont{N.}~\bibnamefont{Huntemann}},
  \bibinfo{author}{\bibfnamefont{T.~E.}
  \bibnamefont{Mehlst$\ddot{\text{a}}$ubler}}, \bibnamefont{and}
  \bibinfo{author}{\bibfnamefont{F.}~\bibnamefont{Riehle}},
  \bibinfo{journal}{Phys. Rev. Lett.} \textbf{\bibinfo{volume}{107}},
  \bibinfo{pages}{030801} (\bibinfo{year}{2011}).

\bibitem[{\citenamefont{Zanon et~al.}(2005)\citenamefont{Zanon, Guerandel,
  de~Clercq, Holleville, Dimarcq, and Clairon}}]{Zanon:2005}
\bibinfo{author}{\bibfnamefont{T.}~\bibnamefont{Zanon}},
  \bibinfo{author}{\bibfnamefont{S.}~\bibnamefont{Guerandel}},
  \bibinfo{author}{\bibfnamefont{E.}~\bibnamefont{de~Clercq}},
  \bibinfo{author}{\bibfnamefont{D.}~\bibnamefont{Holleville}},
  \bibinfo{author}{\bibfnamefont{N.}~\bibnamefont{Dimarcq}}, \bibnamefont{and}
  \bibinfo{author}{\bibfnamefont{A.}~\bibnamefont{Clairon}},
  \bibinfo{journal}{Phys. Rev. Lett.} \textbf{\bibinfo{volume}{94}},
  \bibinfo{pages}{193002} (\bibinfo{year}{2005}).

\bibitem[{\citenamefont{Chen et~al.}(2010)\citenamefont{Chen, Yang, Wang, , and
  Zhan}}]{Chen:2010}
\bibinfo{author}{\bibfnamefont{X.}~\bibnamefont{Chen}},
  \bibinfo{author}{\bibfnamefont{G.-Q.} \bibnamefont{Yang}},
  \bibinfo{author}{\bibfnamefont{M.-S.} \bibnamefont{Wang}}, ,
  \bibnamefont{and} \bibinfo{author}{\bibfnamefont{J.}~\bibnamefont{Zhan}},
  \bibinfo{journal}{Chin. Phys. Lett.} \textbf{\bibinfo{volume}{27}},
  \bibinfo{pages}{113201} (\bibinfo{year}{2010}).

\bibitem[{\citenamefont{Blanshan et~al.}(1991)\citenamefont{Blanshan,
  Rochester, Donley, and Kitching}}]{Blanshan:2015}
\bibinfo{author}{\bibfnamefont{E.}~\bibnamefont{Blanshan}},
  \bibinfo{author}{\bibfnamefont{S.~M.} \bibnamefont{Rochester}},
  \bibinfo{author}{\bibfnamefont{E.~A.} \bibnamefont{Donley}},
  \bibnamefont{and} \bibinfo{author}{\bibfnamefont{J.}~\bibnamefont{Kitching}},
  \bibinfo{journal}{Phys. Rev. A} \textbf{\bibinfo{volume}{91}},
  \bibinfo{pages}{041401(R)} (\bibinfo{year}{1991}).

\bibitem[{\citenamefont{Hemmer et~al.}(1989)\citenamefont{Hemmer, Shahriar,
  Natoli, and Ezekiel}}]{Hemmer:1989}
\bibinfo{author}{\bibfnamefont{P.~R.} \bibnamefont{Hemmer}},
  \bibinfo{author}{\bibfnamefont{M.~S.} \bibnamefont{Shahriar}},
  \bibinfo{author}{\bibfnamefont{V.~D.} \bibnamefont{Natoli}},
  \bibnamefont{and} \bibinfo{author}{\bibfnamefont{S.}~\bibnamefont{Ezekiel}},
  \bibinfo{journal}{J. Opt. Soc. Am. B} \textbf{\bibinfo{volume}{6}},
  \bibinfo{pages}{1519} (\bibinfo{year}{1989}).

\bibitem[{\citenamefont{Micalizio et~al.}(2012)\citenamefont{Micalizio,
  Calosso, Godone, and Levi}}]{Micalizio:2012}
\bibinfo{author}{\bibfnamefont{S.}~\bibnamefont{Micalizio}},
  \bibinfo{author}{\bibfnamefont{C.~E.} \bibnamefont{Calosso}},
  \bibinfo{author}{\bibfnamefont{A.}~\bibnamefont{Godone}}, \bibnamefont{and}
  \bibinfo{author}{\bibfnamefont{F.}~\bibnamefont{Levi}},
  \bibinfo{journal}{Metrologia} \textbf{\bibinfo{volume}{49}},
  \bibinfo{pages}{425} (\bibinfo{year}{2012}).

\bibitem[{\citenamefont{Sanner et~al.}(2017)\citenamefont{Sanner, Huntemann,
  Lange, Tamm, and Peik}}]{Sanner:2017}
\bibinfo{author}{\bibfnamefont{C.}~\bibnamefont{Sanner}},
  \bibinfo{author}{\bibfnamefont{N.}~\bibnamefont{Huntemann}},
  \bibinfo{author}{\bibfnamefont{R.}~\bibnamefont{Lange}},
  \bibinfo{author}{\bibfnamefont{C.}~\bibnamefont{Tamm}}, \bibnamefont{and}
  \bibinfo{author}{\bibfnamefont{E.}~\bibnamefont{Peik}},
  \bibinfo{journal}{arXiv:1707.02630v1}  (\bibinfo{year}{2017}).

\bibitem[{\citenamefont{Tabatchikova et~al.}(2013)\citenamefont{Tabatchikova,
  Taichenachev, and Yudin}}]{Tabatchikova:2013}
\bibinfo{author}{\bibfnamefont{K.~S.} \bibnamefont{Tabatchikova}},
  \bibinfo{author}{\bibfnamefont{A.~V.} \bibnamefont{Taichenachev}},
  \bibnamefont{and} \bibinfo{author}{\bibfnamefont{V.~I.} \bibnamefont{Yudin}},
  \bibinfo{journal}{JETP Lett.} \textbf{\bibinfo{volume}{97}},
  \bibinfo{pages}{311} (\bibinfo{year}{2013}).

\bibitem[{\citenamefont{Tabatchikova et~al.}(2015)\citenamefont{Tabatchikova,
  Taichenachev, Dmitriev, and Yudin}}]{Tabatchikova:2015}
\bibinfo{author}{\bibfnamefont{K.~S.} \bibnamefont{Tabatchikova}},
  \bibinfo{author}{\bibfnamefont{A.~V.} \bibnamefont{Taichenachev}},
  \bibinfo{author}{\bibfnamefont{A.~K.} \bibnamefont{Dmitriev}},
  \bibnamefont{and} \bibinfo{author}{\bibfnamefont{V.~I.} \bibnamefont{Yudin}},
  \bibinfo{journal}{JETP Lett.} \textbf{\bibinfo{volume}{120}},
  \bibinfo{pages}{203} (\bibinfo{year}{2015}).

\bibitem[{\citenamefont{Allen and Eberly}(1975)}]{Allen:1975}
\bibinfo{author}{\bibfnamefont{L.}~\bibnamefont{Allen}} \bibnamefont{and}
  \bibinfo{author}{\bibfnamefont{J.~H.} \bibnamefont{Eberly}},
  \emph{\bibinfo{title}{Optical Resonance and Two-Level Atoms}}
  (\bibinfo{publisher}{John Wiley and Sons, Inc.}, \bibinfo{address}{New York},
  \bibinfo{year}{1975}).

\bibitem[{\citenamefont{Berman and Malinovsky}(1975)}]{Berman:2011}
\bibinfo{author}{\bibfnamefont{P.~R.} \bibnamefont{Berman}} \bibnamefont{and}
  \bibinfo{author}{\bibfnamefont{V.~S.} \bibnamefont{Malinovsky}},
  \emph{\bibinfo{title}{Principles of Laser Spectroscopy and Quantum Optics}}
  (\bibinfo{publisher}{Princeton University Press}, \bibinfo{address}{Princeton
  New Jersey}, \bibinfo{year}{1975}).

\bibitem[{\citenamefont{Zanon-Willette
  et~al.}(2017)\citenamefont{Zanon-Willette, Lefevre, Taichenachev, and
  Yudin}}]{Zanon-Willette:2017}
\bibinfo{author}{\bibfnamefont{T.}~\bibnamefont{Zanon-Willette}},
  \bibinfo{author}{\bibfnamefont{R.}~\bibnamefont{Lefevre}},
  \bibinfo{author}{\bibfnamefont{A.~V.} \bibnamefont{Taichenachev}},
  \bibnamefont{and} \bibinfo{author}{\bibfnamefont{V.~I.} \bibnamefont{Yudin}},
  \bibinfo{journal}{Phys. Rev. A} \textbf{\bibinfo{volume}{96}},
  \bibinfo{pages}{023408} (\bibinfo{year}{2017}).

\bibitem[{\citenamefont{Schoemaker}(1978)}]{Schoemaker:1978}
\bibinfo{author}{\bibfnamefont{R.~L.} \bibnamefont{Schoemaker}},
  \emph{\bibinfo{title}{Coherent transient infrared spectroscopy, in Laser and
  Coherence Spectroscopy}} (\bibinfo{publisher}{Plenum}, \bibinfo{address}{New
  York}, \bibinfo{year}{1978}).

\bibitem[{\citenamefont{Huntemann
  et~al.}(2012{\natexlab{b}})\citenamefont{Huntemann, Okhapkin, Lipphardt,
  Weyers, Tamm, and Peik}}]{Huntemann:2012a}
\bibinfo{author}{\bibfnamefont{N.}~\bibnamefont{Huntemann}},
  \bibinfo{author}{\bibfnamefont{M.}~\bibnamefont{Okhapkin}},
  \bibinfo{author}{\bibfnamefont{B.}~\bibnamefont{Lipphardt}},
  \bibinfo{author}{\bibfnamefont{S.}~\bibnamefont{Weyers}},
  \bibinfo{author}{\bibfnamefont{C.}~\bibnamefont{Tamm}}, \bibnamefont{and}
  \bibinfo{author}{\bibfnamefont{E.}~\bibnamefont{Peik}},
  \bibinfo{journal}{Phys. Rev. Lett.} \textbf{\bibinfo{volume}{108}},
  \bibinfo{pages}{090801} (\bibinfo{year}{2012}{\natexlab{b}}).

\bibitem[{\citenamefont{Vutha}(2015)}]{Vutha:2015}
\bibinfo{author}{\bibfnamefont{A.}~\bibnamefont{Vutha}}, \bibinfo{journal}{New
  J. Phys.} \textbf{\bibinfo{volume}{17}}, \bibinfo{pages}{063030}
  (\bibinfo{year}{2015}).

\bibitem[{\citenamefont{Loeb and Maoz}(2015)}]{Loeb:2015}
\bibinfo{author}{\bibfnamefont{A.}~\bibnamefont{Loeb}} \bibnamefont{and}
  \bibinfo{author}{\bibfnamefont{D.}~\bibnamefont{Maoz}},
  \bibinfo{journal}{arXiv:1501.00996}  (\bibinfo{year}{2015}).

\bibitem[{\citenamefont{Kolkowitz et~al.}(2016)\citenamefont{Kolkowitz,
  Pikovski, Langellier, Lukin, Walsworth, and Ye}}]{Kolkowitz:2016}
\bibinfo{author}{\bibfnamefont{S.}~\bibnamefont{Kolkowitz}},
  \bibinfo{author}{\bibfnamefont{I.}~\bibnamefont{Pikovski}},
  \bibinfo{author}{\bibfnamefont{N.}~\bibnamefont{Langellier}},
  \bibinfo{author}{\bibfnamefont{M.~D.} \bibnamefont{Lukin}},
  \bibinfo{author}{\bibfnamefont{R.~L.} \bibnamefont{Walsworth}},
  \bibnamefont{and} \bibinfo{author}{\bibfnamefont{J.}~\bibnamefont{Ye}},
  \bibinfo{journal}{Phys. Rev. D} \textbf{\bibinfo{volume}{94}},
  \bibinfo{pages}{124043} (\bibinfo{year}{2016}).

\bibitem[{\citenamefont{Bollen et~al.}(1992)\citenamefont{Bollen, Kluge, Otto,
  Savard, and Stolzenberg}}]{Bollen:1992}
\bibinfo{author}{\bibfnamefont{G.}~\bibnamefont{Bollen}},
  \bibinfo{author}{\bibfnamefont{H.-J.} \bibnamefont{Kluge}},
  \bibinfo{author}{\bibfnamefont{T.}~\bibnamefont{Otto}},
  \bibinfo{author}{\bibfnamefont{G.}~\bibnamefont{Savard}}, \bibnamefont{and}
  \bibinfo{author}{\bibfnamefont{H.}~\bibnamefont{Stolzenberg}},
  \bibinfo{journal}{Nucl. Instrum. Meth.} \textbf{\bibinfo{volume}{B70}},
  \bibinfo{pages}{490} (\bibinfo{year}{1992}).

\bibitem[{\citenamefont{Kretzschmar}(2007)}]{Kretzschmar:2007}
\bibinfo{author}{\bibfnamefont{M.}~\bibnamefont{Kretzschmar}},
  \bibinfo{journal}{Int. J. Mass Spectrom.} \textbf{\bibinfo{volume}{264}},
  \bibinfo{pages}{122} (\bibinfo{year}{2007}).

\bibitem[{\citenamefont{Eibach et~al.}(2011)\citenamefont{Eibach, Beyer, Blaum,
  Block, Eberhardt, Herfurth, Ketelaer, Nagy, Neidherr, N\"ortersh\"auser
  et~al.}}]{Eibach:2011}
\bibinfo{author}{\bibfnamefont{M.}~\bibnamefont{Eibach}},
  \bibinfo{author}{\bibfnamefont{T.}~\bibnamefont{Beyer}},
  \bibinfo{author}{\bibfnamefont{K.}~\bibnamefont{Blaum}},
  \bibinfo{author}{\bibfnamefont{M.}~\bibnamefont{Block}},
  \bibinfo{author}{\bibfnamefont{K.}~\bibnamefont{Eberhardt}},
  \bibinfo{author}{\bibfnamefont{F.}~\bibnamefont{Herfurth}},
  \bibinfo{author}{\bibfnamefont{J.}~\bibnamefont{Ketelaer}},
  \bibinfo{author}{\bibfnamefont{S.}~\bibnamefont{Nagy}},
  \bibinfo{author}{\bibfnamefont{D.}~\bibnamefont{Neidherr}},
  \bibinfo{author}{\bibfnamefont{W.}~\bibnamefont{N\"ortersh\"auser}},
  \bibnamefont{et~al.}, \bibinfo{journal}{Int. J. Mass Spectrom.}
  \textbf{\bibinfo{volume}{303}}, \bibinfo{pages}{27} (\bibinfo{year}{2011}).

\bibitem[{\citenamefont{Torrey}(1949)}]{Torrey:1949}
\bibinfo{author}{\bibfnamefont{H.~C.} \bibnamefont{Torrey}},
  \bibinfo{journal}{Phys. Rev.} \textbf{\bibinfo{volume}{76}},
  \bibinfo{pages}{1059} (\bibinfo{year}{1949}).

\end{thebibliography}
\end{document}